\documentclass[12pt]{article}
\usepackage{epsfig}
\usepackage{hhline}
\usepackage{amsmath,amssymb}
\usepackage{boxedminipage}

\begin{document}

\newtheorem{theo}{Theorem}[section]
\renewcommand{\thetheo}{\thesection.\arabic{theo}}
\newtheorem{lemm}[theo]{Lemma}
\renewcommand{\thelemm}{\thesection.\arabic{lemm}}
\newtheorem{coro}{Corollary}[theo]
\renewcommand{\thecoro}{\thetheo.\arabic{coro}}
\newtheorem{obse}[theo]{Observation}
\newtheorem{prop}[theo]{Proposition}
\newtheorem{defi}[theo]{Definition}
\renewcommand{\thedefi}{\arabic {defi}.}
\newtheorem{rema}[theo]{Remark}
\newtheorem{exam}[theo]{Example}
\newtheorem{conj}[theo]{Conjecture}
\newtheorem{claim}{Claim}[theo]
\renewcommand{\theclaim}{\thetheo.\arabic{claim}}
\newcommand{\proofmark}{\ \rule{2.3mm}{2.3mm} \smallskip}
\renewcommand{\topfraction}{.9}
\renewcommand{\bottomfraction}{.9}
\renewcommand{\theequation}{\thesection.\arabic{equation}}
\title{On the Sum-of-Squares Algorithm for Bin Packing}
\author{\sc Janos Csirik
        \thanks{
        {\tt  csirik@inf.u-szeged.hu}.
        Dept. of Computer Sciences,
        University of Szeged, Szeged, Hungary.
        }
        \and
        \sc David S. Johnson
        \thanks{
        {\tt dsj@research.att.com}.
        AT\&T Shannon Labs, Room C239, 180 Park Avenue,
        Florham Park, NJ 07932, USA.
        }
        \and
        \sc Claire Kenyon
        \thanks{
        {\tt Claire.Kenyon@lri.fr}.
        Laboratoire de Recherche en Informatique,
        B\^atiment 490, Universit\'e Paris-Sud, 91405 Orsay Cedex, France.
        }
	\and
	\sc James B. Orlin
	\thanks{
	{\tt jorlin@mit.edu}.
	Sloan School of Management,
	Massachusetts Institute of Technology,
	Cambridge, MA, 02139, USA
	}
        \and
        \sc Peter W. Shor
        \thanks{
        {\tt shor@research.att.com}.
        AT\&T Shannon Labs, Room C239, 180 Park Avenue,
        Florham Park, NJ 07932, USA.
        }
        \and
        \sc Richard R. Weber
        \thanks{
        {\tt rrw1@cam.ac.uk}.
        Statistical Laboratory, University of Cambridge,
        Cambridge CB3 0WB, England.
        }
}

\maketitle

\begin{abstract}
In this paper we present a theoretical analysis of the deterministic
on-line {\em Sum of Squares}
algorithm ($SS$) for bin packing
introduced and studied experimentally in \cite{CJK99},
along with several new variants.
$SS$ is applicable to any instance
of bin packing in which the bin capacity $B$ and item sizes $s(a)$ are integral
(or can be scaled to be so), and runs in time $O(nB)$.
It performs remarkably well from an average case point of view:
For any discrete distribution in which the optimal expected waste
is sublinear, $SS$ also has sublinear expected waste.
For any discrete distribution where the optimal expected waste is
bounded, $SS$ has expected waste at most $O(\log n)$.
In addition, we discuss several interesting variants on $SS$,
including a randomized $O(nB\log B)$-time on-line algorithm
$SS^*$, based on $SS$, whose expected behavior is essentially optimal
for all discrete distributions.
Algorithm $SS^*$ also depends on a new linear-programming-based
pseudopolynomial-time algorithm for solving the NP-hard problem of
determining, given a discrete distribution $F$, just what is the
growth rate for the optimal expected waste.
This article is a greatly expanded version of the conference paper
\cite{sumsq2000}.
\end{abstract}

\pagebreak

\section{Introduction}
\setcounter{equation}{0}
\setcounter{page}{2}

\noindent
In the classical one-dimensional bin packing problem, we are given a
list $L = (a_1, ... , a_n)$ of items, a bin capacity $B$, and a size
$s(a_i) \in (0,B]$ for each item in the list.  We wish to pack the
items into a minimum number of bins of capacity $B$, i.e., to partition
the items into a minimum number of subsets such that the sum of the
sizes of the items in each subset is $B$ or less.
Many potential applications, such as packing small
information packets into somewhat larger fixed-size ones,
involve integer item sizes, fixed and relatively small values of $B$,
and large values of $n$.

The bin packing problem is NP-hard, so research has concentrated on
the design and analysis of polynomial-time approximation algorithms for
it, i.e., algorithms that construct packings that use relatively few
bins, although not necessarily the smallest possible number.  Of
special interest have been \emph{on-line} algorithms, i.e., ones that
must permanently assign each item in turn to a bin without knowing
anything about the sizes or numbers of additional items, a requirement
in many applications.
In this paper we shall analyze the {\em Sum of Squares}
algorithm, an on-line bin packing algorithm recently
introduced in \cite{CJK99} that is applicable to
any instance whose item sizes are integral (or can be scaled to be so),
and is surprisingly effective.

\subsection{Notation and Definitions}

\noindent
Let $P$ be a packing of list $L$
and for $0 \leq h \leq B$ let $N_P(h)$ be the number of partially-filled
bins in $P$ whose contents have total size equal to $h$.
We shall say that such a bin has {\em level} $h$.
Note that by definition $N_P(0) = N_P(B) = 0$.
We call the vector $\langle N_P(1),N_P(2),\ldots,N_P(B-1)\rangle$
the {\em profile} of packing $P$.

\begin{defi}
The \emph{sum of squares} $ss(P)$ for packing $P$ is $\sum_{h=1}^{B-1}
N_P(h)^2$.
\end{defi}

The \emph{Sum-of-Squares Algorithm} ($SS$) introduced in \cite{CJK99}
is an on-line algorithm that packs each item according to the following
simple rule: Let $a$ be the next item to be packed and let $P$ be the
current packing.  A \emph{legal} bin for $a$ is one that is either
empty or has current level no more than $B - s(a)$.
Place $a$ into a legal bin so as to yield the minimum possible value
of $ss(P')$ for the resulting packing $P'$, with ties broken in favor of
the highest level, and then in favor of the newest bin with that level.
(Our results for $SS$ hold for any choice of
the tie-breaking rule, but it is useful to have a
completely specified version of the algorithm.)

Note that in deciding where to place an item of size $s$ under $SS$,
the explicit calculation of $ss(P)$ is not required, a consequence of
the following lemma.

\begin{lemm}\label{deltalemm}
Suppose an item of size $s$ is added to a bin of level $h$ of packing $P$,
thus creating packing $P'$, and that $N_P(h+s) - N_P(h) = d$.
Then
$$
ss(P') - ss(P) ~=~ \left\{ \begin{array}{ll}
2d + 1, & \mbox{if $h = 0$ or $h=B-s$}\\
2d + 2, & \mbox{otherwise}
\end{array} \right.
$$
\end{lemm}

\noindent
{\bf Proof.}
Straightforward calculation using the facts that $d = N_P(h+s)$
when $h=0$ and $d = -N_P(h)$ when $h = B-s$.  \proofmark

Thus to find the placement that causes the least increase in $ss(P)$ 
one simply needs to find that $i$ with $N_P(i)\neq 0$ that minimizes
$N_P(i+s) - N_P(i)$, $0 \leq i \leq B-s$ under the convention that $N_P(0)$
and $N_P(B)$ are re-defined to be $1/2$ and $-1/2$ respectively.
We currently know of no significantly more efficient way to do this
in general than to try all possibilities, so the running time for $SS$ is $O(nB)$
overall.

In what follows, we will
be interested in the following three measures of $L$ and $P$.

\begin{defi}
The \emph{size} $s(L)$ of a list $L$ is the sum of the sizes of all
the items in $L$.
\end{defi}

\begin{defi}
The \emph{length} $|P|$ of a packing $P$ is the number of nonempty bins in $P$.
\end{defi}

\begin{defi}
The \emph{waste} $W(P)$ of packing $P$ is $\sum_{h=1}^{B-1}N_P(h)\cdot
\frac{B-h}{B} = |P| - s(L)/B$.
\end{defi}

\noindent
Note that these quantities are related since
$|P| \geq s(L)/B$ and hence $W(P) \geq 0$.

\medskip
We are in particular interested in the average-case behavior of $SS$
for discrete distributions.  A \emph{discrete distribution} $F$
consists of a bin size $B \in {\mathbb{Z}}^+$, a sequence of positive integral
sizes $s_1 < s_2 < \cdots < s_J \leq B$, and an associated vector
$\bar{p}_F = \langle p_1, p_2, \ldots, p_J\rangle$ of nonnegative rational
probabilities such that $\sum_{j=1}^J p_j = 1$.
(Allowing for the possibility that some $p_j$'s are 0 will be notationally useful
later in the paper.)
In a list generated according to this distribution, the $i$th
item $a_i$ has size $s(a_i) = s_j$ with probability $p_j$,
independently for each $i \geq 1$.
We consider two key measures of average-case algorithmic performance.
For any discrete distribution $F$ and any algorithm $A$, let
$P_n^A(F)$ be the packing resulting from applying $A$ to a random list
$L_n(F)$ of $n$ items generated according to $F$.
Let $OPT$ denote an algorithm that always produces an optimal packing.
We then have

\begin{defi}
The \emph{expected waste rate} for algorithm $A$ and distribution $F$ is
\[
EW_n^A (F) \equiv E \left [ W \left ( P_n^A(F) \right ) \right ]\,.
\]
\end{defi}

\begin{defi}
The \emph{asymptotic expected performance ratio} for $A$ and $F$ is
\[
ER_\infty^A (F) \equiv \limsup_{n \rightarrow \infty} \left ( E
\left [ \frac{\left | P_n^A(F) \right |} {\left | P_n^{OPT}(F) \right
|} \right ] \right )\,.
\]
\end{defi}

\subsection{Our results}

\noindent
Let us say that a distribution $F$ is \emph{perfectly packable} if
$EW_n^{OPT}(F) = o(n)$ (in which case almost all of the bins in
an optimal packing are perfectly packed).
By a result of Courcoubetis and Weber
\cite{CW90} that we shall describe in more detail later, the possible
growth rates for $EW_n^{OPT}(F)$ when $F$ is perfectly packable are quite
restricted: the only possibilities are $\Theta\left(\sqrt{n}\right)$ and $O(1)$.
In the latter case we say $F$ is not only perfectly packable
but is also a {\em bounded waste} distribution.
In this paper, we shall present the following results.

\begin{enumerate}

\item \label{r1} For any perfectly packable distribution $F$, the
Sum-of-Squares algorithm is almost perfect:
$EW_n^{SS}(F) = O(\sqrt{n})$ [Theorem \ref{orlintheo}].

\item \label{r1a} If $F$ is a bounded waste distribution, then
$EW_n^{SS}(F)$ is either $O(1)$ or $\Theta(\log n)$ and there is a
simple combinatorial property that $F$ must satisfy for the first
case to hold
[Theorems \ref{boundedtheo} and \ref{lognupperbound}].
In particular, $EW_n^{SS}(F) = O(1)$ for the discrete uniform
distributions $U\{j,k\}$, $j\leq k-1$, of
\cite{AM98,CCG91,CCG98,CJS93,KRS98}, which are
the main discrete distributions studied to date.

\item \label{r1c} There is a simple $O(nB)$-time deterministic
variant $SS'$ on $SS$ that has bounded expected waste for all bounded waste
distributions and $O(\sqrt{n})$ waste for all perfectly packable distributions
[Theorem \ref{sprimecor}].

\item \label{r2} There is a linear-programming (LP) based approach that,
in time polynomial
in $B$ and the number of bits required to describe the probability vector
$\bar{p}_F$, determines whether $F$ is perfectly packable.
If so, it determines whether $F$ is also a bounded waste distribution.
If not, it computes the value of $\limsup_{n\rightarrow\infty}(EW_n^{OPT}/n)$
[Theorems \ref{EWtheo}, \ref{limsuptheo}, and \ref{lptheo2}].
Note that since the running time is polynomial in $B$ rather than in $\log B$,
the algorithm technically runs in pseudopolynomial time.
We cannot hope for a polynomial time algorithm unless P = NP since
the problem solved is NP-hard \cite{CCG98}.
Moreover, all previous LP-based approaches took time
{\em exponential} in $B$.

\item \label{r1d} For the case where $F$ is not perfectly packable,
there are lower bound examples and upper bound theorems showing that
$1.5 \leq \max_F ER_\infty^{SS}(F) \leq 3$,
and that for all lists $L$, we have $SS(L) \leq 3OPT(L)$, where $A(L)$ is the
number of bins used when algorithm $A$ is applied to list $L$
[Theorems \ref{1.5theo} and \ref{worstcaseUB}].

\item \label{r3} For any fixed $F$, there is a randomized
$O(nB)$-time on-line algorithm $SS_F$ such that
$EW_n^{SS_F}(F) \leq EW_n^{OPT}(F) + O(\sqrt{n})$
and hence $ER_{\infty}^{SS_F}(F) = 1$.  
Algorithm $SS_F$ is based on $SS$ and, given $F$, can be constructed
using the algorithm of (\ref{r2}) above
[Theorem \ref{tunedtheo1}].

\item \label{r4} There is a
randomized $O(nB)$-time on-line algorithm $SS^*$ that for
any $F$ with bin capacity $B$ has
$EW_n^{SS^*}(F) = \Theta(EW_n^{OPT}(F))$ and also
$EW_n^{SS^*}(F) \leq EW_n^{OPT}(F) + O(n^{1/2})$, the latter implying
that $ER_{\infty}^{SS^*}(F) = 1$.
This algorithm works by learning the distribution and using the algorithms
of (\ref{r2}) and (\ref{r3})
[Theorem \ref{learningtheo}].

\item $SS$ can maintain its good behavior even in
the face of a non-oblivious adversary who gets to choose the item size
distribution at each step (subject to appropriate restrictions)
[Theorems \ref{Padversary} and \ref{Badversary}].

\item The good average case behavior of $SS$ is at least partially
preserved under many (but not all)
natural variations on its sum-of-squares objective function and
the accuracy with which it is updated.
Moreover, there is a variant of $SS$
that runs in time $O(n \log B)$ instead of $\Theta(nB)$
and has the same qualitative
behavior as specified for $SS'$ above in (3)
[Theorems \ref{wilsonalgs} through \ref{approxSStimetheo}].

\end{enumerate}

Several of these results were conjectured based on experimental evidence
in \cite{CJK99}, which also introduced the main linear program of
(\ref{r2}).
This linear program turns out to be essentially equivalent to one
previously introduced by Val\'{e}rio de Carvalho in his
{\em arc flow} model for bin packing \cite{Carvalho99}, but has
not previously been adapted to questions of average case behavior.

\subsection{Previous results}

\noindent
The relevant previous results can be divided into two classes:
(1) results for practical algorithms on specific distributions, and
(2) more general (and less practical) results about the existence of algorithms.
We begin with (1).

The average case behavior under discrete distributions for standard heuristics
has been studied in
\cite{AM98,CCG91,CCG98,CJS93,CJS97,KRS98}.
These papers concentrated on the discrete uniform distributions $U\{j,k\}$
mentioned above, where the bin capacity $B=k$ and the item sizes are
$1,2,\ldots,j < k$, all equally likely.
If $j=k-1$, the distribution is symmetric
and we have by earlier results that the optimal packing
and the off-line First and Best Fit Decreasing
algorithms (FFD and BFD) all have $\Theta(\sqrt{n})$ expected waste,
as do the on-line First Fit (FF) and Best Fit (BF) algorithms
\cite{CCG91,CJS97}.

More interesting is the case when $1 \leq j \leq k-2$.
Now the optimal expected waste is $O(1)$ \cite{CCG91,CCG98,CCG02}, and the
results for traditional algorithms do not always match this.
In \cite{CCG91} it was shown that BFD and FFD have $\Theta(n)$ waste for
$U\{6,13\}$, and \cite{CJS3} identifies a wide variety of other
$U\{j,k\}$ with $j<k-1$ for which these algorithms have linear waste.
For the on-line algorithms FF and BF, the situation is no better.
Although they can be shown to have $O(1)$ waste when $j = O(\sqrt{k})$
\cite{CCG91}, when $j=k-2$ \cite{AM98,KRS98}, and (in the case of
BF) for specified pairs $(j,k)$ with $k \leq 14$ \cite{CJS93},
for most values of $(j,k)$ it appears experimentally
that their expected waste is linear.
This has been proved for BF and the
pairs $(8,11)$ and $(9,12)$ \cite{CJS93} as well as all pairs $j,k$ with
$j/k \in [0.66,2/3)$ when $k$ is sufficiently large \cite{KM00}.
In contrast, $EW_n^{SS}(U\{j,k\}) = O(1)$ whenever $j<k-1$.
On the other hand, our current best implementation of basic $SS$
runs in time $\Theta(nB)$ compared to $O(n\log B )$ for BF,
$O(n+B\log^2B)$ for FFD, and $O(n+B\log B)$ BFD \cite{CJS3}.
(The fastest known implementation of FF is $\Theta(n \log n)$
and so FF is asymptotically slower than $SS$ for fixed $B$.) 

Turning to less distribution-specific results, the first relevant results
concerned off-line algorithms.
In the 1960's, Gilmore and Gomory in \cite{GG61,GG63} introduced a
deterministic approach
to solving the bin packing problem that used linear programming, column
generation, and rounding to find a packing that for any list $L$
with $J$ or fewer distinct item sizes is guaranteed to use
no more than $OPT(L) + J-1$ bins.
Since $J < B$ for any discrete distribution,
this implies an average-case performance that is at least as good as that
specified for $SS^*$ in (7) of the previous section, and is in some cases
better.
However, although the approach often seems to work well in practice, its
worst-case running time is conceivably exponential in $B$, since the
basic LP involved in the approach has that many (implicit) variables.

A packing obeying a similar bound can be constructed in time polynomial
in $B$ by using the ellipsoid method to solve
the basic LP of (4) above and then greedily extracting a packing from
the variables of a basic optimal solution, as explained in \cite{ABDJS02}.
A simplistic analysis of the running time
yields a running time bound of $O(n+(JB)^{4.5} \log^2 n)$,
which is linear but with an additive constant that for many distributions
would render the algorithm impractical.
However, if one uses the simplex
method rather than the ellipsoid method to solve the LP's, this approach
too seems to work well in practice.

Theoretically the best approach along these lines is the off-line deterministic
algorithm of Karmarkar and Karp that for any list $L$ never
uses more than $OPT(L) + O(\log^2 J)$ bins and take time
$O(n+J^8 \log J \log^2 n)$.
Although these guarantees are asymptotically stronger than those for
the previous two approaches, the Karmarkar-Karp algorithm is
substantially more complicated and inherently
requires the performance of ellipsoid method steps.
(This Karmarkar-Karp algorithm is closely related to the more famous
one from the same paper that guarantees a packing within
$OPT(L) + O(\log^2 (OPT))$ for {\em all} lists $L$, independent of
the number of distinct item sizes, but for which the
best current running time bound is $O(n^8 \log^3 n)$.)

For on-line algorithms, the most general results
are those of Rhee and Talagrand \cite{Rhe88,RT93b,RT93c}.
In \cite{RT93b},
Rhee and Talagrand proved that for any distribution $F$ (discrete or
not) there exists an $O(n\log n)$ on-line randomized algorithm $A_F$
satisfying $EW_n^{A_F}(F) \leq EW_n^{OPT}(F) + O(\sqrt{n}\log^{3/4}n)$ and hence
$ER_\infty^{A_F}(F) = 1$.
(For distributions with irrational sizes and/or probabilities,
their results assume a real-number RAM model of computation.)
This is a more general result than
(\ref{r3}) above, and although the additive error term is worse than
the one in (\ref{r3}), the extra factor of $\log^{3/4} n$ appears to
reduce to a constant depending only on $B$ when $F$ is a discrete
distribution, making the two bounds comparable.  Unfortunately, Rhee
and Talagrand only prove that such algorithms \emph{exist}.
The details of the algorithms depend on a non-constructive
characterization of $F$ and its packing properties given in
\cite{Rhe88}.

\begin{sloppypar}
In \cite{RT93c}, Rhee and Talagrand present a single (constructive)
on-line randomized algorithm $A$ that works for all distributions $F$
(discrete or not) and has
$EW_n^A(F) \leq EW_n^{OPT}(F) + O(\sqrt{n}\log^{3/4}n)$,
again with the $\log^{3/4}n$ factor likely to reduce to a function of
$B$ for discrete distributions.  Even so, for discrete distributions
this algorithm is not quite as good as our algorithm $SS^*$, which
itself has
$EW_n^{SS^*}(F) \leq EW_n^{OPT}(F) + O(\sqrt{n})$ for all discrete distributions
and in addition gets bounded waste for bounded waste distributions.
Moreover, the algorithm of \cite{RT93c} is unlikely to be practical
since it uses the Karmarkar-Karp algorithm (applied to the items
seen so far) as a subroutine.
\end{sloppypar}

The fastest on-line algorithms previously known that guarantee an
$O\left(\sqrt{n}\right)$ expected waste rate for perfectly packable
discrete distributions
are due to Courcoubetis and Weber, who used them in the proof of their
characterization theorem in \cite{CW90}.  These algorithms are
distribution-dependent, but for fixed $F$ run in linear time.
At each step, the algorithm must solve a linear program whose
number of variables is potentially exponential in $B$, but for
fixed $F$ this takes constant time, albeit potentially a large constant.
Moreover, for bounded waste distributions,
the Courcoubetis-Weber algorithms have $EW_n^A(F) = O(1)$, whereas
the Rhee-Talagrand algorithms cannot provide any guarantee better than
$O\left(\sqrt{n}\right)$.
On the other hand, the Rhee-Talagrand algorithms of
\cite{RT93b,RT93c} guarantee $ER_\infty^A(F) = 1$ for all
distributions, while Courcoubetis and Weber in \cite{CW90} only do this
for those distributions in which $EW_n^{OPT}(F) = O\left(\sqrt{n}\right)$.\smallskip

Thus, although these earlier general approaches rival the packing effectiveness
of $SS$ and its variants, and in the case of the offline algorithms actually
can do somewhat better, none are likely to be as widely usable in practice
(certainly none of the online rivals will be), and none
has the elegance and simplicity of the basic $SS$ algorithm.

\subsection{Outline of the Paper}

\noindent
The remainder of this paper is organized as follows.
In Section 2 we present the details of the Courcoubetis-Weber
characterization theorem and prove our result about the behavior of $SS$ under
perfectly packable distributions.
In Section 3 we prove our results for bounded waste distributions.
Section 4 covers our linear-programming-based algorithm for characterizing
$EW_n^{OPT}(F)$ given $F$.
In Section 5 we discuss our results about the behavior of
$SS$ under linear waste distributions.
In Section 6 we discuss our results about how $SS$ can be modified
so that its expected behavior is asymptotically optimal for such distributions.
Section 7 presents our results about how $SS$
behaves in more adversarial situations.
Section 8 covers our results about the effectiveness of algorithms
that use variants on the sum-of-squares objective function or trade
accuracy in measuring that function for improved running times.
We conclude in Section 9 with a discussion or open problems and related
results, such as the recent extension of the Sum-of-Squares algorithm
to the bin covering problem in \cite{CJK01}.

\section{Perfectly Packable Distributions}
\setcounter{equation}{0}

In order to explain why the Sum-of-Squares algorithm works so well,
we need first to understand the characterization theorem of Courcoubetis
and Weber \cite{CW90}, which we now describe.

Given a discrete distribution $F$, a {\em perfect packing configuration}
is a length-$J$ vector $\bar{b} = \langle
b_1,b_2,\ldots,b_J \rangle$ of nonnegative integers
such that $\sum_{j=1}^J b_js_j = B$.
Such a configuration corresponds to a way of completely filling
a bin with items from $F$.  That is, if we take $b_i$ items of size $s_i$,
$1 \leq i \leq J$, we will precisely fill a bin of capacity $B$.
Let $\Lambda_F$ be the rational cone generated by the set of all perfect packing
configurations for $F$, that is, the closure under rational convex combinations
and positive rational scalar multiplication of the set of all such configurations.
\begin{defi} \label{interior}
A rational vector $\bar{x} = \langle x_1,\ldots,x_J \rangle$ is \emph{in the
interior} of a cone $\Lambda$ if and only if there exists an $\epsilon
> 0$ such that all nonnegative rational vectors
$\bar{y} = \langle y_1,\ldots,y_J \rangle$
satisfying $|\bar{x} - \bar{y}| \equiv \sum_{i=1}^J|x_i-y_i| \leq
\epsilon$ are in $\Lambda$.
\end{defi}

\noindent
\textbf{Theorem (Courcoubetis-Weber \cite{CW90}).}\label{cwtheo}
Let $\bar{p}_F$ denote the vector of size
probabilities $\langle p_1,p_2,\ldots,p_J\rangle$ for a discrete distribution $F$.
{\em
\begin{itemize}
\item[{\rm (a)}] $EW_n^{OPT}(F) = O(1)$ if and only if $\bar{p}_F$ is in the {\rm interior} of $\Lambda_F$.
\item[{\rm (b)}] $EW_n^{OPT}(F) = \Theta\left(\sqrt{n}\right)$ if and only if
$\bar{p}_F$ is on the {\rm boundary} of $\Lambda_F$,
i.e., is in $\Lambda_F$ but not in its interior.
\item[{\rm (c)}] $EW_n^{OPT}(F) = \Theta(n)$ if and only if
$\bar{p}_F$ is outside $\Lambda_F$.
\end{itemize}
}

The Courcoubetis-Weber Theorem can be
used to prove the following lemma, which is key to many
of the results that follow:
\begin{lemm} \label{orlinlemm}
Let $F$ be a perfectly packable distribution
with bin size $B$, $P$ be an arbitrary packing into bins of size $B$,
$x$ be an item randomly generated according to $F$, and $P'$ be the packing
resulting if $x$ is packed into $P$ according to $SS$.  Then
$E[SS(P')|P] < ss(P) +2$.
\end{lemm}

\noindent
{\bf Proof.}
The proof relies on the following claim.

\medskip
\begin{claim} \label{orlinclaim}
If $F$ is a perfectly packable distribution with bin size $B$,
then there is an algorithm $A_F$ such that given any packing $P$ into bins
of size $B$, $A_F$ will pack an item randomly generated according to $F$
in such a way that for each bin level $h$ with $N_P(h) > 0$,
$1 \leq h \leq B-1$, the probability that $N_P(h)$ increases is no more
than the probability that it decreases.
\end{claim}

\medskip
\noindent
{\bf Proof of Claim.}
The algorithm $A_F$ depends on the details
of the Courcoubetis-Weber Theorem.
Since $F$ is perfectly packable, $\bar{p}_F$ must be in $\Lambda_F$
and so there must exist some number $m$ of length-$J$ nonnegative
integer vectors $\bar{b}_i$ and corresponding positive rationals
$\alpha_i$ satisfying
\begin{eqnarray}
\sum_{j=1}^J (b_{i,j} \cdot s_j) \ = & B, & 1 \leq i \leq m
\label{cw1}\\ \sum_{i=1}^m (\alpha_i \cdot b_{i,j}) \ = & p_j, & 1
\leq j \leq J \label{cw2}
\end{eqnarray}
Now since the $\alpha_i$ and $p_j$ are all rational, there exists
an integer $Q$ such that $Q \cdot \alpha_i$ and $Q \cdot p_j$ are
integral for all $i$ and $j$. Consider the {\em ideal packing} $P^*$ which
has $Q\alpha_i$ copies of bins of type $\bar{b}_i$.
We will use $P^*$ to define $A_F$.
Note that by (\ref{cw2}) $P^*$ contains $Qp_j$ items of size
$j$, $1\leq j \leq J$, and hence a total of $Q$ items.
Let $L_F = \{x_1,x_2,\ldots,x_Q\}$ denote the $Q$ items packed into $P^*$,
and denote the bins of $P^*$ as $Y_1,Y_2,\ldots ,Y_{|P^*|}$.

Now let $P$ be an arbitrary packing of integer-size items into bins of size $B$.
We claim that for each bin $Y$ of the packing $P^*$, there is an ordering
$y_1,y_2,\ldots,y_{|Y|}$ of the items contained in $Y$
and a special threshold index $last(Y)<|Y|$
such that if we set $S_i \equiv \sum_{j=1}^is(y_j)$,
$0 \leq i \leq |Y|$, then the following holds:
\begin{enumerate}

\item $P$ has partially filled bins with each level
$S_1 < S_2 < \cdots < S_{last(Y)}$.

\item $P$ has no partially filled bin of level
$S_{last(Y)}+s(y_i)$ for any $i > last(Y)$.

\end{enumerate}

\noindent
That such an ordering and threshold index always exist can be seen from
Figure \ref{permute}, which presents a greedy procedure that, given the current
packing $P$, will compute them.
Assume we have chosen such an ordering and threshold index for
each bin in $P^*$.
Note that $S_{|Y|} = B$ for all such bins $Y$, since each is by
definition perfectly packed.

\begin{figure}[b]
{\small
\begin{center}
\begin{boxedminipage}[t]{4.5in}
\begin{enumerate}
\item Let the set $U$ of as-yet-unordered items
initially\\
be set to $Y$ and let $S = 0$ be the initial total size\\
of ordered items.
\item While $U \neq \emptyset$ and $last(Y)$ is undefined, do the following:
\begin{enumerate}
\item[2.1] If there is an item $x$ in $U$ such that $P$ has\\
a partially filled bin of level $S+x$ 
\begin{enumerate}
\item[2.1.1] Choose such an $x$, put it next in the ordering,\\
and remove it from $U$
\item[2.1.2] Set $S=S+s(x)$.
\end{enumerate}
\item[2.2] Otherwise, set $last(Y)$ to be the number of\\
items ordered so far and exit {\em While} loop.
\end{enumerate}
\item Complete the ordering by appending the remaining\\
items in $U$ in arbitrary order.
\end{enumerate}
\vspace{.025in}
\end{boxedminipage}
\end{center}
\caption{Procedure for ordering items in bin $Y$ given a packing $P$} \label{permute}
}
\end{figure}
Our algorithm $A_F$ begins the processing of an item $a$ by first
randomly identifying it with an appropriate element $r(a) \in L_F$.
In particular, if $a$ is of size $s_j$, then $r(a)$
is one of the $Q \cdot p_j$ items in $L_F$ of size $s_j$,
with all such choices being equally likely.
Note that this implies that for each $i$, $1 \leq i \leq Q$,
the probability that a randomly generated item $a$
will be identified with $x_i$ is $1/Q$.

Having chosen $r(a)$, we then determine the bin into which we should place $a$
as follows.
Suppose that in $P^*$, item $r(a)$ is in bin $Y$ and has index $j$
in the ordering of items in that bin.

\begin{enumerate}
\item[(i)] If $j = 1$, place $a$ in an empty bin, creating a new
bin with level $s(a) = S_1$.
\item[(ii)] If $1 < j \leq last(Y)$, place $a$ in a bin with level
$S_{j-1}$, increasing its level to $S_j$.
\item[(iii)] If $j > last(Y)$, place $a$ in a bin of
size $S_{last(Y)}$ (or in a new bin if $last(Y)=0$).
\end{enumerate}

For example, suppose that the items in $Y$, in our constructed order,
are of size 2, 3, 2, and 4 and $last(Y) = 2$.
Then $S_1=2$, $S_2=5$, $S_3=7$, $S_4=B=11$,
$N_P(2),~N_P(5) > 0$, and $N_P(7) = N_P(9) = 0$.
If $r(a) \in Y$, then it is with equal probability the first 2,
the 3, the second 2, or the 4.
In the first case it starts a new bin, creating a bin of level 2
and increasing $N_P(2)$ by 1.
In the second it goes in a
bin of level 2, converting it to a bin of level 5,
thus decreasing $N_P(2)$ by 1 and increasing $N_P(5)$ by 1.
In the third and fourth cases it goes in a bin of level 5, converting it to a
bin of level 7 or 9, depending on the case, and decreasing $N_P(5)$ by 1.
Thus when $r(a) \in Y$, the only positive level counts that can change
are those for $h \in \{2,5\} = \{S_1,S_2=S_{last(Y)}\}$, counts can
only change by 1,
and each count is at least as likely to decline as to increase.

More generally,
for any bin $Y$ in $P^*$, if $a$ is randomly generated according to $F$
and $r(a) \in Y$, then by the law of conditional probabilities
$r(a)$ will take on each of the values $y_i$, $1 \leq i \leq |Y|$ with probability
$p = 1/|Y|$.
Thus if $r(a) \in Y$ the probability that the count for level $S_i$
increases equals the probability that it decreases when
$1 \leq i < last((Y)$.
The probability that the count for $S_{last(Y)}$ decreases
is at least as large as the probability that it increases (greater
if $last(Y) \leq |Y|-2$).
And for all other levels with positive counts,
the probability that a change occurs is 0.
Since this is true for all bins $Y$ of the ideal packing $P^*$,
the Claim follows.  \proofmark

Claim \ref{orlinlemm} is used to prove Lemma \ref{orlinlemm} as follows.
Note that the claim implies a bound on the expected increase in
$ss(P)$ when a new item is packed under $A_F$.
For any level count $x>0$, the expected increase in $ss(P)$ given that
this particular count changes is, by the claim, at most
$$
\frac{1}{2}\Bigl((x+1)^2 - x^2 \Bigr) + \frac{1}{2}\Bigl((x-1)^2 - x^2\Bigr) = 1
$$
More trivially, the expected increase in $ss(P)$
given that a 0-count changes is also at most 1.
Since a placement changes at most two counts, this means that the
expected increase in $ss(P)$ using algorithm $A_F$ is at most 2.
Since $SS$ explicitly chooses the placement of each item so as to minimize
the increase in $ss(P)$, we thus must also have that the expected increase in
$ss(P)$ under $SS$ is at most 2 at each step.  \proofmark

\medskip
Lemma \ref{orlinlemm} is exploited using the following result.
\begin{lemm}\label{cauchylemm}
Suppose $P$ is a packing of a randomly generated list $L_n(F)$,
where $F$ is a discrete distribution with bin size $B$ and $n>0$.
Then
$$E[W(P)] \leq \sqrt{(B-1)E[ss(P)]}.$$
\end{lemm}

\noindent
{\bf Proof.}
For $1 \leq i \leq n$ let $C_i = \sum_{h=1}^{B-1} p[N_P(h)=i]$,
i.e., the expected number of levels whose count in $P$ equals $i$.
Then $\sum_{i=1}^{n} C_i = B-1$ and
\begin{equation} \label{cl2n}
E[ss(P)] = \sum_{h=1}^{B-1}E\left[N_P(h)\right] = \sum_{i=1}^n C_i \cdot i^2
\end{equation}

\noindent
We now apply the Cauchy-Schwartz inequality, which says that
$$\left (\sum x_i y_i \right )^2 \leq
\left (\sum x_i^2 \right ) \left (\sum y_i^2 \right )$$

\noindent
Let $x_i = \sqrt{C_i}$ and $y_i = i\sqrt{C_i}$, $1 \leq i \leq n$.
We then have
$$\left (\sum_{i=1}^n C_i \cdot i \right )^2 \leq
\left (\sum_{i=1}^n C_i \right ) \left (\sum_{i=1}^n C_i i^2 \right ).$$

\noindent
Taking square roots and using (\ref{cl2n}), we get
\begin{equation} \label{ebound}
E \left [ \sum_{h=1}^{B-1} N_P(h) \right ] \leq \sqrt{(B-1)E[ss(P)]}.
\end{equation}

\noindent
Since no partially full bin has more than $(B-1)/B < 1$ waste, the
claimed result follows.  \proofmark

\begin{theo} \label{orlintheo}
Suppose $F$ is a discrete distribution satisfying $EW_n^{OPT}(F) =
O\left(\sqrt{n}\right)$.  Then $EW_n^{SS}(F) < \sqrt{2nB}$.
\end{theo}

\noindent
{\bf Proof.}
By Lemma \ref{orlinlemm} and the linearity of expectations, we have
\begin{equation*} \label{el2n}
E[ss(P_n^{SS}(F))] \leq 2n.
\end{equation*}
The result follows by Lemma \ref{cauchylemm}.  \proofmark

\section{Bounded Waste Distributions} \label{boundedwastesection}

\noindent
In order to distinguish the broad class of bounded waste distributions under which
$SS$ performs well, we need some new definitions.
If $F$ is a discrete distribution, let $U_F$ denote the set of sizes with
positive probability under $F$.

\begin{defi}
A level $h$, $1 \leq h \leq B-1$,
is a {\em dead-end} level for $F$ if there is some collection of items
with sizes in $U_F$ whose total size is $h$,
but there is no such collection whose total is $B-h$.
\end{defi}

In other words, if $h$ is a dead-end level then it is possible to pack a bin
to level $h$ with items from $U_F$, but once such a bin has been created, it
is impossible to fill it completely.
Note that the dead-end levels for $F$ depend only on $U_F$ and can
be identified in time $O(|U_F|B)$ by dynamic programming.

\begin{obse}\label{deadendobse}
For future reference, note the following easy consequences of the definition
of dead-end level.

\renewcommand{\theenumi}{\alph{enumi}}     
\renewcommand{\labelenumi}{\rm (\alph{enumi})} 
\begin{enumerate}
\item The algorithms $A_F$ of Claim
\ref{orlinclaim} in the proof of Lemma \ref{orlinlemm}
never create bins that have dead-end levels.
(This is because the levels of the bins they create are always the sums
of item sizes from a perfectly packed bin.)
\item If $F$ is a perfectly packable distribution, then for no $s_j \in U_F$
is $s_j$ a dead-end level.
(Otherwise, no bin containing items of size $s_j$ could be perfectly packed.
Since the expected number of such bins in an optimal packing
is at least $np_j/B$, this means
that the expected waste would have to be at least $np_j/B^2$ and hence linear,
contradicting the assumption that $F$ is a perfectly packable distribution.)
\item No distribution with $1 \in U_F$ can have a dead-end level, so
that in particular the $U\{j,k\}$ do not have dead-end levels.
\end{enumerate}
\end{obse} 

A simple example of a distribution that does have dead-end levels is
any $F$ that has $B = 6$ and $U_F = \{2,3\}$.
Here 5 is a dead-end level for $F$ while 1,2,3,4 are not.
There is a sense, however, in which this distribution is still fairly
benign. 

\begin{defi}
A level $h$ is {\em nontrivial} for a
distribution $F$ if there is some list $L$ with item sizes from $U_F$ such that
the $SS$ packing $P$ of $L$ has $N_P(h)>1$.
\end{defi}

It is easy to verify that there are {\em no} nontrivial levels,
dead-end or otherwise,
in the above $B = 6$ example.

\smallskip
We shall divide this section into three parts.
In subsection \ref{boundedSS} we show that $SS$ has bounded expected waste
for bounded waste distributions with no nontrivial dead-end levels.
In subsection \ref{boundedSS'} we show that a simple variant on $SS$
has bounded expected waste for {\em all} bounded waste distributions.
In subsection \ref{lognSS} we characterize the behavior of $SS$ for
bounded waste distributions that do have nontrivial dead-end levels.

\subsection{A bounded expected waste theorem for $SS$}\label{boundedSS}

\begin{theo}\label{boundedtheo}
If $F$ is a bounded waste distribution with no nontrivial dead-end levels, then
$EW_n^{SS}(F) = O(1)$.
\end{theo}

To prove this result we rely on the Courcoubetis-Weber Theorem, Lemma
\ref{orlinlemm}, and the following specialization of
a result of Hajek \cite{Haj82}.

\medskip
\noindent
{\bf Hajek's Lemma.}
{\em
Let $S$ be a state space and let
${\cal F}_k$, $k \geq 1$, be a sequence of functions, where ${\cal F}_k$
maps $S^{k-1}$ to probability distributions over $S$.
Let $X_1,X_2,\ldots$ be a sequence of random variables over $S$
generated as follows:  $X_1$ is chosen according to ${\cal F}_1 (\cdot)$
and $X_k$ is chosen according to ${\cal F}_k(X_1,\ldots ,X_{k-1})$.
Suppose there are constants $b > 1$, $\Delta < \infty$, $D > 0$,
and $\gamma > 0$ and a function $\phi$ from $S$ to $[0,\infty)$
such that
\begin{enumerate}
\renewcommand{\theenumi}{\alph{enumi}}     
\renewcommand{\labelenumi}{\rm (\alph{enumi})} 
\item {\rm [Initial Bound Hypothesis]}.
$~E \left[b^{\phi(X_1)} \right] < \infty$.
\item {\rm [Bounded Variation Hypothesis]}. For all $N \geq 1$,
$\displaystyle{|\phi(X_{N+1}) - \phi(X_N)| \leq \Delta.}$
\item {\rm [Expected Decrease Hypothesis]}. For all $N \geq 1$,
$$E[\phi(X_{N+1}) - \phi(X_N)|\phi(X_N) > D] \leq - \gamma.$$
\end{enumerate}

\noindent
Then there are constants $c>1$ and $T>0$ such that for all $N \geq 1$,
$E \left[c^{\phi(X_N)} \right] < T$.
}

\medskip
Note that the conclusion of this lemma implies that there is also a constant
$T'$ such that $E[{\phi(X_N)}] < T'$ for all $N$.
A weaker version of the lemma was used in the analyses of
the Best and First Fit bin packing heuristics in
\cite{AM98,CJS93,KRS98}.
The added strength is not needed for Theorem \ref{boundedtheo},
but will be used in the proof of Theorem \ref{lognupperbound}.

We prove Theorem \ref{boundedtheo} by applying Hajek's Lemma
with the following interpretation.
The state space $S$ is the set of all length-$(B-1)$ vectors
of non-negative integers $\bar{x} = \langle x_1,x_2,\ldots ,x_{B-1}\rangle$,
where we view $\bar{x}$ as the profile of a packing that has $x_i$
bins with level $i$, $1 \leq i \leq B-1$.
$X_0$ is then the profile of the empty packing and $X_{i+1}$ is the profile
of the packing obtained by generating a random item according to $F$ and
packing it according to $SS$ into a packing with profile $X_i$.
The potential function is
$$\phi(\bar{x}) = \sqrt{\sum_{i=1}^{B-1}x_i^2}.$$

Note that if the hypotheses of Hajek's Lemma are
satisfied under this interpretation, then the lemma's conclusion
would say that there is a $T'$ such that for all $N$,
$$E\left[\sqrt{\sum_{i=1}^{B-1}x_{N,i}^2}~~\right] < T'$$
which implies that $E[x_{N,i}]$ is bounded by $T'$ as well,
$1 \leq i \leq B-1$.
Thus the expected waste is less than the constant $BT'$ and Theorem
 \ref{boundedtheo} would be proved.

Hence all we need to show is that the three hypotheses of Hajek's lemma
apply.
The Initial Bound Hypothesis applies
since the profile of an empty packing is all 0's and hence $\phi(X_0) = 0$.
The following lemma implies that Bounded Variation Hypothesis also holds.

\begin{lemm} \label{supportlemm}
Let $\bar{x}$ be the profile of a packing into bins of size $B$,
and let $\bar{x}'$ be the profile of the packing obtained from $\bar{x}$
by adding an item to the packing in any legal way.
Then\medskip
$$|\phi(\bar{x}') - \phi(\bar{x})| \leq 1$$
\end{lemm}

\noindent\textbf{Proof.}
Consider the case when $\phi(\bar{x}') > \phi(\bar{x})$ and
suppose that $i$ is the level whose count increases when the item is packed
is level $i$.
We have

\begin{align*}
\phi(\bar{x}') - \phi(\bar{x}) & \leq
\sqrt{\phi(\bar{x})^2+(x_i+1)^2-x_i^2}-\phi(\bar{x})\\[.08in]
& = \frac{\left(\sqrt{\phi(\bar{x})^2+2x_i+1} -  \phi(\bar{x})\right)
\left(\sqrt{\phi(\bar{x})^2+2x_i+1} +  \phi(\bar{x})\right)}{\sqrt{\phi(\bar{x})^2+2x_i+1} +  \phi(\bar{x})}\\[.05in]
&= \frac{ 2x_i+1 }
{\sqrt{\phi(\bar{x})^2+2x_i+1} + \phi(\bar{x}) }\\[.05in]
&\leq \frac{ 2x_i+1  }{\sqrt{x_i^2+2x_i+1} + x_i } = 1\,.
\end{align*}

\noindent
A similar argument handles the case when
$\phi(\bar{x}') < \phi(\bar{x})$.  \proofmark

To complete the proof of the theorem, we need to show that the Expected
Decrease Hypothesis
of Hajek's Lemma applies.  For this we need the following
three combinatorial lemmas.

\begin{lemm} \label{squarelemm}
Suppose $y$ be any number and $a>0$.  Then
$$y-a \leq \frac{y^2-a^2}{2a}.$$
\end{lemm}

\noindent
{\bf Proof.}
Note that $y-a = (y^2-a^2)/(y+a)$, and then observe that no matter whether
$y \geq a$ or $y < a$, this is less than or equal to
$(y^2-a^2)/2a$.  \proofmark

\begin{lemm} \label{goodmovelemm}
Let $F$ be a distribution with no nontrivial dead-end levels
and let $P$ be any packing that can be created by applying $SS$
to a list of items all of whose sizes are in $U_F$.
If $\bar{x}$ is the profile of $P$ and $\phi(\bar{x}) > 2B^{3/2}$,
then there is a size $s \in U_F$ such that if an item of size $s$ is packed
by $SS$ into $P$, the resulting profile
$\bar{x}'$ satisfies
$$\phi(\bar{x}')^2 \leq \phi(\bar{x})^2 - \frac{\phi(\bar{x})}{B^{3/2}}$$.
\end{lemm}

\noindent
{\bf Proof.}
Suppose $\bar{x}$ is as specified and let
$h$ be the index for a level at which $\bar{x}$ takes on its maximum value.
It is easy to see that
\begin{equation}\label{maxsbound}
x_h \geq \phi(\bar{x})/\sqrt{B}\,.
\end{equation}

\noindent
Thus $x_h > 2B > 1$ and so by definition $h$ cannot be a dead-end level for $F$.
Hence there must be a sequence of levels
$h = \ell_0 < \ell_1 < \cdots < \ell_m = B$, $m \leq B$,
such that for $1 \leq i \leq m$, $\ell_i - \ell_{i-1} \in U_F$.
Taking $x_B = 0$ by convention, we have
\begin{equation}\label{collapsesum}
x_h = \sum_{i=0}^{m-1}(x_{\ell_i} - x_{\ell_{i+1}})\,.
\end{equation}

\noindent
Let $q$, $0 \leq q < m$ be an index which yields the maximum value $\Delta$ for
$x_{\ell_i} - x_{\ell_{i+1}}$, and let $s = \ell_{q+1}-\ell_q$.
Then by (\ref{collapsesum}) we have $\Delta \geq x_h/m \geq x_h/B \geq
\phi(\bar{x})/B^{3/2}$, where the last inequality follows from (\ref{maxsbound}).
By Lemma \ref{deltalemm} this means that if an item of size
$s$ arrives, $\phi(\bar{x})^2$ must decline by at least
$$2(\Delta-1) \geq 2 \left( \frac{\phi(\bar{x})}{B^{3/2}} - 1 \right)
\geq \frac{\phi(\bar{x})+2B^{3/2}}{B^{3/2}} - 2
\geq  \frac{\phi(\bar{x})}{B^{3/2}}$$

\noindent
as claimed.  \proofmark

\begin{lemm} \label{epsilonlemm}
Let $F$ be a bounded waste distribution with $U_F = \{s_1,s_2,\ldots ,s_J\}$.
For each $i$, $1 \leq i \leq J$ and $\epsilon > 0$,
let $F[i,\epsilon]$ be the distribution which
decreases $p_i$ to $p'_i=(p_i-\epsilon )/(1-\epsilon)$ and increases all other
probabilities $p_j$ to $p'_j=p_j/(1-\epsilon)$.
Then there is a constant $\epsilon_0 > 0$ such that $F[i,\epsilon]$
is a perfectly packable distribution for all $i$, $1 \leq i \leq J$,
and $\epsilon$, $0 < \epsilon \leq \epsilon_0$.
\end{lemm}

{\bf Proof.}
Since $F$ is a bounded waste distribution and $p_i>0$, $1 \leq i \leq J$,
this follows from the Courcoubetis-Weber theorem, part (a).  \proofmark

We can now prove that the Expected Decrease Hypothesis of Hajek's Lemma applies,
which will complete the proof of the Theorem \ref{boundedtheo}.
Let $F$ be a bounded waste distribution with no nontrivial dead-end levels,
and let $\epsilon_0$ be the value specified for $F$ by Lemma \ref{epsilonlemm}.
Without loss of generality we may assume that $\epsilon_0 < 2$.
Let $P$ be a packing as specified in Lemma \ref{goodmovelemm}
but with profile $\bar{x}$ satisfying
$\phi(\bar{x}) > 4B^{3/2}/\epsilon_0 > 2B^{3/2}$.
Let $i$ be the index of the size $s \in U_F$ whose existence is proved in
Lemma \ref{goodmovelemm}, and let $F_i$ be the distribution that always generates
an item of size $s_i$.

Consider the two-phase item generation process that first randomly chooses
between distributions $F_i$ and $F[i,\epsilon_0]$, the first choice being
made with probability $\epsilon_0$ and the second with probability $1-\epsilon_0$.
It is easy to see that this process is just a more complicated way of
generating items according to distribution $F$.
Now consider what happens when this process is used to add one item to
packing $P$.
If $F_i$ is chosen, then by Lemma \ref{goodmovelemm}, the value of
$\phi^2$ declines by at least $\phi(\bar{x})/B^{3/2}$.
If $F[i,\epsilon_0]$ is chosen, the expected value of $\phi^2$ increases
by less than 2 by
Lemma \ref{orlinlemm} and the fact that $F[i,\epsilon_0]$
is a perfectly packable distribution (Lemma \ref{epsilonlemm}).
Thus if $\bar{x}'$ is the resulting profile, we have by
applying  Lemma \ref{squarelemm} for $a = \phi(\bar{x})$ 
and taking expectations
\begin{align*}
E\left[\phi(\bar{x}') - \phi(\bar{x})\right]
& ~~<~~ (1-\epsilon_0)(2)\left(\frac{1}{2\phi(\bar{x})}\right)
+ \epsilon_0 \left(-\frac{\phi(\bar{x})}{B^{3/2}}\right)\left(\frac{1}{2\phi(\bar{x})}\right)\\
& ~~<~~ \frac{1}{\phi(\bar{x})}
- \frac{\epsilon_0}{2B^{3/2}}\\
& ~~<~~ -\frac{\epsilon_0}{4B^{3/2}}
\end{align*}
since $\phi(\bar{x}) > 4B^{3/2}/\epsilon_0$.
Thus  the Expected Decrease Hypothesis of Hajek's Lemma holds with
$D = 4B^{3/2}/\epsilon_0$
and $\gamma = \epsilon_0/4B^{3/2}$, and so Hajek's Lemma applies.
Thus $EW_n^{SS}(F) = O(1)$, the conclusion of Theorem \ref{boundedtheo}.  \proofmark

\subsection{Improving on SS for bounded waste distributions}\label{boundedSS'}

Unfortunately, although $SS$ has bounded expected waste for bounded
waste distributions with no nontrivial dead-end levels,
it doesn't do so well for {\em all} bounded waste distributions.
Consider the distribution $F$ with $B = 9$, $J=2$,
$s_1=2$, $s_2=3$, and $p_1=p_2=1/2$.
It is easy to see that $F$ is a bounded waste distribution,
since $3$'s by themselves can pack perfectly, and only one 3 is needed for
every three 2's in order that the 2's can go into perfectly
packed bins.
Note, however, that 8 is a nontrivial dead-end level for $F$, so
Theorem \ref{boundedtheo} does not apply.
In fact, $EW_n^{SS}(F) = \Omega(\log n)$, as the following
informal reasoning suggests: It is likely that somewhere within a sequence
of $n\log n$ items from $F$ there will be $\Omega (\log n)$ consecutive 2's.
These are in turn likely to create $\Omega (\log n)$ bins of
level 8, and hence, since 8 is a dead-end level, $\Theta (\log n)$ waste.

Fortunately, this is the worst possible result for $SS$
and a bounded waste distribution, as we shall
see below in Theorem \ref{lognupperbound}.
First, however, let us show how a simple modification to $SS$ yields a
variant with the same running time that has $O(1)$ expected
waste for all bounded waste distributions.

Like $SS$, this variant ($SS'$) is on-line.
It makes use of a parameterized variant $SS_D$ on the packing rule of $SS$,
where $D$ is a set of levels.
In $SS_D$, we place items so as to minimize $ss(P)$ subject to the
constraint that no bin with level in $D$ may be created unless this is
unavoidable.  In the latter case we start a new bin.
$SS'$ works as follows.
Let $U$ be the set of item sizes seen so far and let $D(U)$ denote the set of
dead-end levels for $U$.
(Initially, $U$ is the empty set.)
Whenever an item arrives, we first check if its size is in $U$.
If not, we update $U$ and recompute $D(U)$.
Then we pack the item according to $SS_{D(U)}$.
A first observation about $SS'$ is the following.

\begin{lemm}\label{nodeadlemm}
If $F$ is a perfectly packable distribution, then $SS'$ will never create
a dead-end level when packing a sequence of items with sizes in $U_F$.
\end{lemm}

\noindent
\textbf{Proof.}
By Observation \ref{deadendobse}(b), starting a bin with an item whose
size is in $U_F$ can never create a dead-end level for $U_F$.
On the other hand, if $SS'$ puts a item in a partially full bin,
it must by definition be the case that the new level is not a dead-end level
for $U$.
Thus, since the new level is attainable using items whose sizes are in $U$,
the resulting gap must be precisely fillable with items whose sizes are
in $U \subseteq U_F$.
Thus the new level is not a dead-end level for $U_F$ either.
\proofmark

\begin{theo}\label{sprimecor}

\noindent
\begin{itemize}
\item[{\rm (i)}]
If $F$ is a perfectly packable distribution, then
$EW_n^{SS'}(F) = O(\sqrt{n})$.
\item[{\rm (ii)}]
If $F$ is a bounded waste distribution, then
$EW_n^{SS'}(F) = O(1)$.
\end{itemize}
\end{theo}

\noindent
\textbf{Proof.}
We begin by bounding the
expected number of items that can arrive before we have seen all
item sizes in $U_F$.
Assume without loss of generality that $U_F = \{s_1,s_2,\ldots ,s_j\}$.
The probability that the $i$th item size does not appear
among the first $h$ items generated is $(1-p_i)^h$.
Thus, if we let $p_{min} = \min\{p_i: 1 \leq i \leq J\}$, the probability
that we have not seen all item sizes after the $h$th item arrives is
at most
\begin{equation*}
\sum_{i=1}^J \left( 1-p_i \right)^h \leq J \left( 1-p_{min} \right)^h
\end{equation*}

Let $t$ be such that $J(1-p_{min})^t \leq 1/2$.
Then for each integer $m \geq 0$, the probability that all the item sizes
have not been seen after $mt$ items have arrived is at most $1/2^m$.
Thus if $M$ is the number of items that have arrived when the last item
size is first seen, we have that for each $m \geq 0$, the probability
that $M \in (mt,(m+1)t]$ is at most $1/2^m$.

For (i), note that if $P$ is the packing that exists immediately after
the last item size is first seen, then $ss(P) \leq M^2$ and
$$E[ss(P)]
\leq \sum_{m=0}^\infty \Bigl( (m+1)t)^2 \cdot p\bigl[M \in (mt,(m+1)t]\bigr]\Bigr)
\leq \sum_{m=0}^\infty \frac{((m+1)t)^2}{2^m} = 12t^2
$$
which is a constant bound depending only $F$.
After all sizes have been seen, $SS'$ reduces to $SS_{D(U_F)}$,
and it follows from Observation \ref{deadendobse}(a) that Lemma \ref{orlinlemm}
applies to the latter.
We thus can conclude that for any $n$ the packing $P_n$ satisfies
$$E\left[ss(P_n)\right] < 12t^2 + 2n$$
which by Lemma \ref{cauchylemm} implies that
$EW_n^{SS'}(F) = O(\sqrt{n})$, so (i) is proved.

The argument for (ii) mimics the proof of Theorem
\ref{boundedtheo}.  Using the same potential function $\phi$ we
show that Hajek's Lemma applies when $SS$ is replaced
by $SS_{D(U_F)}$, $F$ is a bounded waste distribution,
and the initial state $\bar{x}$ is taken to be the profile of the packing $P$
that exists immediately after the last item size is first seen by $SS'$.

To see that the Initial Bound Hypothesis is satisfied,
we must show that there exists a
constant $b > 1$ such that $E\left[b^{\phi(\bar{x})}\right]$ is bounded.
To prove this, let $M$ be the number of items in packing $P$.
It is immediate that $\phi(\bar{x}) = \sqrt{\sum_{i=1}^{B-1} x_i^2} \leq M$.
Thus if we take $b = 2^{1/(2t)}$ and exploit the analysis used for (i)
above we have
$$E\left[b^{\phi(\bar{x})}\right] \leq E\left[b^M\right]
\leq \sum_{m=0}^\infty b^{(m+1)t}\cdot\frac{1}{2^m}
= \sum_{m=0}^\infty \frac{2^{(m+1)/2}}{2^m}
= \sqrt{2}\sum_{m=0}^\infty \frac{1}{\sqrt{2}^m}
= \frac{2}{\sqrt{2}-1} < 4.83.$$

Thus the Initial Bound Hypothesis is satisfied.
The Bounded Variation Hypothesis again
follows immediately from Lemma \ref{supportlemm}.
To prove the Expected Decrease Hypothesis, we need the facts that
Lemmas \ref{orlinlemm} and \ref{goodmovelemm}
hold when $SS$ is replaced by $SS_{D(U_F)}$.
We have already observed that Lemma \ref{orlinlemm} holds.
As to Lemma \ref{goodmovelemm}, the properties of $SS$ were used in
only two places.
First, we needed the fact that $SS$ could never create a packing where
the count for a dead-end level exceeded 1, an easy observation there
since we assumed there were no nontrivial dead-end levels.
Here there can be nontrivial dead-end levels, but this is not
a problem since by Lemma
\ref{nodeadlemm} $SS'$ can never create a packing where the count
for a dead-end level is nonzero.

The other property of $SS$ used in proving Lemma \ref{goodmovelemm}
was simply that, in the terms of the proof
of that lemma, it could be trusted to pack an item of size
$s=\ell_{q+1}-\ell_q$ in such a way as to reduce $ss(P)$ by
at least as much as it would be reduced by placing the item in
a bin of level $\ell_q$.
$SS_{D(U_F)}$ will clearly behave as desired, since level $\ell_{q+1}$,
as it is constructed in the proof, is not a dead-end level,
and so bins of level $\ell_q$ are legal placements for items of
size $\ell_{q+1}-\ell_q$ under $SS_{D(U_F)}$.

We conclude that Lemma \ref{goodmovelemm} holds
when $SS_{D(U_F)}$ replaces $SS$, and so the Expected Decrease Hypotheses
of Hajek's Lemma is satisfied.
Thus the latter Lemma applies, and the proof of bounded
expected waste can proceed exactly as it did for $SS$.
\proofmark

\subsection{The worst behavior of SS for bounded waste distributions}\label{lognSS}

\begin{theo}\label{lognupperbound}
If $F$ is a bounded waste distribution that has nontrivial dead-end
levels, then $EW_n^{SS}(F) = \Theta(\log n)$.
\end{theo}

We divide the proof of this theorem into separate upper and lower bound
proofs.
These are by a substantial margin the most complicated proofs in the paper,
and readers may prefer to skip this section on a first reading of the paper.
None of the later sections depend on the details of these proofs.

\noindent
\subsubsection{Proof of the $O(\log n)$ Upper Bound}

For this result we need to exploit more of the power of Hajek's Lemma
(which surprisingly is used in proving the lower bound as well as
the upper bound).  We will also need a more complicated potential function.
Let ${\cal D}_F$ denote the set of dead-end levels for $F$ and
let ${\cal L}_F$ denote the set of levels that are not dead-end levels for $F$.
We shall refer to the latter as {\em live} levels in what follows.
For a given profile $\bar{x}$, define
$\tau_D(\bar{x}) = \sum_{i\in{\cal D}_F} x_i^2$ and
$\tau_L(\bar{x}) = \sum_{i\in{\cal L}_F} x_i^2$.
Note that $\phi(\bar{x}) = \sqrt{\tau_D(\bar{x}) + \tau_L(\bar{x})}$.
Our new potential function $\psi$ must satisfy two key properties.
\begin{enumerate}
\item Hajek's Lemma applies with the potential function $\psi$ and, as before,
$X_i$ representing the profile after $SS$ has packed $i$ items
generated according to $F$.
\item For any live level $h$,
\begin{equation}\label{countbound}
\psi(\bar{x}) > \sqrt{\tau_L(\bar{x})} \geq x_h.
\end{equation}
\end{enumerate}

Let us first show that the claimed upper bound will follow if we can
construct a potential function $\psi$ with these properties.
Since Hajek's Lemma applies,
there exist constants $c>1$ and $T>0$ such that for all $N>0$,
\begin{equation}\label{Tbound}
E\left[ c^{\psi(X_N)}\right] \leq T.
\end{equation}
We can use (\ref{Tbound}) to separately bound the sums of the counts
for live and dead levels.
For each live level $h$, the component $X_{n,h}$ of the final
packing profile $X_n$ satisfies
$X_{n,h} \leq \psi(X_n) < c^{\psi(X_n)}/\log_ec$,
and so we have
\begin{equation}\label{livebound}
E\left[ \sum_{h \in {\cal L}_F} X_{n,h} \right ]
\leq E \Biggl[B\frac{c^{\psi(X_n)}}{\log_ec}\Biggr]
\leq \frac{BT}{\log_ec} = O(1)
\end{equation}
In other words, the expected sum of the counts for live levels is bounded
by a constant.

To handle the dead-end levels, we begin by noting
that (\ref{Tbound}) also implies that for all $N$ and all $\alpha > 1$,
\[
P\left[ c^{\psi(X_N)} > \alpha T \right] < \frac{1}{\alpha},
\]

\noindent
so if we take logarithms base $c$ and set $\alpha = n^2/T$ we get
\begin{equation}\label{logprob}
P\left[ \psi(X_N) > 2\log_c n \right] < \frac{T}{n^2}.
\end{equation}

Say that a placement is a {\em major uphill move} if it increases
$ss(P)$ by more than $4\log_cn+1$.
By Observation \ref{deadendobse}(b) and (\ref{countbound}),
we know that whenever an item is generated according to $F$ and packed
by $SS$, one option will be to start a new bin with a live level
and hence, no matter where the item is packed,
the increase in $ss(P)$ will be bounded by $2\psi(X_N)+1$.
Using (\ref{logprob}), we thus can conclude that at any point in the packing
process, the probability that the next placement is a major
uphill move is at most $T/n^2$.
Thus, in the process of packing $n$ items, the expected number of
major uphill moves is at most $T/n$ by the linearity of expectations.

Now let us consider the dead-end levels.
Suppose the count for dead-end level $h$ is $2B(\log_c n+1)$ or greater and
a bin $b$ with level less than $h$ receives an additional item that
brings its level up to $h$.
We claim that bin $b$, in the process of attaining this level from the
time of its initial creation, must
have at one time or another experienced an item placement that was
a major uphill move.

To see this, let us first recall the tie-breaking rule used by $SS$
when it must choose between bins with a given level for packing the next item.
Although the rule chosen has no effect on the amount of waste created,
our definition of $SS$ specified a particular rule, both so the algorithm
would be completely defined and because the particular rule chosen facilitates
the bookkeeping needed for this proof.
The rule says that when choosing which bin of
a given level $h$ to place an item in, we always pick the bin which
most recently attained level $h$.
In other words, the bins for each level will act as a stack, under the
``last-in, first-out'' rule.
Now consider the bin $b$ mentioned above.
In the process of reaching level $h$, it received less than $B$ items,
so it changed levels fewer than $B$ times.
Note also that by our tie-breaking rule above, we know that every time the bin
left a level, that level had the same count that it had when the
bin arrived at the level.
Thus at least one of the steps in packing bin $b$ must have involved
a jump from a level $i$ to a level $j$ such that
$N_P(j) \geq N_P(i) + 2(\log_c n +1)$.
By Lemma \ref{deltalemm} this means that the move caused
$ss(P)$ to increase by at least $4(\log_c n + 1) + 1 > 4\log_c n + 1$
and hence was a major uphill move.
We conclude that
\begin{eqnarray*}
E\Biggl[ \sum_{h \in {\cal D}_F} \left( X_{n,h} \right.\Biggr.
&-& \Biggl.\left. 2B(\log_c n+1) \right) \Biggr]\\
&\leq& \sum_{h \in {\cal D}_F}
E\left[ \Bigl(\left( X_{n,h} - 2B(\log_c n+1) \right):
X_{n,h} > 2B(\log_cn+1)\Bigr) \right]\\
&\leq& E\left[\hbox{\em Number of major uphill moves}\right] ~~\leq~~ \frac{T}{n}
\end{eqnarray*}

\noindent
and consequently
\begin{equation}\label{deadbound}
E\left[ \sum_{h \in {\cal D}_F} X_{n,h} \right] < 2B^2(\log_c n+1) + \frac{T}{n}
= O(\log n)
\end{equation}

\noindent
for fixed $F$.
Combining (\ref{livebound}) with (\ref{deadbound}), we conclude that
\[
EW_n^{SS}(F) < E\left[ \sum_{h \in {\cal D}_F} X_{n,h} \right]
+ E\left[ \sum_{h \in {\cal L}_F} X_{n,h} \right ] = O(\log n).
\]

Thus all that remains is to exhibit a potential function $\psi$
that obeys (\ref{countbound}) and the three hypotheses of Hajek's Lemma.
Our previous potential function
$\phi(\bar{x}) = \sqrt{\tau_L(\bar{x})+\tau_D(\bar{x})}$
obeys (\ref{countbound}) and the Initial Bound and Bounded Variation
Hypotheses.
Unfortunately, it doesn't obey the Expected Decrease Hypothesis for all
bounded waste distributions $F$ with nontrivial dead-end levels.
There can exist realizable packings in which the count for the
largest dead-end level is arbitrarily large (and hence so is
$\phi(\bar{x})$), and yet any
item with size in $U_F$ will cause $\phi(\bar{x})$ to increase.
One can avoid such obstacles by taking instead the potential function $\psi$
to be $\sqrt{\tau_L(\bar{x})}$, the variant on
$\phi$ that simply ignores the dead-end level counts.
This function unfortunately fails to obey the Expected Decrease Hypothesis
for a different reason.
There are relevant situations in which any item with a size in $U_F$
will either cause an increase in $\tau_L(\bar{x})$ or else go in a bin
with a dead-end level and thus leave $\tau_L(\bar{x})$ unchanged.


Thus our potential function must somehow deal with the effects of items
going into dead-end level bins.
Let us say that a profile $\bar{x}'$ is {\em constructible} from a profile
$\bar{x}$ under $F$ if there is a way of adding items with sizes in
$U_F$ to dead-end level bins of a packing with profile $\bar{x}$ so that a packing
with profile $\bar{x}'$ results.
Let
\begin{equation}
\tau_0(\bar{x}) = \min \{\tau_D(\bar{x}'):
\bar{x}' \hbox{ is constructible from $\bar{x}$ under $F$}  \}\\
\end{equation}
Note for future reference that $\tau_0(\bar{x})$ can never decrease as
items are added to the packing.  Now let
\begin{equation}
r_D(\bar{x}) = \tau_D(\bar{x})-\tau_0(\bar{x})
\end{equation}
Thus $r_D(\bar{x})$
is the amount by which we can reduce $\tau_D(\bar{x})$ by adding items
with sizes in $U_F$ into bins with dead-end levels.
Our new potential function is
\begin{equation}
\psi(\bar{x}) = \sqrt{\tau_L(\bar{x}) + r_D(\bar{x})}
\end{equation}

Note that since we must always have $r_D(\bar{x}) \geq 0$,
we have $\psi(\bar{x}) \geq \sqrt{\tau_L(\bar{x})}$ and so
(\ref{countbound}) holds for $\psi$.
It remains to be shown that Hajek's Lemma applies to $\psi$.
This is significantly more difficult than showing it
applies to $\phi$ when $F$ has no dead-end levels.

First we prove a technical lemma that will help us understand
the intricacies of the $r_D(\bar{x})$ part of our potential function $\psi$.
Recall that if $r_D(\bar{x}) = t$,
then there is some list $L$ of items
with sizes in $U_F$ that we can add to the dead-end level
bins of a packing with profile $\bar{x}$
to get to one with a profile $\bar{y}$ such that
$\tau_D(\bar{y}) = \tau_D(\bar{x}) - t$,
and no such list of items can yield a profile $\bar{y}'$ with
$\tau_D(\bar{y}') < \tau_D(\bar{x}) - t$.
In what follows, we will use an equivalent graph-theoretic formulation
based on the following definition.

\begin{defi}\label{multigraph}
A {\em reduction graph} $G$ for $F$ is a directed multigraph
whose vertices are the dead-end levels for $F$ and
for which each arc $(h,i)$ is such that $i-h$ can be decomposed into a
sum of item sizes from $U_F$.
Such a graph $G$ is {\em applicable} to a profile $\bar{x}$ if
$outdegree_G(i) \leq x_i$ for all dead-end levels $i$.
The profile $G[\bar{x}]$ derived from applying $G$ to $\bar{x}$
is the vector $\bar{y}$ that has
$y_i = x_i + indegree_G(i) - outdegree_G(i)$ for
all dead-end levels and $y_i = x_i$ for all live levels.
We say that $G$ {\em verifies $t$ for $\bar{x}$} if
$\tau_D(\bar{x}) - \tau_D(G[\bar{x}]) \geq t$.
\end{defi}

Note that $r_D(\bar{x})$ equals the maximum $t$
verified for $\bar{x}$
by some applicable reduction graph $G$.
The list $L$ corresponding to $G$, i.e., the one that can be added to
$\bar{x}$ to obtain $G[\bar{x}]$, is a union of sets of items of total size $i-h$
for each arc $(h,i)$ in $G$.

\begin{lemm}\label{tau0lemm}
Let $G$ be a reduction graph with the minimum possible number of arcs that
verifies $r_D(\bar{x})$ for $\bar{x}$.
Then the following three properties hold:
\begin{enumerate}
\renewcommand{\theenumi}{\alph{enumi}} 
\renewcommand{\labelenumi}{(\roman{enumi})}
\item No vertex in $G$ has both a positive indegree and a positive outdegree.
\item Suppose that the arcs of $G$ are ordered arbitrarily as $a_1,a_2,\ldots,a_m$,
that we inductively define a sequence of profiles
$\bar{y}[0] = \bar{x}, \bar{y}[1], \ldots \bar{y}[m]$
by saying that $\bar{y}[i+1]$ is derived by applying the graph consisting
of the single arc $a_i$ to $\bar{y}[i-1]$, $1 \leq i \leq m$, and that
we define $\Delta[i] = \tau_D(\bar{y}[i-1]) - \tau_D(\bar{y}[i])$,
$1 \leq i \leq m$.
Then
\begin{align}
\sum_{i=1}^m\Delta[i] & = r_D(\bar{x})~and\label{deltasum}\\
\Delta[i] & > 0,~1 \leq i \leq m.\label{deltamin}
\end{align}

\item $G$ contains fewer than $\psi(\bar{x})$ copies of any arc $(h,i)$.
\end{enumerate}
\end{lemm}

\noindent\textbf{Proof.}
If (i) did not hold, there would be a pair of arcs $(h,i)$ and $(i,j)$
in $G$ for some $h,i,j$.
But note that then the graph $G'$ with these two arcs replaced by
$(h,j)$ would also verify $r_D(\bar{x})$ for $\bar{x}$,
and would have one less arc, contradicting our minimality assumption.

For (ii), equality (\ref{deltasum}) follows from a collapsing sum
argument and the fact that $\bar{y}[m] = G[\bar{x}]$.
The proof of (\ref{deltamin}) is a bit more involved.
Suppose there were some $k$ such that $\Delta[k] \leq 0$.
We shall show how this leads to a contradiction.
Consider the result of deleting arc $a_k = (h,j)$ from $G$,
thus obtaining new graph $G'$
and new sequences $\bar{y}[i]'$ and $\Delta_i'$, $1 \leq i \leq m-1$.
We will show that $G'$ also verifies $r_D(\bar{x})$
for $\bar{x}$, contradicting our minimality assumption.

Note that $\bar{y}[i]' = \bar{y}[i]$, $1 \leq i < k$,
and hence $\Delta[i]' = \Delta[i]$ for $1 \leq i < k$.
Thereafter the only difference between $\bar{y}[i]$ and $\bar{y}[i]'$ is that
$y[i]_h' = y[i+1]_h+1$ and $y[i]_j' = y[i+1]_j-1$.
Suppose $i \geq k$ and that $a_i = (r,q)$.
Note that by (i), $r \neq j$ and $q \neq h$.
Thus we have
$y[i]_r' \geq y[i+1]_r$ and $y[i]_q' \leq y[i+1]_q$ and by Lemma \ref{deltalemm}
(noting that $\Delta[i]$ as defined is $-1$ times the quantity evaluated in
that lemma),
\[
\Delta[i]' = 2\left(y[i]_r' - y[i]_q' -1\right)
\geq 2\left(y[i+1]_r - y[i+1]_q -1\right) = \Delta[i+1].
\]

\noindent
Thus we have by (\ref{deltasum})
\[
\sum_{i=1}^{m-1}\Delta[i]' \geq \sum_{i=1}^{m}\Delta[i] - \Delta[k]
\geq  \sum_{i=1}^{m}\Delta[i] = r_D(\bar{x}),
\]

\noindent
and so $G'$ verifies $r_D(\bar{x})$ for $\bar{x}$.
Since $G'$ has one less arc than $G$, this violates our assumption about
the minimality of $G$ and so yields
our desired contradiction, thus proving (\ref{deltamin}).

Finally, let us consider (iii).
Suppose there were $\psi(\bar{x})$ copies of some arc $(h,i)$
in $G$.
By (ii) we may assume that these are arcs $a_1,a_2,\ldots,a_{\psi(\bar{x})}$,
and that each yields an improvement in $\tau_D$.
Thus when the last is applied, the count for level $h$ must have been
at least 2 more than the count for level $j$, and inductively, when
arc $a_{\psi(\bar{x})+1-i}$ was applied, the difference in counts
had to be at least $2i$.
Now by Lemma \ref{deltalemm},
if the count for level $h$ exceeds that for level $j$
by $\delta$, then the decrease in $\tau_D$ caused by applying
the arc is $2\delta-2$.
Thus by (ii) we have
\begin{equation*}
\psi(\bar{x})^2 \geq r_D(\bar{x})
\geq \sum_{i=1}^{\psi(\bar{x})} (4i-2) = 2\psi(\bar{x})^2,
\end{equation*}

\noindent
a contradiction.
Thus (iii) and Lemma \ref{tau0lemm} have been proved. \proofmark\medskip

\medskip
Now let us turn to showing that Hajek's Lemma applies when $\psi$
plays the role of $\phi$.
Since the initial state is the empty packing, for which $\psi(\bar{x}) = 1$,
the Initial Bound Hypothesis is trivially satisfied.
For the Bounded Variation Hypothesis we must show that there is a
fixed bound $\Delta$ on $|\psi(\bar{x}')-\psi(\bar{x})|$, where $\bar{x}$
is any profile that can occur with positive probability in an $SS$ packing
under $F$ and $\bar{x}'$ is any profile that can be obtained by adding an
item with size $s \in U_F$ to a packing with profile $\bar{x}$ using $SS$.
We will show this for $\Delta = 10B$.
We may assume without loss of generality that $B \geq 2$,
as otherwise $EW_n^{SS}(F) = 0$ for all $n$.

There are two cases, depending on whether $\psi(\bar{x}') \geq \psi(\bar{x})$.
First suppose $\psi(\bar{x}') \geq \psi(\bar{x})$.
By Lemma \ref{squarelemm} it suffices to prove that
$\psi(\bar{x}')^2-\psi(\bar{x})^2 \leq 2\Delta\psi(\bar{x})
=20B\psi(\bar{x})$.
By Observation \ref{deadendobse}(b) we know that $s$ is not
a dead-end level and hence by (\ref{countbound}) $x_s \leq \psi(\bar{x})$.
Thus by the operation of $SS$ and the
fact that $\tau_0(\bar{x})$ cannot decrease,
the increase in $\psi(\bar{x})^2$ is at most
$(x_s+1)^2 - x_s^2 = 2x_s+1.$
If $x_s = 0$, this is clearly less than $10B$.
Otherwise, we have $\psi_L(\bar{x}) \geq \tau_L(\bar{x}) > 1$,
and so $2x_s+1 \leq 3\psi(\bar{x}) \leq 20B\psi(\bar{x})$,
as desired.

Suppose on the other hand that $\psi(\bar{x}') < \psi(\bar{x})$,
a significantly more difficult case.
We need to show that $\psi(\bar{x}) - \psi(\bar{x}') \leq \Delta = 10B$.
Lemma \ref{squarelemm} again applies, but now requires that we show
$\psi(\bar{x})^2 - \psi(\bar{x}')^2 \leq 2\Delta\psi(\bar{x}')$,
where the bound is in terms of the resulting profile $\bar{x}'$
rather than the initial one $\bar{x}$.
To simplify matters, we shall first show that the former
is within a constant factor of the latter.
This is not true in general, but we may restrict attention to
a case where it provably {\em is} true. 
In particular we may assume without loss of generality that
$\psi(\bar{x}) \geq 10B$, since otherwise it is obvious that any
placement will reduce $\psi(\bar{x})$ by at most $10B$.

\begin{lemm}\label{bigenoughlemm}
Suppose $F$ is a bounded waste distribution with $B\geq 2$, $\bar{x}$ is a profile
with $\psi(\bar{x}) \geq 10B$, and $\bar{x}'$ is the profile resulting
from using $SS$ to place an item of size $s \in U_F$ into a packing with
profile $\bar{x}$.  Then $\psi(\bar{x}') \geq \psi(\bar{x})/2$.
\end{lemm}

\noindent\textbf{Proof.}
By hypothesis, $\tau_L(\bar{x}) + r_D(\bar{x}) \geq 100B^2$.
We break into cases depending on the relative values of $\tau_L(\bar{x})$
and $r_D(\bar{x})$.

Suppose $\tau_L(\bar{x}) \geq r_D(\bar{x})$,
in which case $\tau_L(\bar{x}) \geq \psi(\bar{x})^2/2 \geq 50B^2$.
If the new item goes into a dead-end level bin,
then $\tau_L(\bar{x})$ remains unchanged and
$\psi(\bar{x}') \geq \sqrt{\psi(\bar{x})^2/2} \geq .707\psi(\bar{x})
> \psi(\bar{x})/2$.
If on the other hand the new item goes into a bin with a live level, say $h$,
then $\tau_L(\bar{x})$ will decline by at most $2x_h-1$.

We now break into two further subcases.
If $2x_h-1 < \tau_L(\bar{x})/2$, then we will have
$\tau_L(\bar{x}') > \tau_L(\bar{x})/2 \geq \psi(\bar{x})^2/4$ and so
$\psi(\bar{x}') > \sqrt{\psi(\bar{x})^2/4} = \psi(\bar{x})/2$.
If $2x_h-1 \geq \tau_L(\bar{x})/2$,
then $x_h > \tau_L(\bar{x})/4 \geq 12.5B^2$.
But this means that
$$\frac{\tau_L(\bar{x}')}{\tau_L(\bar{x})} \geq \frac{(x_h-1)^2}{x_h^2}
\geq \frac{\left(12.5B^2-1\right)^2}{\left(12.5B^2\right)^2}
\geq \left(\frac{49}{50}\right)^2 \geq .96$$
Thus $\tau_L(\bar{x}') \geq .96\tau_L(\bar{x}) \geq .48\psi(\bar{x})^2$
and $\psi(\bar{x}') \geq \sqrt{.48\psi(\bar{x})^2} \geq .69\psi(\bar{x})
> \psi(\bar{x})/2$.
Thus when $\tau_L(\bar{x}) \geq r_D(\bar{x})$
we have $\psi(\bar{x}') \geq \psi(\bar{x})/2$ in all cases.

Now suppose that $\tau_L(\bar{x}) < r_D(\bar{x})$,
in which case
$r_D(\bar{x}) \geq \psi(\bar{x})^2/2 \geq 50B^2$.
Consider the bin in which the new item is placed.
If its new level is a live level,
then so must have been its original level.
Thus $r_D(\bar{x})$ is unchanged,
and we have 
$\psi(\bar{x}') \geq \sqrt{\psi(\bar{x})^2/2} \geq .707\psi(\bar{x})
> \psi(\bar{x})/2$.

The only case remaining is when
$r_D(\bar{x}) \geq \psi(\bar{x})^2/2 \geq 50B^2$
and the new item increases the level of the bin that receives it
to a dead-end level.
Thus the count for one dead-end level increases by 1.
Let us denote this level by $h^+$.
If the item was placed in a bin with a live level, that is the only
change in the dead-end level counts.
Otherwise, an additional
one of those counts (the one corresponding to the original
level of the bin into which the item was placed) will decrease by 1.
Let $h^-$ denote this level if it exists.

In the terms of Lemma \ref{tau0lemm},
let $G$ be a minimum-arc graph that verifies $r_D(\bar{x})$ for
$\bar{x}$.
Let $G'$ equal $G$ if $h^-$ doesn't exist or if
$outdegree_G(h^-) < x_{h^-}$.
Otherwise let $G'$ be a graph obtained by deleting one of the out-arcs
leaving $h^-$ in $G$.
In both cases, $G$ will be applicable to $\bar{x}'$.
Order the arcs of $G$ so that the deleted arc (if it exists)
comes last, preceded by all the other arcs out of $h^-$ (if they exist),
preceded by the arcs into $h^+$ (if they exist), preceded by all remaining arcs,
and let the arcs of $G'$ occur in the same order as they do in $G$.
Let us now see what happens when we apply $G'$ to $\bar{x}'$,
and how this differs from what happens when we apply $G$ to $\bar{x}$.

Let $\delta(a)$ be the change in $\tau_D$ due to the application
of arc $a$ when $G$ is being applied to $\bar{x}$, and let $\delta'(a)$
be the change when $G'$ is applied to $\bar{x}'$.
By Lemma \ref{tau0lemm} and the definition of $\tau_0$ we have
\begin{eqnarray}
\label{deltaa}r_D(\bar{x}) &=& \sum_{a \in G}\delta(a)\\
\label{deltaaprime}r_D(\bar{x}') &\geq& \sum_{a \in G'}\delta'(a).
\end{eqnarray}
Thus to complete the proof of Lemma \ref{bigenoughlemm}, it will suffice to show
that
\begin{equation}
\sum_{a \in G'}\delta'(a) \geq c\sum_{a \in G}\delta(a).
\end{equation}
for an appropriate constant $c$.

Consider an arc $a=(i,j)$ in $G$ and let $n_i(a)$ and $n_j(a)$
($n_i'(a)$ and $n_j'(a)$)
be the corresponding level counts when $a$ is applied during the
course of applying $G$ to $\bar{x}$ ($G'$ to $\bar{x}'$).
By Lemma \ref{deltalemm} and the fact that since $i$ and $j$
are dead-end levels neither can be 0 or $B$, we have
$\delta(a) = 2(n_i(a)-n_j(a))-2$ 
and $\delta'(a) = 2(n_i'(a)-n_j'(a))-2$.

Let $h$ be one of $i,j$.
Observe that if $h \notin \{h^+,h^-\}$, then $n_h'(a) = n_h(a)$,
if $h = h^+$ then $n_h'(a) = n_h(a)+1$, and if
$h = h^-$ then $n_h'(a) = n_h(a)-1$.
Thus the only arcs $a=(i,j)$ for which $\delta'(a) < \delta(a)$
are those with $i=h^-$, $j=h^+$, or both.
If only one of the two holds, then $\delta'(a) = \delta(a) -2$.
If both hold then $\delta'(a) = \delta(a) -4$.
As a notational convenience, let $A^*$ denote the set of deleted arcs.
(Note that $A^*$ will either be empty or contain a single arc.)
Then we have
\begin{equation} \label{GprimeShortfall}
\sum_{a \in G'}\delta'(a) \geq \sum_{a \in G}\delta(a)
- 2\left(indegree_G(h^+) + outdegree_G(h^-)\right) - \sum_{a \in A^*}\delta_G(a).
\end{equation}
where $outdegree_G(h^-)$ is taken by convention to be 0 if $h^-$ does not
exist.

Let us deal with that last term first.
If there is an arc $a^* = (i,j)$ in $A^*$
then by our ordering of arcs in $G$ it is the last arc.
Suppose $\delta_G(a*) = 2(n_i(a)-n_j(a)) -2 > 4$.
Then we $n_i(a)-n_j(a) > 3$.
But this means that after the arc is applied we will have
$N_P(i) - N_P(j) \geq 2$, and so it would be possible to
apply an additional arc $(i,j)$, and this would further
decrease $\tau_D$ by at least 2.
But this contradicts our choice of $G$ as a graph whose application
to $\bar{x}$ yielded the maximum possible decrease in $\tau(\bar{x})$.
So we can conclude that
\begin{equation}\label{deltaa*}
\delta_G(a^*) \leq 4 \leq 2B.
\end{equation}

Now let us consider the rest of the right hand side of (\ref{GprimeShortfall}).
Let $M = indegree_G(h^+) + outdegree_G(h^-)$.
If $M \leq 10B$, then
\begin{equation*}
\sum_{a \in G}\delta(a) - \sum_{a \in G'}\delta'(a)
\leq 2M+2B \leq 22B \leq .11\psi(\bar{x})^2
\end{equation*}
since by assumption $\psi(\bar{x})^2 \geq 100B^2 \geq 200B$.
Thus by (\ref{deltaa}), (\ref{deltaaprime}), and our assumption that
$r_D(\bar{x}) \geq \psi(\bar{x})^2/2$,
\begin{equation*}
r_D(\bar{x}') \geq .39\psi(\bar{x})^2
\end{equation*}
and hence $\psi(\bar{x}') \geq .624\psi(\bar{x}) > \psi(\bar{x})/2.$

Thus we may assume that $M > 10B$.
Let $A_h$ denote the multiset of arcs in $G$ with $i=h^-$ or $j=h^+$
or both, and let us say that a pair $<i,j>$ of dead-end levels is
a {\em valid pair} if $A_h$ contains at least one arc $(i,j)$.
Note that there can be at most $B-1$ valid pairs,
since by Lemma \ref{tau0lemm} no vertex in $G$ can have both
positive indegree and positive outdegree.

Suppose $<i,j>$ is a valid pair
and there are $m$ copies of arc $(i,j)$ in $A_h$.
By Lemma \ref{tau0lemm} each copy must decrease $\tau_D$ when
it is applied, so if we let the last copy of $(i,j)$ in our defined order
be $a_1$, the next-to-last by $a_2$, etc., we will have $\delta_G(a_k) \geq 2$,
$1 \leq k \leq m$.
Moreover, since an application of an arc $(i,j)$ reduces
$N_P(i)-N_P(j)$ by at least 1, and since by Lemma \ref{tau0lemm}
applications of other arcs cannot increase $N_P(i)$ or
decrease $N_P(j)$, we must in fact have $\delta_G(a_k) \geq \delta_G(a_{k-1})+2$,
$2 \leq k \leq m$.
If $(i,j) = (h^-,h^+)$ then each application reduces $N_P(i)-N_P(j)$ by 2,
and so in this case $\delta_G(a_k) \geq \delta_G(a_{k-1})+4$,
$1 \leq k \leq m$.
Thus
$$
\sum_{k=1}^m\delta_G(a_k) \geq \left\{
\begin{array}{ll}
\sum_{k=1}^m(2k) = m(m+1), & i = h^- \hbox{ or } j = h^+ \hbox{ but not both}\\
&\\
\sum_{k=1}^m(4k-2) = 2m^2, & i = h^- \hbox{ and } j = h^+
\end{array}
\right.
$$
Since $2m^2 \geq m(m+1)$ for all $m \geq 1$, we thus have
\begin{equation}\label{MslashBfloor}
\sum_{a \in A_h}\delta_G(a) \geq (B-1)\left\lfloor\frac{M}{B-1}\right\rfloor
\left(\left\lfloor\frac{M}{B-1}\right\rfloor + 1\right) \geq
\frac{M^2}{B-1} - M
\end{equation}
\noindent
Then by (\ref{GprimeShortfall}), (\ref{deltaa*}), (\ref{MslashBfloor}),
and our assumption that $M > 10B$, we have
\begin{eqnarray}
\frac{\sum_{a \in G}\delta(a) - \sum_{a \in G'}\delta'(a)}
{\sum_{a \in G}\delta(a)}
&\leq& \frac{2M+2B}{\frac{M^2}{B-1}-M}
= \frac{2+\frac{B}{M}}{\frac{M}{B-1}-1}\\ \nonumber
&&\\ \nonumber
&\leq& \frac{2.1(B-1)}{M-B+1} \leq \frac{2.1(B-1)}{9B+1}
\leq \frac{2.1}{19} \leq .111
\end{eqnarray}
Thus by (\ref{deltaa}) and (\ref{deltaaprime}) and our assumption
that $r_D(\bar{x}) \geq \psi(\bar{x})/2$, we have
\begin{equation*}
r_D(\bar{x}') \geq
.889r_D(\bar{x}) > .444\psi(\bar{x})^2
\end{equation*}
And hence $\psi(\bar{x}') \geq \sqrt{r_D(\bar{x}')}
> .666\psi(\bar{x}) > \psi(\bar{x})/2$.
Thus in all cases we have $\psi(\bar{x}') \geq \psi(\bar{x})/2$
and Lemma \ref{bigenoughlemm} is proved.  \proofmark

\medskip
Returning to the proof that Hajek's Lemma applies,
recall that we are in the midst of proving that the Bounded Variation
Hypothesis holds, and are left with the task of showing
that $\psi(\bar{x}) - \psi(\bar{x}') \leq 10B$ in the case where
$\psi(\bar{x}') < \psi(\bar{x})$.
By Lemma \ref{squarelemm} it will suffice to show that
$\psi(\bar{x})^2 - \psi(\bar{x}')^2 \leq 20B\psi(\bar{x}')$
when $\psi(\bar{x}) \geq 10B$,
which by Lemma \ref{bigenoughlemm} will follow if we can show that
\begin{equation}\label{downbound}
\psi(\bar{x})^2 - \psi(\bar{x}')^2 \leq 10B\psi(\bar{x})
\end{equation}

As in the proof of Lemma \ref{bigenoughlemm}, we divide the
difference $\psi(\bar{x})^2 - \psi(\bar{x}')^2$ into two parts
that we will treat separately:
$\tau_L(\bar{x}) - \tau_L(\bar{x}')$
and $r_D(\bar{x})-r_D(\bar{x}')$.

We begin by bounding the first part.
If the item being packed goes in an empty bin, then a live level gets
increased and no dead-end level is changed, so $\psi(\bar{x})$ {\em increases},
contrary to hypothesis.
If the item being packed goes into a bin with a dead-end level,
then $\tau_L(\bar{x})$ remains unchanged.
If the item goes into a bin with a live level $h$, then
by (\ref{countbound}) we have that $x_h \leq \psi(\bar{x})$,
so by Lemma \ref{deltalemm} the decrease in $\tau_L$ is at most
$2x_h-1 < 2\psi(\bar{x}) \leq B\psi(\bar{x})$.
Thus to prove (\ref{downbound}) it will suffice to prove that
$r_D(\bar{x})-r_D(\bar{x}') \leq 9B\psi(\bar{x})$.

To bound this second difference, note first that the hypotheses
of Lemma \ref{bigenoughlemm} hold.
So as in the proof of that Lemma,
let $G$ be a graph that verifies $r_D(\bar{x})$.
If the placement of the item changes no dead-end level counts,
there is nothing to prove, so we again may assume that there
is a dead-end level $h^+$ that increases by 1 and (possibly)
a dead-end level $h^-$ that decreases by 1.
As in the proof of the Lemma we have
\begin{equation}\label{GprimeShortfall2}
r_D(\bar{x})-r_D(\bar{x}')
\leq 2\bigl(indegree_G(h^+)+outdegree_G(h^-)\bigr) + 2B
\end{equation}
where by convention $outdegree_G(h^-)$ is taken to be 0 if $h^-$ doesn't
exist.

Also, as in the proof of Lemma \ref{bigenoughlemm}, there are
at most $B-1$ distinct pairs $<i,j>$ such that $(i,j)$ is an arc of $G$
and $i = h^-$, $j = h^+$, or both.
But then by Lemma \ref{tau0lemm}(iii) we have fewer than
$\psi(\bar{x})$ copies of each.
Given that arcs $(h^-,h^+)$ will be double counted in
$indegree_G(h^+)+outdegree_G(h^-)$, we thus have
$$indegree_G(h^+)+outdegree_G(h^-) < B\psi(\bar{x})$$
Combining this with (\ref{GprimeShortfall2}) we conclude that
$$r_D(\bar{x})-r_D(\bar{x}') \leq 2B\psi(\bar{x}) + 2B < 9B\psi(\bar{x})
$$
We thus conclude (\ref{downbound}) holds
and hence so does the Bounded Variation Hypothesis.

\medskip
To complete the proof that Hajek's Lemma applies, all that remains
is to show that the Expected Decrease Hypothesis holds.
Essentially the same proof that was used when there were no nontrivial
dead-end levels will work, except that Lemma \ref{goodmovelemm}
needs to be modified to account for the possibility of such levels
and we need to show that both it and Lemma \ref{orlinlemm} hold for
$\psi(\bar{x})^2$.

This is straightforward for Lemma \ref{orlinlemm}, which essentially
says that assuming $F$ is a perfectly packable distribution, the expected
increase in $\phi(\bar{x})^2$ that can result from using $SS$ to pack an item
generated according to $F$ is less than 2.
This will hold for $\psi(\bar{x})^2$ as well since by definition
\begin{eqnarray*}
\psi(\bar{x})^2 &=& \tau_L(\bar{x}) + r_D(\bar{x})\\
&=& \tau_L(\bar{x})+\tau_D(\bar{x})-\tau_0(\bar{x})\\
&=& \phi(\bar{x})^2 - \tau_0(\bar{x}),
\end{eqnarray*}
and by definition $\tau_0(\bar{x})$ can never decrease.

As to Lemma \ref{goodmovelemm}, we need only modify it by increasing the
two key constants involved.
The precise values of these constants are not
relevant to satisfying the Expected Decrease Hypothesis.
In particular, we can prove the following variant on Lemma \ref{goodmovelemm}.

\begin{lemm}\label{goodmovelemm2}
Let $F$ be a bounded waste distribution
and let $P$ be any packing that can be created by applying $SS$
to a list of items all of whose sizes are in $U_F$.
If $\bar{x}$ is the profile of $P$ and $\psi(\bar{x}) > 2\sqrt{2}B^{3/2}$,
then there is a size $s \in U_F$ such that if an item of size $s$ is packed
by $SS$ into $P$, the resulting profile
$\bar{x}'$ satisfies
$$\psi(\bar{x}')^2 \leq \psi(\bar{x})^2 - \frac{\psi(\bar{x})}{B^2}.$$
\end{lemm}

\noindent\textbf{Proof.}
Since $\tau_0(\bar{x})$ can never decrease, the result will follow if we can
show that there exists an item size $s$ such that if an item
of size $s$ is packed by $SS$, $\psi(\bar x)^2 - \tau_0(\bar x)
= \tau_L(\bar x)+\tau_D(\bar x) = ss(P)$
will decline by at least $\psi(\bar{x})/B^2$.

Suppose $\tau_L(\bar{x}) \geq \psi(\bar{x})^2/2$.
Then as in the argument used in the proof of Lemma \ref{goodmovelemm}
there has to be a live level $h$ with $x_h \geq \sqrt{\tau_L(\bar{x})/B}
\geq \psi(\bar{x})/(\sqrt{2B})$
and hence a size $s$ that will
cause $ss(P)$ to decline by at least
$$2\left(\frac{x_h}{B} -1\right) \geq 2\frac{\psi(\bar{x})}{\sqrt{2}B^{3/2}}-2
\geq \frac{\psi(\bar{x})+2\sqrt{2}B^{3/2}}{\sqrt{2}B^{3/2}} - 2
= \frac{\psi(\bar{x})}{\sqrt{2}B^{3/2}}
\geq \frac{\psi(\bar{x})}{B^2}.$$

Suppose on the other hand that $\tau_L(\bar{x}) < \psi(\bar{x})^2/2$.
In this case we must have
$r_D(\bar{x}) > \psi(\bar{x})^2/2$.
Let $G$ be a minimum-arc reduction graph that verifies
$r_D(\bar{x}) \geq \psi(\bar{x})^2/2$,
and suppose $G$ contains $m$ arcs, ordered as $a_1,a_2,\ldots,a_m$.
By Lemma \ref{tau0lemm}(i),(iii), we know that $m < (B-1)\psi(\bar{x})$.
Thus by Lemma \ref{tau0lemm}(ii) we know that for some $i$, $1 \leq i \leq m$,
\[
\Delta[i] > \frac{r_D(\bar{x})}{m}
> \frac{\psi(\bar{x})^2}{2m}
> \frac{\psi(\bar{x})^2}{2B\psi(\bar{x})} = \frac{\psi(\bar{x})}{2B} \geq
\frac{\psi(\bar{x})}{B^2},
\]
where recall that $\Delta[i]$ is defined to be the reduction in $\tau_D$
when the arc $a_i$ is applied to the intermediate profile $\bar{y}[i-1]$,
created by the application of earlier arcs in sequence to $\bar{x}$.
Suppose arc $a_i = (h,j)$.
Now by Lemma \ref{tau0lemm}(i), the fact that $h$ is the source of arc
$a_i$ means that it cannot have been a sink of a previous arc,
so we must have $y[i-1]_h \leq x_h$.
Similarly the fact that $j$ is the sink of arc $a_i$ means that it
cannot be the source of any previous arc, so $y[i-1]_j \geq x_j$.
But then the reduction in $\tau_D$ that would be obtained
if $a_i$ were applied directly to $\bar{x}$, i.e., if an item
of size $j-h$ is placed in a bin of level $h$, is by Lemma \ref{deltalemm}

\[
2(x_h - x_j - 1) \geq 2(y[i-1]_h - y[i-1]_j -1) = \Delta[i]
> \frac{\psi(\bar{x})}{B^2}.
\]

\noindent
Thus, $SS$ will place an item of size $s=j-h$ in such a way as to
reduce $ss(P)$ by at least this much.  \proofmark

The remainder of the proof that
Expected Decrease Hypothesis
is satisfied by $\psi(\bar{x})$ proceeds just as the
proof for $\phi(\bar{x})$ did when there were no nontrivial
dead-end levels.
Thus Hajek's Lemma applies and the upper bound of Theorem
\ref{lognupperbound} is proved.  \proofmark

\subsubsection{Proof of the $\Omega(\log n)$ Lower Bound.}
We begin the proof with a sequence of lemmas.

\begin{lemm} \label{maxup}
Suppose $s$ is a divisor of the bin size $B$.
Then if an item of size $s$ is placed into a packing $P$ using $SS$, the value
of $ss(P)$ can increase by at most 1.
\end{lemm}

\noindent\textbf{Proof.}
If there is a bin of level $B-s$, then placing an item of size $s$ into that
bin would decrease $ss(P)$.
If there is no bin of level $s$, then starting a new bin with an item
of size $s$ will increase $ss(P)$ by 1.
Otherwise, let $h_s = \max\{h: s|h \mathrm{\ and\ } N_P(h) > 0\}$,
and note by assumption that $h \leq B-2s$.
By Lemma \ref{deltalemm}, placing an item of size $s$ in one of the bins with
level $h_s$ increases $ss(P)$
by at most $2(N_P(h_s+s)-N_P(h_s))+2 \leq 0$.
Thus in every case there is a way to increase $ss(P)$ by 1 or less, and so
$SS$ must choose a move that increases $ss(P)$ by at most 1.
\proofmark

Let us say that a level $h$ is {\em divisible for $F$} if any set of
items with sizes in $U_F$ that has total size $h$ can contain
only items whose sizes are divisors of $B$.

\begin{lemm} \label{divisorlemm}
If $h$ is a nontrivial dead-end level for $F$ then $h$ is not divisible for $F$.
\end{lemm}

\noindent\textbf{Proof.}
Let $\cal H$ be the set of all levels $i$, $1 \leq i \leq B-1$, that
are divisible for $F$ and assume, for the sake of contradiction, that
$h \in \cal H$.
Since $h$ is a nontrivial dead-end level for $F$, there
is some list $L$ that under
$SS$ yields a packing containing at least two bins with level $h$.
Consider the first time during the packing of $L$ that
a level $i \in \cal H$ had its count $N_P(i)$ increase from 1 to 2,
and let $s$ be the size of the item $x$ whose placement caused this to happen.
By definition of {\em divisible level}, $s$ must be a divisor of $B$,
and so by Lemma \ref{maxup}, the placement of
$x$ can have increased $ss(P)$ by at most 1.
But this is impossible:  If $i = s$ then the insertion
of $x$ would have increased $ss(P)$
by $2^2-1^2=3$.
On the other hand, suppose $i>s$.
Since $i$ is a divisible level, so is $i-s$.
Thus $N_P(i-s) = N_P(i) = 1$ just before $x$ was packed:
Neither count can exceed 1 by our choice of $i$, the latter must be 1 if it is
to increase to 2 after the placement of $x$, and the former must be 1 since
$x$ can only create a bin with level $i$ if there is a bin of level
$i-s$ into which it can be placed.
However, this means that $ss(P)$ increases by 2, contradicting Lemma \ref{maxup}.
So $h \notin \cal H$, as desired.  \proofmark

\begin{lemm}\label{stairlemm}
Suppose $s$ is an item size that does not evenly divide the bin capacity
$B$ and we are asked to pack an
arbitrarily long sequence of items of size $s$ using $SS$.
Let $d_i = is$, $0 \leq i \leq \lfloor B/s \rfloor$.
For all $m>0$, the packing in existence just before the
first time $N_P(d_1) > m$ must have
$N_P(d_i) = mi$ for every $d_i$.
\end{lemm}

\noindent
{\bf Proof.}
Let us say that $mi$ is the {\em target} for level $d_i$.
We first show that it must be the case that
$N_P(d_i)$ is no more than its target, $1 < i \leq \lfloor B/s \rfloor$,
so long as $N_P(d_1)$ has never yet exceeded its target.
Suppose not, and consider the packing just before the first one of these counts,
say $N_P(d_i)$, exceeded its target.
In this packing we must have $N_P(d_i) = mi$.
Let $\Delta_h = N_P(d_h) - N_P(d_{h-1})$,
$1 \leq h \leq \lfloor B/s \rfloor$, where by convention $N_P(0) = 0$ and so
$\Delta_1 = N_P(d_1)$.
Since $\Delta_1$ by hypothesis is $m$ or less, Lemma \ref{deltalemm}
implies that $\Delta_i < \Delta_1 \leq m$.
But then we must have $N_P(d_{i-1}) \geq (i-1)m+1$,
contradicting our assumption that level $d_i$ was the first to
have its count exceed its target.

For the lower bound, note that in the packing just before $N_P(d_1)$ first
exceeds $m$, it must be the case that $\Delta_1 = N_P(d_1) = m$.
Since this was the preferred move under $SS$, it must be the case
by Lemma \ref{deltalemm} that
$\Delta_i \geq m$, $2 \leq i \leq \lfloor B/s \rfloor$.
The result follows.  \proofmark

\begin{lemm}\label{Hlemm}
Suppose $F$ is a fixed discrete distribution with at least one
nontrivial dead-end level $h$ and $H$ is a positive constant.
Then there is a list $L_{H}$ of length $O(H)$ consisting solely of
items with sizes in $U_F$, such that
the packing resulting from using $SS$ to pack $L_{H}$
contains at least $H$ bins with dead-end levels.
\end{lemm}

\noindent
{\bf Proof.}
By Lemma \ref{divisorlemm} there must be a set $S = \{x_0,x_1,\ldots,x_t\}$
of items with sizes in
$U_F$ whose total size is $h$, and for which $s(x_0)$ is not a divisor of $B$.
Let us also assume that all items of any given size appear contiguously
in the sequence $s(x_0),s(x_1),\ldots,s(x_t)$.
Note that we may assume that $s(x_i) \geq 2$, $0 \leq i \leq t$, since if
1 were in $U_F$ there could be no dead-end levels.
Let $h_i = \sum_{j=0}^i s(x_j)$, $0 \leq i \leq t$.
Note that $h_t = h$.
Further, let $k = \lfloor B/s(x_0) \rfloor$ and $d_i = i\cdot s(x_0)$,
$0 \leq i \leq k$.

Our list $L_H$ will consist of a sequence of $t+1$ (possibly empty)
segments, the first of which (Segment 0)
consists of $H3^t\sum_{i=1}^ki^2 = H3^tk(k+1)(2k+1)/6$ items of size $s(x_0)$.
In the packing $P$ obtained by using $SS$ to pack these items, we
will have by Lemma \ref{stairlemm} that level $i\cdot s_1$ will have count
$iH3^t$, $1 \leq i \leq k$, and in particular level $h_0 = s(x_0)$ will
have level $H3^t$.
In what follows we use ``$P$'' generically to denote the current packing.
Note that after Segment 0 has been packed, $P$ contains
$H3^t\sum_{i=1}^ki = H3^tk(k+1)/2$ partially filled bins.

Segment 1 consists of the shortest possible sequence
of items of size $s(x_1)$ that, when added to $P$ using $SS$,
will cause the count for level $h_1 = s(x_0)+s(x_1)$ to equal or exceed
$H3^{t-1}$.  A sequence of this sort must exist for the following reasons:
If $N_P(h_1)$ is itself $H3^{t-1}$ or greater, as for instance it would be
if $s(x_1)=s(x_0)$, then the empty segment will do.
Otherwise, suppose $N_P(h_1) < H3^{t-1}$.
So long as $N_P(h_0) \geq 2H3^{t-1}$ and $N_P(h_1) < H3^{t-1}$,
placing an item of size $s(x_1)$ in a bin with level $h_0$
would cause a greater reduction in $ss(P)$ than placing it in a
bin of level $h_1$ could, and so would be the preferred move.
Since we can place $H3^{t-1}$ items in bins of level $h_0$
before $N_P(h_0) \leq 2H3^{t-1}$, and
each such placement would increase $N_P(h_1)$ by 1,
this means we will eventually
have placed enough to increase $N_P(h_1)$ to the desired target.
Note that we will eventually be forced to place items in bins of level
$h_0$ rather than some level other than $h_0$ or $h_1$, since
the existence of moves that decrease $ss(P)$ means that
no new bins are being created.

We complete our argument by induction.
In general, we start Segment $j$, $2 \leq j \leq t$ with a packing
in which $N_P(h_{j-1}) \geq H3^{t-j+1}$ and no new bins have been
created since Segment 0.
The segment then consists of the shortest possible sequence of
items of size $s(x_{j})$ that will cause the count for level
$h_j = h_{j-1}+s(x_{j})$ to equal or exceed $H3^{t-j}$.
An argument analogous to that for Segment 1 says that this must
eventually occur without any additional bins being started.
Thus at the end of Phase $t$ we have $H$ bins with level $h_t = h$.
Given that all the $s(x_j)$ are 2 or greater,
the total number of items included in Segments 1 through $t$, none
of which started a new bin, is no more than $BH3^tk(k+1)/4$
and so the total number of items in our overall list $L_H$
is at most
$$\frac{H3^tk(k+1)(2k+1)}{6} + \frac{BH3^tk(k+1)}{4} < B^33^BH = O(H)$$
for fixed $F$, as required.
\proofmark

For future reference, note that since 1 cannot be in $U_F$ if $F$
has dead-end levels, the number of segments in $L_H$ is less than $B/2$.

\begin{lemm} \label{profilelemm}
Suppose $P$ and $Q$ are two packings for which
$$|P-Q| \equiv \sum_{h=1}^{B-1} |N_P(h)-N_Q(h)| = M$$
and $L$ is a list consisting
entirely of items of the same size $s \geq 2$.  Then the packings $P'$ and $Q'$
resulting from using $SS$ to pack $L$ into $P$ and $Q$
satisfy $|P' - Q'| \leq BM.$
\end{lemm}

\noindent\textbf{Proof.}
We prove the lemma for the special case of $M=1$.
The general result then follows by repeated applications of
this $M=1$ case.  So assume $|P-Q| = 1$.

Let $g$ denote the level that has different counts under $P$ and $Q$
and suppose without loss of generality that $N_P(g) = N_Q(g)+1$.
Let $P_i$ and $Q_i$ denote the packings that result after the first $i$
items of $L$ have been packed into $P$ and $Q$ respectively.
We will say that a triple $(i,j,\ell)$, $0 \leq i,j \leq |L|$ and
$0 \leq \ell \leq B$, is a {\em compatible} triple if either
\begin{enumerate}
\item $P_i=Q_j$ and $\ell \in \{0,B\}$, or
\item $|P_i-Q_j| = 1$, and $\ell$ is the unique bin level such that
$1 \leq \ell \leq B-1$, $N_{P_j}(\ell) = N_{Q_i}(\ell)+1$.
\end{enumerate}
Note that by this definition $(0,0,g)$ is a compatible triple.

\begin{claim}
If $(i,j,\ell)$ is a compatible triple with $i,j <|L|$ then
one of the following three triples must also be compatible:
$$(i+1,j+1,\ell),~~(i+1,j,\ell+s),~~(i,j+1,\ell-s).$$
\end{claim}

\noindent\textbf{Proof of Claim.}
Consider the packings $P_i$ and $Q_j$. 
Suppose $SS$ would place an item of size $s$ in bins with
the same level in both $P_i$ and $Q_j$, as for instance it
must if $\ell \in \{0,B\}$ and hence the two packings have
identical level counts.
Then the same bins counts would be changed in the same
way for $P_i$ and $Q_j$ and so $(i+1,j+1,\ell)$ is a compatible triple.

Otherwise suppose $SS$ would place an item of
size $s$ in bin $h_P$ for $P_i$ and in $h_Q$ for $Q_j$, with
$h_P\neq h_Q$.
In this case $P_i$ and $Q_j$ must be different, and we are in case 2
of compatibility. 
Let $\Delta_Q(h)$ (resp. $\Delta_P(h)$) denote the net reduction in the
sum of squares if an item of size $s$ is placed in a bin of level $h$ in
$Q_j$ (resp. $P_i$), assuming such a placement is legal.
Since the
bin counts $N_{P_i}(h)$ and $N_{Q_j}(h)$ are equal for every $h$ other than
$\ell$, it follows that $\Delta_P(h)=\Delta_Q(h)$ for all $h$'s other than
$\ell$ and $\ell-s$. Since $SS$ makes different choices for $P_i$ and 
for $Q_j$, it must be that at
least one of $h_Q, h_P$ is either  $\ell$ or $\ell-s$.  
By hypothesis we have $N_{P_i}(\ell) = N_{Q_j}(\ell)+1$ and all other counts
are equal, so $\Delta_{P}(\ell-s) < \Delta_{Q}(\ell-s)$ 
(if $\ell-s\geq 0$), $\Delta_{P}(\ell) > \Delta_{Q}(\ell)$
(if $\ell+s \leq B$), and for all other values of $h$,
$\Delta_{P}(h) = \Delta_{Q}(h)$.

Thus if $h_P = \ell-s$ we must have $h_Q = \ell-s=h_P$, given
that it is even more valuable to place an item of size $s$ into a bin of
level $\ell-s$ in $Q_j$ than in $P_i$.
Similarly, if $h_Q = \ell$ then we must have $h_P = \ell$.
Since by assumption $h_P \neq h_Q$, this means that either
$h_P = \ell$ or $h_Q = \ell-s$.

In the first case, $h_P=\ell$, we must have $\ell+s \leq B$.
Packing an item of size $s$ into a bin with level $\ell$ in $P_i$
reduces $N_{P_i}(\ell)$ by 1, so that $N_{P_{i+1}}(\ell)=N_{Q_j}(\ell)$.
If $\ell+s = B$, i.e. we fill up a bin, then $|P_{i+1}-Q_j|=0$,
and so $(i+1,j,\ell+s=B)$ is a compatible triple.
If $\ell+s < B$ then $N_{P_i}(\ell+s)$ will increase by 1 and we will
have $N_{P_{i+1}}(\ell+s)=N_{P_i}(\ell)+1=N_{Q_{j}}(\ell+s)+1$, while
all other levels now have the same counts.
Thus $(i+1,j,\ell+s)$ is again a compatible triple.

In the second case, $h_Q = \ell-s$, we must have $\ell -s\geq 0$.  
Packing an item of size $s$ into a bin with level $\ell -s$ in
$Q_j$  increases $N_{Q_j}(\ell)$ by 1, so that 
$N_{P_{i}}(\ell)=N_{Q_{j+1}}(\ell)$.
If $\ell-s = 0$, i.e. we pack $s$ into a new bin, then 
$|P_i-Q_{j+1}|=0$, and so $(i,j+1,\ell-s=0)$ is a compatible triple.
If $\ell-s >0$ then $N_{Q_j}(\ell-s)$ will decrease by 1 and we will have
$N_{Q_{j+1}}(\ell-s)=N_{Q_{j}}(\ell-s)-1=N_{P_i}(\ell-s)-1$, while
all other levels now have the same counts.
Thus $(i,j+1,\ell-s)$ is again a compatible triple.

This completes the proof of the Claim. \proofmark

\smallskip
Given the Claim and the fact that $(0,0,g)$ is a compatible triple,
we have by induction that at least one of the three following scenarios
must hold:
\begin{enumerate}
\item $(|L|,|L|,g)$ is a compatible triple, or
\item There is an integer $a$, $1 \leq a \leq (B-g)/s$ such that
$(|L|,|L|-a,g+as)$ is a compatible triple, or
\item There is an integer $b$, $1 \leq b \leq g/s$ such that
$(|L|-b,|L|,g-bs)$ is a compatible triple.
\end{enumerate}

In the first case we have $|P'-Q'| = 1$, which clearly satisfies the
Lemma's conclusion.  In the second we have $|P' - Q_{|L|-a}| = 1$,
but to get $Q'$ from $Q_{|L|-a}$ we will need to add $a$ additional
items of size $s$, and each addition will change one or two level
counts by 1.
Since $s \geq 2$ and $g \geq 1$, we must have $a \leq (B-g)/s \leq (B-1)/2.$
Thus we can conclude that $|P'-Q'| \leq 1 + B-1 = B$ as desired.
The third case follows analogously and the Lemma is proved.  \proofmark

\begin{lemm}\label{XDlemm}
Suppose $F$ is a fixed discrete distribution with at least one
nontrivial dead-end level and $X$ is a positive constant.
Then for any $D > X$ there is a list $L_{X,D}$ of length $O(D)$
consisting solely of items with sizes in $U_F$, such that for any
packing $P$ with no live-level count exceeding $X$,
the packing $Q$ resulting from using $SS$ to add $L_{X,D}$ into $P$
contains at least $D$ bins with dead-end levels.
\end{lemm}

\noindent\textbf{Proof.}
We may assume that $P$ contains fewer than $D$ bins with dead-end
levels, because the number of bins with dead-end levels can never
decrease and if we already had $D$ such bins any list will do for
$L_{X,D}$.
Let $h$ be a nontrivial dead-end level for $F$.
For our list we simply let $L_{X,D}$ be the list $L_H$ derived
for $h$ using Lemma \ref{Hlemm},
with $H = (XB+D)B^{B/2} + D = O(D)$ for fixed $F$.
By Lemma \ref{Hlemm} the length of $L_H$ will by $O(H) = O(D)$.

If $P_0$ denotes the empty packing, we know by Lemma \ref{Hlemm}
that if $SS$ is used to pack $L_H$ into $P_0$
it will create a packing $P_0'$ with at least $H$ bins having
the dead-end level $h$.
Let $P'$ denote the packing that would result if we used $SS$ to add
$L_H$ to $P$.
Note that $|P-P_0| = \sum_{i=1}^{B-1}N_P(i) \leq X(B-1)+D-1$.
Thus by applying Lemma \ref{profilelemm} once for each segment of
$L_H$ and using the fact that $L_H$ contains less than $B/2$
segments, we have that $|P'-P_0'| < B^{B/2}(XB+D)$.
But this means that for dead-end level $h$ we must have
$N_{P'}(h) > H - B^{B/2}(XB+D) = D$ and so $P'$ contains at least the
desired number of bins with dead-end levels.  \proofmark

\begin{lemm} \label{Xlemm}
Let $P_N$ be the packing after $N$ items generated according to $F$
have been packed by $SS$.
There is a constant $X$, depending only on $F$, such that for
any $N>0$
\begin{equation}
p[N_{P_N}(i) \leq X \mbox{ for all live levels } i] \geq \frac{1}{2}
\end{equation}
\end{lemm}

\noindent\textbf{Proof.}
Recall from the inequality (\ref{Tbound}) of the proof of the $O(\log n)$
upper bound on the expected waste of $SS$ that for any $N > 0$,
if $X_N$ is the profile after packing $N$ items, then there are a
constants $c$ and $T$, depending only on $F$, such that
$$
E \left[ c^{\psi(X_N)}\right] \leq T
$$
This meant that
$$
p \left[ c^{\psi(X_N)} > 2T \right] \leq \frac{1}{2}
$$
and hence that
$$
p\left[ \psi(X_N) > \log_c(2T) \right] \leq \frac{1}{2}
$$
Since as we have repeatedly observed
$\psi(\bar x) \geq x_h$ for every live level $h$,
this in turn means that the probability is at least $1/2$ that no
live level count exceeds $\log_c(2T)$.
Thus the Lemma holds with $X = \log_c(2T)$.  \proofmark

\medskip
We are now in a position to prove our $\Omega(\log n)$ lower bound on
$EW_n^{SS}(F)$ when $F$ has nontrivial dead-end levels.
We may assume without loss of generality that all the sizes $s_1,\ldots,s_J$
specified by $F$ are in $U_F$, i.e., that $p_j > 0$, $1 \leq j \leq J$.
Consider the lists $L_{X,D}$ specified by Lemma \ref{XDlemm} for the
value of $X$ given by Lemma \ref{Xlemm}, and let $\ell_D$ denote
the length of $L_{X,D}$.  Since the value of $X$ depends
only on $F$, Lemma \ref{XDlemm} implies that there is a constant $c$,
depending only on $F$,
such that for all $D>X$, $\ell_D < cD$.

Now suppose we have a random list $L$ of length $cD$ of items generated
according to $F$.
The probability that $L_{X,D}$ is a prefix of $L$ is at least
$\epsilon^{\ell_D}$, where $\epsilon = \min\{p_j:~1\leq j \leq J\}$.
Let $a = \log_2(1/\epsilon)$.
Then the probability that $L_{X,D}$ is {\em not} a prefix of $L$
is at most $(1 - (1/2)^{acD})$.

Now consider a random list $L^*$ of length $cD2^{acD}$, viewed as a sequence
of $2^{acD}$ random segments of length $cD$.  The probability that none
of these segments has $L_{X,D}$ as a prefix is
$$\left( 1 - \frac{1}{2^{acD}} \right)^{2^{acD}}
< ~~\frac{1}{e} ~~<~~ \frac{1}{2}.$$

In other words, the probability that at least one of these segments has
$L_{X,D}$ as a prefix exceeds $1/2$.
Consider the {\em last} segment that has $L_{X,D}$ as a prefix (should
any such segments exist),
and the packing $P$ that exists just before this copy of $L_{X,D}$ is packed.
Note that by choosing the last such segment, we do not condition in any way
the list that precedes this copy or the packing $P$.
Hence by Lemma \ref{Xlemm}, with probability at least $1/2$ the packing
$P$ has no live level count exceeding $X$, and
by Lemma \ref{XDlemm}, after the segment is added to the packing,
the new packing (and all subsequent ones) will contain at least $D$ bins
with dead-end levels.
Thus the expected number of bins with dead-end levels after all of
$L^*$ is packed is at least $(1/2)(1/2)D = D/4 = \Omega(\log |L^*|)$.
The lower bound follows.  \proofmark

\section{SS and Linear Waste Distributions} \label{worstsect}

The implication of Theorem \ref{orlintheo} that $ER_\infty^{SS} (F) =
1$ for all perfectly packable distributions $F$
unfortunately does not carry over to the case where $EW_n^{OPT} =
\Theta(n)$.

\begin{theo}\label{1.5theo}
There exist distributions $F_k$, $1 \leq k \leq \infty$,
such that
$$\limsup_{k\rightarrow\infty}ER_\infty^{SS}(F_k) = 1.5~.$$
\end{theo}

\noindent\textbf{Proof.}
Let $F_k$ be the distribution in which
the bin size is $B = 2k+1$ and the single item size 2 occurs with
probability 1.
Consider an $n$-item list $L_n$ generated according to $F_k$ where
$n$ is divisible both by $k$ and by $\sum_{i=1}^k i^2 = k(k+1)(2k+1)/6$.
Then $OPT(L_n) = n/k$ and
by Lemma \ref{stairlemm}, we have
$$SS(L_n) = \left(\frac{n}
{\sum_{i=1}^k i^2}\right) \left( \sum_{i=1}^k i\right) = n \cdot
\frac{\left(\frac{k(k+1)}{2}\right)}{\left(\frac{k(k+1)(2k+1)}{6}\right)} =
\frac{3n}{2k+1}.$$
Thus $ER_\infty^{SS}(F_k)$, which is defined as a $\limsup$, equals
$3k/(2k+1)$ and the Theorem follows.  \proofmark

We conjecture that $3/2$ is the worst possible value for $ER_\infty^{SS}(F)$
over all discrete distributions $F$,
although at present the best upper bound we can prove is 3, which
is implied by the following worst-case result.

\begin{theo}\label{worstcaseUB}
For all lists $L$, $SS(L) \leq 3\lceil s(L)/B\rceil \leq 3OPT(L).$
\end{theo}

\noindent\textbf{Proof.}
Let $x$ be the last item of size less than $B/3$ that starts a new bin
and let $s$ be the size of $x$.
(If no such $x$ exists, then all bins are at least $B/3$ full in the
final packing and we are done.)
Let $P$ be the packing just before $x$ was packed.
It is sufficient to show that the average bin content in the bins of $P$
is at least $B/3$.
If that is so, then the packing of subsequent items
cannot reduce the average bin content in the bins not containing
$x$ to less than $B/3$.
Consequently if $m$ is the final number of bins in the packing,
we must have $s(L)/B > (m-1)/3$ and hence
$OPT(L) \geq \lceil s(L)/B \rceil \geq m/3$ and the theorem follows.

So let us show that that the average bin content in the bins of $P$
is at least $B/3$.
For $1 \leq j\leq s$, let $\ell_j$ as the greatest integer such that
$j+\ell_j s<B$ and let $\Omega_j$ denote the set of bins with contents
$j,j+s,\ldots,j+\ell_j s$.
Note that $\Omega_1,\ldots,\Omega_{s}$ is a
partition of the bins of $P$ into $s$ sets, and if we can show that the average
contents of the bins in each nonempty $\Omega_j$ is at least $B/3$,
we will be done.
Fix $j$ and suppose $k$ is the least integer such that either
$N_P(j+ks)>0$ or $j+ks\geq B/3$.
If $j+ks\geq B/3$ then {\em every} bin in $\Omega_j$ has contents at least $B/3$
and so $\Omega_j$ behaves as desired.
So suppose $j+ks<B/3$, in which case we must have $k< \ell_j$.
Since $SS$ places $x$ in a new bin, we must by Lemma \ref{deltalemm} have
\[
0 \leq N_P(s) \leq N_P(j+hs+s)-N_P(j+hs),\quad h=k,\ldots,\ell_j-1
\]
and hence $N_P(j+\ell_j s)\geq \cdots \geq N_P(j+ks)$.
This means that if we let $t = j+ks$ the
average contents of the bins in $\Omega_j$ is at least
\[
\frac{t + (t+s) + \cdots + (t+(\ell_j -k) s)}{\ell_j -k+1}
= \frac{2t+(\ell_j -k)s}{2}
> \frac{j+\ell_j s}{2} \geq \frac{B-s}{2} > \frac{B}{3}\,.~~\proofmark
\]

\section{Identifying Perfectly Packable Distributions} \label{lpsect}
\setcounter{equation}{0}

Given the observations of the previous section,
it would be valuable to be able to identify those distributions
$F$ that satisfy the hypotheses of Theorem \ref{orlintheo}, i.e., those
for which $EW_n^{OPT}(F) = O(\sqrt{n})$ and hence $ER_\infty^{SS}(F) = 1$
is guaranteed.
This task is unfortunately NP-complete, as it would require us to solve
the PARTITION problem \cite{GJ79}.
Fortunately, however, the problem is not NP-complete in the strong sense, and
as we shall now see, can be solved in time pseudo-polynomial in $B$ via
linear programming, as was claimed but not proved in \cite{CJK99}.

Suppose our discrete distribution is as described above,
with a bin capacity $B$, integer item sizes $s_1,s_2,\ldots,s_J$, and
rational probabilities $p_1,p_2,\ldots,p_J$.
We may assume without loss of generality that all these probabilities
are positive.
Our linear program, which for future reference we shall call the
``Waste LP for $F$,'' will have $JB$ variables $v(j,h)$, $1 \leq j \leq J$
and $0 \leq h \leq B-1$, where $v(j,h)$ represents the rate at which
items of size $s_j$ go into bins whose current level is $h$.
The constraints are:

\vspace{.1in}
\begin{eqnarray}
v(j,h)  \geq  0, && 1 \leq j \leq J, ~~0 \leq h \leq B-1 \label{lp1}\\
v(j,h)  =  0, && 1 \leq j \leq J, ~~s_j > B-h \label{lp1'}\\
\sum_{h=0}^{B-1} v(j,h)  =  p_j, && 1 \leq j \leq J \label{lp2}\\
\sum_{j=1}^J v(j,h)  \leq 
\sum_{j=1}^J v(j,h-s_j), && 1 \leq h \leq B-1 \label{lp3}
\end{eqnarray}

\noindent
where by definition the value of $v(j,h-s_j)$ when $h-s_j < 0$ is taken to be 0
for all $j$.
Constraints (\ref{lp1'}) say that no item can go into a bin that is
too full to have room for it.
Constraints (\ref{lp2}) say that all items must be packed.
Constraints (\ref{lp3}) say that bins with a given level are created at least as fast
as they disappear.
The goal is to minimize
\begin{equation}
c(F) \equiv \sum_{h=1}^{B-1} \left ( (B-h) \cdot \left ( \sum_{j=1}^J v(j,h-s_j) - 
\sum_{j=1}^J v(j,h) \right ) \right ) \label{lp4}
\end{equation}
Note for future reference that by definition we must have $c(F) < B-1$.

\medskip
In what follows, $c(F)$ will always denote the optimal solution value for
the Waste LP for $F$, and $ES(F)$ will denote the expected item size under
$F$, i.e., $\sum_{i=1}^Jp_js_j$.

\begin{lemm} \label{lplemm}
Suppose $F$ is a discrete distribution and let $L_n(F)$ be a random
$n$-item list generated according to $F$.
\begin{enumerate}
\item For all $n>0$, $\displaystyle{
\left| EW_n^{OPT}(F) - \frac{nc(F)}{B} \right| \leq O(\sqrt{n}).}$
\item There exist constants $b$ and $N^*$ such that for all $n \geq N^*$
$$P \left[ \; \left| OPT(L_n(F)) - \frac{n}{B}\Bigl(ES(F)+c(F)\Bigr) \right|
 >  bn^{2/3} \right]\leq \frac{1}{n^{1/6}}~.$$
\end{enumerate}
\end{lemm}

This lemma, which we shall prove shortly, implies the following three results.

\begin{theo} \label{limsuptheo}
Suppose $F$ is a discrete distribution.
Then
$$\limsup_{n\rightarrow\infty}\left(\frac{EW_n^{OPT}(F)}{n}\right) = \frac{c(F)}{B}~.$$
\end{theo}

\begin{theo} \label{EWtheo}
Suppose $F$ is a discrete distribution.
Then $EW_n^{OPT}(F) = O(\sqrt{n})$ if and only if $c(F) = 0$.
\end{theo}

\begin{lemm} \label{ERlemm}
Suppose $F$ is a discrete distribution and
$A$ is a (possibly randomized) bin packing algorithm for which
$E[A(L)]/OPT(L) \leq b$ for some fixed constant $b$ and all lists $L$.  Then
$$ER_\infty^A(F) = 
\frac{ES(F) + B\cdot\limsup_{n\rightarrow\infty}EW_n^A(F)/n}{ES(F)+c(F)}~.$$
\end{lemm}

Theorems \ref{EWtheo} and \ref{limsuptheo} are immediate consequences
of claim (1) of Lemma \ref{lplemm}.
Lemma \ref{ERlemm} follows from claim (2).
Basically, it says that $ER_\infty^A(F)$, which is defined in terms
of expected ratios, can actually be computed in terms of ratios
of expectations.  It follows because (2) implies that we can divide
the set of lists $L$ of length $n$ generable according to $F$ into
two sets.  For the first set, which has cumulative probability $1-1/n^{1/6}$,
we have
\begin{equation}\label{EWcoro1}
E\left[ \frac{A(L)}{OPT(L))} \right] =
\left(\frac{nES(F)+B\cdot EW_n^A(F)}{nES(F)+nc(F)}\right)
\left(1 + O\left(\frac{1}{n^{1/3}}\right)\right)
\end{equation}
For the second set, which has cumulative probability $1/n^{1/6}$,
$E[A(L)/OPT(L)] \leq b$.
Thus this set contributes at most $b/n^{1/6}$ to the overall expected
ratio for $L_n(F)$, meaning that (\ref{EWcoro1}) holds with
$L$ replaced by $L_n(F)$ and $1/n^{1/3}$ replaced by $1/n^{1/6}$.
Lemma \ref{ERlemm} follows.
We now turn to the proof of Lemma \ref{lplemm}.

\medskip
\noindent
{\bf Proof.}
Consider the values $v(j,h)$ of the variables in
an optimal basic solution to the LP.
Since all the coefficients and right-hand sides of the LP are rational,
all these variable values must be rational as well, and there exists
a positive integer $N$ such that $Nv(j,h)$ is an integer,
$1 \leq j \leq J$ and $0 \leq h \leq B$.
For each positive integer $k$, let $L_k$ be a list consisting of
$k\sum_{h=0}^{B-1}Nv(j,h)$ items of
size $s_j$, $1 \leq j \leq J$.
By (\ref{lp2}) $L_k$ will contain $kNp_j$ items of size $s_j$ for
each $j$, for a total of $kN$ items.
We will thus have $s(L_k) = kN \cdot ES(F)$.

Note that we can construct a packing of $L_k$ simply by following
the instructions provided by the variable values in the solution to the LP.
That is, for each $j$, start $Nv(j,0)$ bins by placing an item of size
$s_j$ into an empty bin.  By (\ref{lp3}), the number of bins of level 1
will now be at least $\sum_{j=1}^JNv(j,1)$.
Thus we can take a set consisting
of $Nv(j,1)$ items of size $s_j$, $1 \leq j \leq J$, and place each of
these items in a distinct bin with level 1.
We can now proceed to pack bins of level 2, and so on.
Let $P_k$ denote the resulting packing.

How many bins does this packing contain?
A bin in $P_k$ that has level $h$ contains items of total
size $h$ by definition, and in addition has a gap of size $B-h$.
Thus the total number of bins is simply the $1/B$ times the
sum of the item sizes and the sum of the gap sizes, that is

$$
\frac{1}{B} \left (kN \cdot ES(F) + kN\sum_{h=1}^{B-1} \left ( (B-h) \cdot \left ( \sum_{j=1}^J v(j,h-s_j) - 
\sum_{j=1}^J v(j,h) \right ) \right ) \right )
$$

\noindent
and hence

\begin{equation} \label{lptheo3a}
\left | P_k \right | ~~=~~ \left(\frac{kN}{B}\right)\Bigl(ES(F) + c(F)\Bigr).
\end{equation}

Now, since $L_k$ is in essence the ``expected value'' of
the random list $L_{kN}(F)$, we can
use the packings $P_k$ as models for packing the
random lists $L_n(F)$, $n > 0$.
We proceed as follows:  Given $n$, find that $k \geq 0$ such that
$kN \leq n < (k+1)N$.
Now note that the packing $P_k$ has $kNp_j$ ``slots'' for items of
size $s_j$, $1 \leq j \leq J$, and $L_n(F)$ is expected to have
between $kNp_j$ and $(k+1)Np_j$ such items.
Place as many items of $L_n$ into the appropriate slots as possible,
and then place the leftover items in additional bins, one per bin.
The total number of bins used will then be $|P_k|$ plus the number
$X_n$ of leftover items, which implies that

\begin{equation} \label{lptheo3b}
OPT(L_n(F)) ~~\leq~~ \left(\frac{n}{B}\right)\Bigl(ES(F) + c(F)\Bigr) + X_n.
\end{equation}

Let $n_j$ denote the number of items of size $s_j$ among the first
$kN$ items of $L_n(F)$ and define
\begin{eqnarray*}
\Delta_j^+ = \max\{0,n_j - kNp_j\}, & 1 \leq j \leq J\\
\Delta_j^- = \max\{0,kNp_j - n_j\}, & 1 \leq j \leq J
\end{eqnarray*}
Thus $\Delta_j^+$ is the {\em oversupply} of items of size $s_j$
among the first $kN$ items and $\Delta_j^-$ is the {\em shortfall}.
The number of leftover items among the first $kN$
items of $L_n$ is hence $\sum_{j=1}^J \Delta_j^+ = \sum_{j=1}^J \Delta_j^-$,
and so $X_n < N + \sum_{j=1}^J \Delta_j^+$.
Since each $n_j$ is a sum of independent Bernoulli variables
when considered by itself,
we have $E[\Delta_j] \leq \sqrt{kNp_j(1-p_j)} < \sqrt{kNp_j}$.
Given that $\sum_{j=1}^J \sqrt{kNp_j}$
is maximized when all the probabilities are equal, we have that
$E[\sum_{j=1}^J \Delta_j] \leq J \sqrt{kN/J}  \leq \sqrt{nJ}$ and so
$E[X_n] \leq N+\sqrt{nJ} = O(\sqrt{n})$ since $N$ and $J$ are constants.

Since $X_n$ is a nonnegative random variable, we thus can conclude
from (\ref{lptheo3b}) that Claim (2) of the lemma holds when the
quantity inside the absolute value signs is positive.
Since $E[s(L_n(F))] = nES(F)$ we can also conclude that

\begin{equation} \label{lptheo3d}
EW_n^{OPT}(F) = E \left[ \; OPT(L_n(F)) - \frac{s(L_n(F))}{B} \; \right]
\leq \frac{nc(F)}{B} + O(\sqrt{n})
\end{equation}

\noindent
and so (1) also holds when the quantity
inside the absolute value signs is positive.

To prove that (1) and (2) hold when the quantities inside the
absolute value signs are negative, first observe that the
packing $P_k$ defined above for $L_k$ must be an optimal packing
for $L_k$.  If not, i.e., if $OPT(L_k) \leq (kN/B)(ES(F)+c(F))$,
then we could use an optimal packing for $L_k$ to define a better
solution to our LP, contradicting
our assumption that $c(F)$ was the optimal solution value for the LP.

Next observe that if we are given a packing $P$ for $L_n(F)$,
we can construct a closely related one for $L_k$
(as defined above, with $k = \lfloor n/N \rfloor$), by a process
of addition.
For each of the at most $\sum_{j=1}^J \Delta_j^- = \sum_{j=1}^J \Delta_j^+$
items in $L_k$ that do not have counterparts of the same size in $L_n$,
we add a new bin to $P$ containing just that item.
This new packing contains at least as many items of each size
as does $L_k$ and so must contain at least $OPT(L_k)$ bins.
Thus by (\ref{lptheo3a}) we must have

\begin{equation} \label{lptheo3e}
OPT(L_n(F)) + \sum_{j=1}^J\Delta_j^+ \geq OPT(L_k) =
\left(\frac{kN}{B}\right)\Bigl(ES(F) + c(F)\Bigr)
\end{equation}

Claims (1) and (2) then follow by the same analysis
of $E[\sum_{j=1}^J\Delta_j]$
as was used when the quantity inside the absolute value signs
was positive.  \proofmark

Thus one can determine whether $EW_n^{OPT}(F)$ is sublinear and,
if it is not,
compute the constant of proportionality on the expected linear waste,
all in the time it takes to construct and solve the Waste LP for $F$.
The worst-case time for this process obeys the following time bound.

\begin{theo}\label{lptimetheo}
Given a description of a discrete distribution $F$ in which all
probabilities are presented as rational numbers with a common
denominator $D \geq B$, the Waste LP for $F$
can be constructed and solved in time
$$O\left( (JB)^{4.5}\log^2D\right) = O\left(B^9\log^2D\right).$$
\end{theo}

\noindent\textbf{Proof.}
Given its straightforward description, the LP can clearly be constructed in
time proportional to its size, so construction time will be dominated by
the time to solve the LP.
For that, the best algorithm currently available is that of Vaidya \cite{V89},
which runs in time $O((M+N)^{1.5}NL^2)$, where $M$ is the larger of the number
of variables and the number of constraints (the latter including the
``$\geq 0$'' constraints), and $N$ is the smaller, and $L$ is a measure of
the number of bits needed in the computation if all operations are to be
performed in exact arithmetic.

Our LP has $JB$ variables and the number of constraints is $\Theta(JB)$.
Thus for our LP the running time is $O((JB)^{2.5}L^2) = O(B^5L^2)$.
To obtain a bound on $L$, note that all coefficients in the constraints
of the LP are $1$, $0$, or $-1$ and the coefficients in the objective function
are all $O(B)$.
The leaves the probabilities $p_j$ to worry about.
Note that we can determine $c(F)$ by solving the LP with each $p_j$
replaced by its numerator (the integer $Dp_j$), and then dividing the
answer by $D$.
If we proceed in this way, then all the ``probabilities'' are integers
bounded by $D$.
Following the precise definition of $L$ given in \cite{V89} we can then
conclude that $L = O(JB \log D)$, giving us the overall running time bound
claimed.  \proofmark

Although this running time bound is pseudopolynomial in $B$, it will
be polynomial if $B$ is polynomially bounded in terms of $J$, which
is true for many of the distributions of interest in practice.
Moreover, much better running times are obtainable in practice by
using commercial primal simplex codes rather than interior point
techniques to solve the LP's.
See \cite{ABDJS02} which details simplex-based methods that can be used to
compute $c(F)$ in reasonable
time for discrete distributions with $J$ and $B$ as large
as 1,000 and 10,000, respectively.

\vspace{.5\baselineskip}
In the remainder of this section, we will show how
we can further distinguish between the cases in which
$EW_n^{OPT}(F) = \Theta(\sqrt{n})$ and those in which $EW_n^{OPT}(F) = O(1)$.
Our goal is to distinguish cases (a) and (b) in the Courcoubetis-Weber theorem,
as described in Section 2.
Thus we need to determine, given that $\bar{p}_F$ is in $\Lambda_F$,
whether it is also in the interior of $\Lambda_F$.
Our approach is based on solving $J$ additional, related LP's.
The total running time will simply be $J+1$ times that for solving
the original LP, and so we will be able to determine whether
$EW_n^{OPT}(F) = O(\sqrt{n})$ and if so, which of the two cases hold,
in total time $O(J^{5.5}B^{4.5}\log^2 D) = O(B^{10}\log^2D)$.

For each $i$, $1 \leq i \leq J$, let $x_i \geq 0$ be a new variable
and let ${\rm LP}_i$ denote the linear
program obtained from the Waste LP for $F$ by (1) changing the
inequalities in (\ref{lp3}) to equalities, (2) replacing (\ref{lp2}) by

\begin{eqnarray}
\sum_{h=0}^{B-1} v(i,h)  & = & p_i + x_i \nonumber\\ 
\sum_{h=0}^{B-1} v(j,h)  & = & p_j,~~~1 \leq j \neq i \leq J, \label{lp2'}
\end{eqnarray}

\noindent
and (3) changing the optimization criterion to ``maximize $x_i$.''
Let $c_i(F)$ denote the optimal objective function value for ${\rm LP}_i$.
Note that ${\rm LP}_i$ is feasible for $x_i = 0$ whenever $c(F) = 0$,
so that $c_i(F)$ is always well-defined and non-negative in this case.

\begin{theo} \label{lptheo2}
If $F$ is a discrete distribution, then $EW_n^{OPT}(F) = O(1)$ if and only
if $c(F) = 0$ and $c_i(F) > 0$, $1 \leq i \leq J$.
\end{theo}

\noindent
{\bf Proof.}
Combining the Courcoubetis-Weber Theorem with Theorem \ref{EWtheo} we know that
for all discrete distributions $F$,

\begin{equation}\label{cf-lambda}
\bar{p}_F \in \Lambda_F \mbox{ if and only if } c(F) = 0.
\end{equation}

Let $\bar{q}(i,\beta)$ denote the vector obtained from
$\bar{p}_F$ by setting $q_i = p_i + \beta$ and
$q_j = p_j, 1 \leq j \neq i \leq J$.
By (\ref{cf-lambda}) and the construction of
the linear programs ${\rm LP}_i$, it is easy to see that
$\bar{q}(i,\beta)$ is in $\Lambda_F$ if and only if ${\rm LP}_i$ is
feasible when $x_i = \beta$.
Thus by convexity, $\bar{q}(i,\beta)$ is in $\Lambda_F$ if and only
if $0 \leq \beta \leq c_i(F)$.

Let us first suppose that the stated properties of $c(F)$ and the $c_i(F)$'s
do not hold.
If $c(F) \neq 0$, then $\bar{p}_F$ is not even in $\Lambda_f$, much less
in its interior.
So suppose $c(F) = 0$ but $c_i(F) = 0$ for some $i$, $1 \leq i \leq J$.
Then for any $\epsilon > 0$ there is a vector $\bar{q}$ with
$|q - \bar{p}_F| \leq \epsilon$ that is not in $\Lambda_F$, namely
$\bar{q}(i,\epsilon)$.  Thus by definition $\bar{p}_F$ is not in the
interior of $\Lambda_F$.

On the other hand, suppose $c(F) = 0$ and $c_i(F) > 0$, $1 \leq i \leq J$.
To show that $\bar{p}_F$ is in the interior of $\Lambda_F$, we
make use of two elementary properties of such cones:

\begin{enumerate}
\item[C1.] \label{cone1} If the vector
$\bar{a} = \langle a_1,\ldots,a_d \rangle$ is in
a cone $\Lambda$, then so is the vector
$r\bar{a} = \langle ra_1,\ldots,ra_d \rangle$ for any $r > 0$.
\item[C2.] \label{cone2} If vectors  $\bar{a} = \langle a_1,\ldots,a_d \rangle$ and
$\bar{b} = \langle b_1,\ldots,b_d \rangle$ are in $\Lambda$, then so is
the vector sum $\bar{a} + \bar{b} =  \langle a_1+b_1,\ldots,a_d+b_d \rangle$.
\end{enumerate}

In other words, any positive linear combination of elements of the cone
is itself in the cone.  Our proof works by showing that there is an
$\epsilon$ such that any $\bar{q}$ with $|\bar{p}_F - \bar{q}| \leq \epsilon$
can be constructed out of a positive linear combination of vectors
$\bar{q}(i,\beta_i)$ with $0 \leq \beta_i \leq c_i(F)$, $1 \leq i \leq J$.
We begin by defining a set of key quantities.

\begin{eqnarray*}
c_{min} & = & \min \{c_i(F): 1 \leq i \leq J\}\\
p_{max} & = & \max \{p_i: 1 \leq i \leq J\}\\
p_{min} & = & \min \{p_i > 0: 1 \leq i \leq J\}\\
\delta & = & \min \left \{ \frac{1}{2},\frac{c_{min}}{4Jp_{max}} \right \}\\
\epsilon & = & \min \left \{ \frac{p_{min}}{4} , \left ( \frac{c_{min}}{8J} \right )
\left ( \frac{p_{min}}{p_{max}} \right ) \right \}
\end{eqnarray*}

Note that by hypothesis $c_{min} > 0$ and since $F$ is a probability distribution
there must be some positive $p_i$'s and so $p_{min} > 0$.
Hence $\delta$ and $\epsilon$ are also positive.
Suppose $\bar{q} = \langle q_1,\ldots,q_J \rangle$ is any vector with
$|\bar{p}_F - \bar{q}| \leq \epsilon$.
We will show that $\bar{q}$ can be constructed out of a positive
linear combination of vectors
$\bar{q}(i,\beta_i)$ as specified above.

Let $\epsilon_i = q_i - (1-\delta)p_i$, $1 \leq i \leq J$.
We first observe that all the $\epsilon_i$ are positive.
This is clearly true for all $i$ such that $q_i \geq p_i$.
Suppose $q_i < p_i$.  In that case $p_i$ cannot be 0, so we must
have $p_i \geq p_{min}$.
If $\delta = 1/2$ we have

\begin{equation} \label{ep1}
\epsilon_i = q_i - p_i + \delta p_i \geq \delta p_i - \epsilon \geq
\frac{p_{min}}{2} - \frac{p_{min}}{4} = \frac{p_{min}}{4}> 0.
\end{equation}

\noindent
If on the other hand $\delta = c_{min}/(4Jp_{max})$, then

\begin{equation} \label{ep2}
\epsilon_i \geq \delta p_i - \epsilon \geq
\left ( \frac{c_{min}}{4Jp_{max}} \right ) p_{min} - \left ( \frac{c_{min}}{8J} \right )
\left ( \frac{p_{min}}{p_{max}} \right ) =  \left ( \frac{c_{min}}{8J} \right )
\left ( \frac{p_{min}}{p_{max}} \right ) > 0.
\end{equation}

\noindent
We next observe that for each $i$, $1 \leq i \leq J$,

\begin{equation} \label{ep3}
\epsilon_i \leq \epsilon + \delta p_i \leq  \frac{c_{min}}{8J}
+ \frac{c_{min}}{4Jp_{max}}p_{max} <  \frac{c_{min}}{2J}.
\end{equation}

Now consider the vectors $\bar{q}(i,\beta_i)$, where
$\beta_i = J\epsilon_i/(1-\delta)$, $1 \leq i \leq J$.
By (\ref{ep1}) through (\ref{ep3}) and the definition of $\delta$, we have

$$
0 < \beta_i = \frac{J\epsilon_i}{1-\delta} \leq 2J \left ( \frac{c_{min}}{2J} \right ) = c_{min},
$$

\noindent
and so all these vectors are in $\Lambda_F$.  Now consider the vector

$$
\bar{r} = \langle r_1,\ldots,r_J \rangle = \frac{1-\delta}{J}\sum_{i=1}^J \bar{q}(i,\beta_i).
$$

Since $\bar{r}$ is a positive linear combination of vectors in $\Lambda_F$,
it is itself in $\Lambda_F$ by (C1) and (C2).
But now note that for $1 \leq i \leq J$, we have

$$
r_i = \left (\frac{1-\delta}{J} \right ) (Jp_i) + \left ( \frac{1-\delta}{J}
\right ) \left ( \frac{J\epsilon_i}{1-\delta} \right )
= (1-\delta)p_i + \epsilon_i = q_i.
$$

Thus $\bar{q} = \bar{r}$ and the latter is in $\Lambda_F$, as claimed.
This implies that $\bar{p}_F$ is in the interior of $\Lambda$ and the
theorem is proved. \proofmark

\section{Handling Non-Perfectly Packable Distributions} \label{tuned}
\setcounter{equation}{0}

In this section we consider the case when $EW_n^{OPT}(F) = \Theta(n)$.
As we saw in Section 4, we can have $ER_\infty^{SS}(F) > 1$ for such $F$.
Fortunately, for each such $F$ one can design a distribution-specific
variant on $SS$ that performs much better.
For notational simplicity in what follows, we shall assume 
without loss of generality that the
size vector $\bar{s}$ for $F$ has $s_1=1$.
(If $1 \notin U_F$ then we simply set $p_1=0$.)
Note also that we must have $B > 1$.

\begin{theo} \label{tunedtheo1}
For any discrete distribution $F$ with $EW_n^{OPT}(F) = \Theta(n)$,
there exists a randomized
variant $SS_F$ of $SS$ such that
$EW_n^{SS_F}(F) = EW_n^{OPT}(F) + O(\sqrt{n})$
and hence $ER_\infty^{SS_F}(F) = 1$ by Lemmas \ref{lplemm} and \ref{ERlemm}.
This algorithm has expected running time $O(nB)$
and can itself be constructed in time polynomial in $B$ and the size
of the description of $F$.
\end{theo}

\noindent
{\bf Proof.}  Algorithm $SS_F$ is based on the solution
to the Waste LP for $F$,
and in particular on the optimal solution value $c(F)$, which
by Theorem \ref{lptimetheo} can be computed in
time polynomial in $B$ and the size of the description of $F$.
The algorithm works by performing a series of steps,
with new steps being taken
so long as an item in $L$ remains to be packed.
At each step we flip a biased coin and according to the outcome proceed
as follows.
\begin{enumerate}
\item With probability $1/(1+c(F))$ we take the next item from
$L$ and pack it according to $SS$.
\item With probability $c(F)/(1+c(F))$ we generate a new ``imaginary''
item of size 1 and pack it according to $SS$.
\end{enumerate}

Let $G_n$ denote the total size of the gaps in the packing of
$L_n(F)$ by this algorithm, and let $I_n$ denote the total size
of the imaginary items in the packing.  Then
\begin{equation}\label{wasteform}
EW_n^{SS_F}(F) = \frac{E[I_n]+E[G_n]}{B}
\end{equation}

It is straightforward to determine $E[I_n]$.  Divide the
packing process into $n$ phases, each phase ending on a step in which
a real rather than imaginary item is packed.
The expected number of imaginary items packed in each phase is
$$\sum_{i=1}^\infty\left(\frac{c(F)}{1+c(F)}\right)^i = c(F).$$
We thus can conclude the expected
total number of imaginary items is $nc(F)$, and since each is
of size 1 we have $E[I_n] = nc(F)$.

Let us now turn to $E[G_n]$.
Note that if we consider both real and imaginary items, we are essentially
packing a list generated by the distribution $F^+$ that has
$p_1^+ = (p_1+c(F))/(1+c(F))$ and $p_i^+ = p_i/(1+c(F))$ for all $i > 1$.

\begin{claim}\label{onesclaim}
$EW_n^{OPT}(F^+) = O(\sqrt{n})$.
\end{claim}

\noindent
{\bf Proof of Claim.}
By Theorem \ref{EWtheo} all we need show is that the solution to
the Waste LP for $F^+$ has $c(F^+) = 0$.
Denote this LP by $LP_{F^+}$ and denote the Waste LP for $F$
by $LP_F$.
Let $v_0(j,h)$ be the variable values in an optimal solution for $LP_F$,
and for $1 \leq h \leq B-1$ define
$$\Delta_h=\sum_j v_0(j,h-s_j)-\sum_jv_0(j,h).$$
Define a new assignment $v$ by
\begin{eqnarray*}
v(j,h) & = & \frac{v_0(j,h)}{1+c(F)},~~ j \neq 1\\
v(1,h) & = & \frac{v_0(1,h)+\sum_{h'=1}^h\Delta_{h'}}{1+c(F)}
\end{eqnarray*}
for $0 \leq h \leq B-1$.

We claim that $v$ satisfies the constraints of $LP_{F^+}$ and achieves
0 for the objective function, this implying that $c(F^+)=0$.
It is easy to see that $v$ satisfies constraints
(5.1), (5.2), and (5.3) for $j\neq 1$. 
For $j=1$, we have
\begin{eqnarray*}
\sum_{h=0}^{B-1}v(1,h)&=&\frac{1}{1+c(F)}\left(p_1+\sum_{h=1}^{B-1}\sum_{h'=1}^h\Delta_{h'}\right)\\[.2in]
&=&\frac{1}{1+c(F)}\left(p_1+\sum_{h'=1}^{B-1}(B-h')\Delta_{h'}\right)\\[.2in]
&=& \frac{1}{1+c(F)}\Bigl(p_1+c(F)\Bigr),
\end{eqnarray*}
as required.  As for the constraints (5.4), we have for each $h$,
$1 \leq h \leq B-1$, that
$$
\sum_j v(j,h-s_j)-\sum_jv(j,h) = \frac{1}{1+c(F)}
\left(\Delta_h+\sum_{h'=1}^{h-1}\Delta_{h'}-\sum_{h'=1}^h\Delta_{h'}\right)~=~0.
$$
Thus $v$ is a feasible solution for $LP_{F^+}$.
Finally, the value of the objective function is
$$
\sum_{h=1}^{B-1}\Bigl(B-h\Bigr)
\left(\sum_j v(j,h-s_j)-\sum_jv(j,h)\right)~=~0.
\mbox{  \proofmark}
$$

Thus $F^+$ is a perfectly packable distribution and by Lemma \ref{orlinlemm}
the expected increase in $ss(P)$ during each step of algorithm
$SS_F$ is less than 2, no matter what the current packing looks like.
For all $i>0$
the expected increase during step $i$ is thus less than 2 times the probability
$SS_F$ takes $i$ or more steps.
Since the expected number of steps by the above argument about $E[I_n]$
is $n(1+c(F))$,
the expected value of $ss(P)$ when the algorithm terminates is thus no more
than $2n(1+c(F))$.
By Lemma \ref{cauchylemm} this implies that $E[G_n] \leq B\sqrt{(B-1)n(1+c(F)})
= O(B^2\sqrt{n})$ since $c(F) \leq B-1$ by definition.
Thus by (\ref{wasteform}) we have
$$EW_n^{SS_F}(F) = \frac{nc(F) + O(B^2\sqrt{n})}{B}$$
which by Lemma \ref{lplemm} is $EW_n^{OPT}(F) + O(\sqrt{n})$, as desired.

All that remains is to show that algorithm $SS_F$ can be implemented
to run in time $O(nB)$.  This is not immediate, since there are
distributions $F$ for which $c(F)$ is as large as $\lceil B/2 \rceil -1$.
Thus the total number of items packed (including imaginary ones)
can be $\Theta(nB)$, and the standard implementation of $SS$ will take
$\Theta(nB^2)$.  We avoid this problem by using a more sophisticated
implementation, that adds an additional data structure to aid with the
packing of the imaginary items.

This data structure is a doubly-linked list of doubly-linked lists $D_d$.
If $P$ is the current packing, define $\delta_h = N_p(h+1)-N_P(h)$,
$0 \leq h \leq B-1$, with $N_P(0)$ and $N_P(B)$ taken by convention to
be $1/2$ and $-1/2$ respectively.  Then we know by Lemma \ref{deltalemm}
and the discussion that follows it that placing an item of size 1 into a bin
of level $h$ will yield a smaller increase (or bigger decrease) in $ss(P)$
than placing it in a bin of level $h'$ if and only if $\delta_h < \delta_{h'}$.
At any given time in the packing process, there is a sublist $D_d$
for each value $d$ taken on by some $\delta_h$, with that sublist containing
representatives for all those $h$ such that $\delta_h = d$ and
annotated by the value of $d$.
The sublists are ordered in the main list by increasing value of $d$.
For each value of $h$, $0 \leq h \leq B-1$, there is a pointer to the
list for $\delta_h$ and to the representative for $h$ in that list.

Given this data structure, we can pack an item of size 1 in constant time:
find the first $h$ in the first list $D_d$ and place the item into a bin
of level $h$.
Note that this choice of $h$ may violate the official tie-breaking
rule for $SS$ which requires that in case of ties, we should choose
the {\em largest} $h$ with $\delta_h = d_1$.
However, as observed when we originally specified the official tie-breaking
rules, none of the performance bounds proved in this paper depend on
the precise tie-breaking rule used.  Thus, we will still have
$ER_\infty^{SS_F}(F) = 1$ if $SS_F$ is implemented this way.

To complete the proof that this implementation takes $O(nB)$ time overall,
we must show how to keep the data structure current with a constant amount
of effort per item packed.
Here we exploit the fact that in packing a single item, only two counts
get changed, and no count changes by more than 1.
Thus at most four $\delta_h$'s will change, and no $\delta_h$ can change
by more than 2.
Thus all we need show is that if $\delta_h$ changes by 2 or less, only
a constant amount of work is required to update the data structure.
But this follows from the fact if $h$ is in $D_d$, then its new sublist
can be at most two sublists away in the overall doubly-linked list,
either in an already-existing sublist to which $h$
can be prepended, or in a new sublist containing only $h$ that can be
created in constant time.  \proofmark

\medskip
An obvious drawback of the algorithms $SS_F$ is that we must know the
distribution $F$ in advance.
Fortunately, we can adapt the approach taken in these algorithms to
obtain a distribution-independent algorithm,
simply by learning the distribution as we go along.
If we engineer this properly, we can get a randomized algorithm that
matches the best expected behavior we have seen in all situations:

\begin{theo}\label{learningtheo}
There is a randomized online algorithm $SS^*$ that for any discrete
distribution $F$ with bin capacity $B$ has the following properties:
\begin{enumerate}
\renewcommand{\theenumi}{\alph{enumi}}     
\renewcommand{\labelenumi}{\rm (\alph{enumi})} 
\item $SS^*$ runs in time $O(nB)$.
\item $EW_n^{SS^*}(F) = EW_n^{OPT}(F) + O(\sqrt{n})$
\item $ER_\infty^{SS^*}(F) = 1$.
\item If $EW_n^{OPT}(F) = \Theta(\sqrt{n})$, then $EW_n^{SS^*}(F) = \Theta(\sqrt{n})$.
\item If $EW_n^{OPT}(F) = O(1)$, then $EW_n^{SS^*}(F) = O(1)$.
\end{enumerate}
\end{theo}

\noindent\textbf{Proof.}
Note that (d) will follow immediately from (b) and that
(c) will follow from (b) via Lemmas \ref{lplemm} and \ref{ERlemm}.
Thus we only need prove (a), (b), and (e), which we will do in that order.

As the basic building blocks of $SS^*$, we will use a class of algorithms
$SS_D^r$, $0 \leq r < 1$ and $D \subset \{1,2,\ldots,B-1\}$,
that capture the essence of the algorithms $SS_F$ of
Theorem \ref{tunedtheo1}, modified slightly so that we can
guarantee (e) above.
Recall from Section \ref{boundedSS'} the algorithm $SS'$ that
guaranteed $EW_n^{SS'}(F) = O(1)$ for all bounded waste distributions.
This algorithm made use of a parameterized packing rule $SS_D$,
which packed so as to minimize $ss(P)$ subject to the constraint that
no bin with a level in $D$ should be created unless this is unavoidable,
in which case we start a new bin.
Algorithm $SS'$ maintained a set $U$ of all the item sizes seen so far,
and used $SS_{D(U)}$ to pack items, where $D(U)$ is the set of dead-end
levels for $U$, and $SS^*$ will do likewise.

Algorithm $SS_D^r$ works in
steps, where in each step we flip a biased coin and proceed as follows: 
\begin{enumerate}
\item With probability $1-r$ we take the next item from
$L$ and pack it according to packing rule $SS_D$.
\item With probability $r$ we generate a new ``imaginary''
item of size 1 and pack it according to $SS_D$.
\end{enumerate}
Note that if $r = c(F)/(1+c(F))$, this is the same as $SS_F$ except for
the modified packing rule.

\medskip
In algorithm $SS^*$ we maintain an auxiliary data structure of counts $X_i$,
$1 \leq i \leq B-1$, where $X_i$ is the number of items of size $i$ so
far encountered in the list.
From this we can derive the set $U$ of
the item sizes actually seen so far, as well as the
current {\em empirical distribution} $F'$, whose probability vector $\bar{p}$
is $\langle X_1/N,X_2/N,\ldots,X_{B-1}/N\rangle$, where $N$ is the number of items
seen so far.
The packing process consists of a sequence of {\em phases},
during each of which we apply the packing rule $SS_{D(U)}^r$,
where $U$ is the set of item sizes seen up to and including the
first item to be packed in the phase and
$r = c(F')/(1+c(F'))$ for the empirical distribution $F'$ at the
beginning of the phase.

We start with a 0-phase.
An $i$-phase terminates when either (a) we
see a new item size and have to update $U$ and recompute $D(U)$
or (b) we have packed a prespecified number of real items during the phase,
where the number is $10B$ for a 0-phase and $30B4^{i-1}$ for an $i$-phase,
$i > 0$.
If an $i$-phase is terminated by the arrival of an item with a previously
unseen size, the next phase is once again a 0-phase.
Otherwise, it is an $(i+1)$-phase.
If the new phase has a different value for $U$ or $r$, we
begin it by closing all open bins.
(A partially filled bin is considered {\em open} until it is closed.
A closed bin can receive no further items and does not contribute to the
count for its level.)
We shall refer to phases that occur before all item sizes have been
seen as {\em false} phases, and ones that occur after as {\em true}
phases.  Note that once the true phases begin, each phase (except possibly
the last) packs 3 times as many items as the total number of items
packed in all previous true phases.

Note that this algorithm will have the claimed running time.
The list-of-lists data structure developed to enable the algorithms $SS_F$
to run in time $O(nB)$ can be adapted to handle the $SS_D^r$ packing rules,
so the cumulative time spent
running $SS_D^r$ for the various values of $D$ and $r$ is $O(nB)$.
In $SS^*$ we have the added cost of re-initializing this data structure
from time to time when we close all open bins, which can take $\Theta(B)$
time, but this can happen no more than $J\log_4(n/10B)$ times.
Thus the overall time for reinitialization is $O(B^2\log B\log n) = o(nB)$
for fixed $B$.
The only other computation time we need to worry about is that
needed to solve the LP's used to compute the values of $c(F')$.
By Theorem \ref{lptimetheo},
the time for the LP computed at the beginning of an $i$-phase
is $O(B^9\log^2D)$ where $D \leq n$.
Since there are no more than $J\log_4(n/10B)$ phases, the
total time spent in solving the LP's is thus $O(B^{10}\log^3 n)$ and
for fixed $B$ is again asymptotically dominated by the time to pack the items.

The proof that $SS^*$ satisfies (b) will proceed via a series of lemmas.
In what follows,
if $\bar{p}$ and $\bar{p}'$ are two length-$J$ vectors,
we will use  $||\bar{p} - \bar{p}'||$ to denote the $L^1$ distance between
them, that is,
$$\|\bar{p} - \bar{p}'\| \equiv \sum_{i=1}^j |p_i - p_i'|.$$

\begin{lemm}\label{errormag}
Suppose $F$ and $F'$ are two distributions over the same set
$\{s_1,\ldots,s_J\}$ of item sizes with probability vectors
$\bar{p}$ and $\bar{p}'$.  Then
\begin{equation}
\left |c(F) - c(F') \right | \leq B\|\bar{p} - \bar{p}'\|. \label{ccdiff}
\end{equation}
\end{lemm}

\noindent\textbf{Proof.}
We show how to convert an optimal solution
to the LP for $F$ to
a solution to the LP for $F'$ for which
the objective function $c$ satisfies
\begin{equation}
c \leq c(F) + B\|\bar{p} - \bar{p}'\|. \label{cc2diff}
\end{equation}
A symmetric argument holds for the situation where the roles of
$F$ and $F'$ are interchanged, and so (\ref{ccdiff}) will follow.

For the purposes of this proof, where $\{s_1,\ldots, s_J\}$
and $B$ are fixed, we can view our LP's as determined simply by the
probability vectors for the distributions, $\bar{p}$
and $\bar{p}'$, and write $c(\bar{p})$ and $c(\bar{p}')$
for $c(F)$ and $c(F')$ respectively.
We will convert an optimal solution to the LP for $\bar{p}$ to a feasible
one for $\bar{p}'$ via a series of steps.

For $0 \leq j \leq J$, let $\bar{p}^j=(p_1^j,\ldots ,p_J^j)$ be the vector
with $p_i^j = p'_i$, $1 \leq i \leq j$ and $p_i^j = p_i$, $j+1 \leq i \leq J$.
Note that $\bar{p}^0 = \bar{p}$ and $\bar{p}^J = \bar{p}'$.
Let ${\rm LP}_j$ denote the LP for $\bar{p}^j$.
Note that these are legitimate LP's even though the intermediate
vectors $\bar{p}^j$, $0 < j < J$, may not have $\sum_{i=1}^Jp_i^j = 1$
and hence need not correspond to probability distributions.
We will show how to convert an optimal solution to ${\rm LP}_{j-1}$
to a feasible one for ${\rm LP}_{j}$, $1\leq j \leq J$, for which
the objective function $c$ satisfies
\begin{equation}
c \leq c(\bar{p}^{j-1}) + B|p_j - p_j'|. \label{c1diff}
\end{equation}
Inequality (\ref{cc2diff}) will then follow by induction.

So consider a feasible solution to ${\rm LP}_{j-1}$.
Note that the only constraint of ${\rm LP}_j$ that is violated is
the constraint of type (\ref{lp2}) for $j$, i.e., the constraint that
says that $\sum_{h=0}^{B-1}v(j,h) = p_j'$.
If $p_j' \geq p_j$, our task is simple.
We simply add $p'_j - p_j$ to $v(j,0)$ and leave all other variables
unchanged.
This will now satisfy the above constraint for $j$ while not causing any of the
others to be violated.
The increase in the objective function will be
$(B-s_j)|p_j' -p_j| \leq B|p_j' -p_j|$, so (\ref{c1diff}) holds, as desired.

For the remaining case, suppose $p_j' < p_j$ and consider an optimal solution to
${\rm LP}_{j-1}$ that maximizes the potential function
$\sum_{h=0}^{B-1}h\cdot v(j,h)$.
We claim that this solution must be such that
\begin{equation}\label{potineq}
\mbox{for all levels $h$,
if $v(j,h) > 0$, then $v(i,h+s_j) = 0$ for all $i \neq j$}
\end{equation}
Suppose not, and hence there is a level $h$ and an integer $i \neq j$
such that $v(j,h) > 0$ and $v(i,h+s_j) > 0$.
This means that a positive amount of size $s_j$ was
placed in bins with level $h$ and then a positive amount of size $s_i$
was placed in bins with the resulting level $h+s_j$.
Let $\Delta = \min\{v(j,h),v(i,h+s_j)\}$, and modify the solution so
that instead of first placing an amount $\Delta$ of $s_j$ in bins of
level $h$ and then adding $\Delta$ of size $s_i$, we do these in
reverse order.
To be specific,
revise $v(j,h)$ to $v(j,h)-\Delta$, $v(i,h)$ to $v(i,h)+\Delta$,
$v(i,h+s_j)$ to $v(i,h+s_j)-\Delta$ and $v(j,h+s_i)$ to
$v(j,h+s_i)+\Delta$.
It is not difficult to see that this will not affect the objective
function or any of the constraints, and so the new set of variable values will
continue to represent an optimal solution to ${\rm LP}_{j-1}$.
Moreover, the potential function will have increased
by $s_i\Delta$, a contradiction.

To convert the above optimal solution to one that is feasible for
${\rm LP}_j$, we proceed as follows.
Let $H^* = \min\{H \leq B: \sum_{h=H}^{B-1}v(j,h) \leq p_j - p_j'$.
Set $v(j,h) = 0$, $H^* \leq h \leq B-1$, and reduce
$v(j,H^*-1)$ by $p_j - p_j' - \sum_{h=H^*}^{B-1}v(j,h)$.
The resulting solution will now satisfy the constraint of
type (\ref{lp2}) for $j$ in ${\rm LP}_j$.
It will continue to satisfy the constraints of type (\ref{lp3}) 
because of (\ref{potineq}).
Finally, the increase in the objective function will be at most
$s_j| p_j - p_j'| \leq B| p_j - p_j'|$ and so (\ref{c1diff}) again holds,
as desired.
\proofmark

\begin{defi}
If $\bar{p}$ is a probability vector and $r \geq 0$, then
$aug(\bar{p},r)$ is the probability vector $\bar{q}$ with
$$
q_j = \left\{ \begin{array}{ll}
\displaystyle{\frac{p_1+r}{1+r}} & \mbox{if $j=1$}\\
&\\
\displaystyle{\frac{p_j}{1+r}} & \mbox{otherwise}
\end{array}
\right.
$$
\end{defi}

\begin{lemm} \label{qdiff}
Suppose $F$ is a discrete distribution with probability vector $\bar{p}$.
Let $r,r'\geq 0$ and define $\bar{q}=aug(\bar{p},r)$ and
$\bar{q}'= aug(\bar{p},r')$.
Then $$||\bar{q}-\bar{q}'||\leq 2|r-r'|.$$
\end{lemm}

\noindent\textbf{Proof.}
By Definition,
\begin{eqnarray*} \label{Fdiff1}
\|q-q'\| &=& \sum_{j=2}^J \left | \frac{p_j}{1+r} - \frac{p_j}{1+r'} \right |
~+~ \left | \frac{p_1+r}{1+r} - \frac{p_1+r'}{1+r'} \right |\\ [.1in]
&\leq& \left | \frac{1}{1+r} - \frac{1}{1+r'} \right |
~+~ \left | \frac{r}{1+r} - \frac{r'}{1+r'} \right |\\ [.2in]
&=& \left | \frac{(1+r')-(1+r)}{(1+r')(1+r)} \right | +
\left | \frac{r(1+r') - r'(1+r)}{(1+r')(1+r)} \right |
\leq 2 \left |r-r' \right | \proofmark.
\end{eqnarray*}

\begin{lemm} \label{pdiff}
Suppose $F$ is a discrete distribution with
$\bar{s} = (s_1,\ldots,s_J)$, and $F'$ is the
empirical distribution measured after sampling $n$ items with sizes chosen
according to $F$ for some $n>0$.
Let $\bar{q} = aug(\bar{p},c(F))$ and  $\bar{q}' = aug(\bar{p},c(F'))$.
Then for all $\beta > 0$,
\begin{quote}
\begin{enumerate}
\renewcommand{\theenumi}{\alph{enumi}}     
\renewcommand{\labelenumi}{\rm (\alph{enumi})} 
\item $~~\displaystyle{P \left ( \left\| \bar{p} - \bar{p}' \right\|
\geq \frac{J\beta}{\sqrt{2n}} \right ) \leq 2Je^{-\beta^2}}$
\item $~~\displaystyle{P \left ( \left| c(F) - c(F') \right|
\geq \frac{JB\beta}{\sqrt{2n}} \right ) \leq 2Je^{-\beta^2}}$
\item $~~\displaystyle{P \left ( \| \bar{q} - \bar{q}' \|
\geq \frac{\sqrt{2}JB\beta }{\sqrt{n}} \right ) \leq 2Je^{-\beta^2}}$
\end{enumerate}
\end{quote}
\end{lemm}

\noindent\textbf{Proof.}
By a straightforward application of the Chernoff bound, as described
for example in \cite[pp.\ 234--236]{AS92}, we have that
for all $j$, $1 \leq j \leq J$, and $\beta > 0$,
$$
P \left ( \left | p_j - p_j' \right | \geq \frac{\beta}{\sqrt{2n}} \right )
\leq 2e^{-\beta^2}
$$
Thus the probability that the bound is exceeded for at least one $j$
is no more than $2Je^{-\beta^2}$.  However, if
$\left\| \bar{p} - \bar{p}' \right\| \geq J\beta/\sqrt{2n}$
then the bound must be exceeded for some $j$.  Hence conclusion (a)
holds.  Conclusions (b) and (c) follow by Lemmas \ref{errormag}
and \ref{qdiff}.  \proofmark

\begin{lemm} \label{wdiff}
Suppose $F$ and $F'$ are discrete distributions over the same
size vector $\bar{s} = (s_1,\ldots,s_J)$,
$\bar{q} = aug(\bar{p},c(F))$, $\bar{q}' = aug(\bar{p},c(F'))$, and
$r'=c(F')/(1+c(F'))$.
Suppose $q_{min}$ is the smallest nonzero entry in $\bar{q}$ and
$\|\bar{q} - \bar{q}'\| < q_{min}$.
Then if the algorithm $SS_{D(U_F)}^{r'}$ is applied to a list $L$ of $n$ items
generated according to $F$, the resulting
packing $P$ of $L$ plus the imaginary items created by
$SS_{D(U_F)}^{r'}$ satisfies
$$
E[W(P)] = O(\max\{n\|\bar{q} - \bar{q}'\|,\sqrt{n}\}).
$$
\end{lemm}

\noindent\textbf{Proof.}
Since $\|\bar{q} - \bar{q}'\| < q_{min}$, we have that for all $j$ with $q_j>0$,
$q_j' > q_j - \|\bar{q} - \bar{q}'\|
\geq q_j(1-\|\bar{q} - \bar{q}'\|/q_{min}) > 0$.
Let $\delta = \|\bar{q} - \bar{q}'\|/q_{min}$.
Then for all $j$ we have $q_j'>(1-\delta)q_j >0$.

Suppose items are generated according to $F$ and we use
$SS_{D(U_F)}^{r'}$ to pack them.
At each step, we will thus be using $SS_{D(U_F)}$ to pack an item that
looks as if it
were generated according to the probability vector $\bar{q}'$.
Let us view the packing process as follows:
When an item of size $s_j$ arrives, randomly classify it as an
{\em good} item with probability $(1-\delta)q_j/q_j'$ and as a {\em bad} item
with probability $1-(1-\delta)q_j/q_j'$.
Note that if one restricts attention to the  good items, they
now arrive as if generated according to $\bar{q}$.
Further note that by Claim \ref{onesclaim} of Theorem \ref{tunedtheo1},
the distribution determined by $\bar{q}$ is a perfectly packable distribution.
Thus for these arrivals we can apply Lemma \ref{orlinlemm},
which we have already shown applies to $SS_{D(U_F)}$ as well as $SS$.
Thus we can conclude that
the expected increase in $ss(P)$ each time a good item is packed is less than 2.

Let $D$ denote the constant $(1+q_{min})/q_{min}$.
The probability that a random item is a bad item is
$$
\sum_{i=1}^J q_j'\left( 1 - \frac{(1-\delta)q_j}{q_j'} \right)
= \sum_{i=1}^{J}\left(q_j' - q_j + \delta q_j \right) \leq
\|\bar{q} - \bar{q}'\| + \delta = D\|\bar{q} - \bar{q}'\|
$$
For bad items, the worst-case
increase in $ss(P)$ is less than $2 \max_j\{N_P(j)\}+2$, an upper bound
by Lemma \ref{deltalemm} on the increase that
would occur if our placement caused the maximum count to increase.
Thus the expected increase in $ss(P)$ is less than
\begin{equation}\label{EdeltaSS}
2 \left( 1+ D\|\bar{q} - \bar{q}'\| \max_j\{N_P(j)\} \right)
\end{equation}

Let $P_i$ be the packing after $i$ items have been packed and
let $i(t)$, $1 \leq t \leq n$, be the index of the packing that results
when the $t$th real item is packed, with $i(0)=0$ by convention.  Define
\begin{eqnarray*}
Max_t &\equiv&\max\{1,N_{P_{i(t)}}(j):~1\leq j \leq J\},~1 \leq t \leq N\\
MaxE &\equiv& \max \{E[Max_t]: 0\leq t \leq n\}\\
\end{eqnarray*}
\begin{claim}\label{maxclaim}
For all $t$, $0\leq t \leq n$, and all $i$, $i(t)\leq i < i(t+1)$,
the maximum level count in $P_i$ is at most $Max_t$.
\end{claim}

\noindent
{\bf Proof of Claim.}
The claim holds by definition for $P_{i(t)}$.
Suppose it holds for packing $P_i$ and $i+1 < i(t+1)$,
i.e., the next item to be packed is imaginary.
Note that the fact that imaginary items (of size 1) can be generated
implies that there are no dead-end levels.
Since $SS^{r'}$ by assumption knows this, this means that it is not
forbidden from making any legal move by its requirement to avoid
creating dead-end levels, and must make an improving move whenever one exists.
Suppose the current packing has a count greater than 0 and $j$ is the level
with the biggest count, ties broken in favor of larger levels.
Then there is at least one bin with level $j$ and placing an item of size 1
into such a bin will decrease $ss(P)$.
Thus $SS^{r'}$ must choose a placement that decreases $ss(P)$.
This cannot increase the largest level count.
Suppose on the other hand that the current packing has no level
count exceeding 0.
Then placing an imaginary item will only increase the maximum level
count from 0 to 1, which is still no more than $Max_{i(t)}$.
In both cases, we are left with a packing in which no count
exceeds $Max_{i(t)}$.
The claim follows by induction.  \proofmark
\begin{claim}\label{Emaxclaim}
For $0\leq t < n$,
$$E\left[ss(P_{t+1}) - ss(P_{t})|P_{t}\right]
\leq 2B\bigl( 1+D\|\bar{q} - \bar{q}'\|MaxE\bigr).
$$
\end{claim}

\noindent
{\bf Proof of Claim.}
For each $k\geq 0$, the probability that there are more than
$k$ items packed in going from $P_{t}$ to $P_{t+1}$ is
$\left(c(F')/(1+c(F'))\right)^k$.
Given that there are more than $k$ items packed, the expected increase
in $ss(P)$ due to the packing of the $k+1$st item is by (\ref{EdeltaSS}),
Claim \ref{maxclaim}, and the definitions of
$Max_t$ and $MaxE$ at most$$2(1+ D\|\bar{q} - \bar{q}'\|E[Max_t]) \leq 2(1+ D\|\bar{q} - \bar{q}'\|MaxE).$$
The total expected increase in going from $P_t$ to $P_{t+1}$ is thus at most
$$\sum_{k=0}^\infty \left(\frac{c(F')}{1+c(F')}\right)^k
2(1+ D\|\bar{q} - \bar{q}'\|MaxE)
= 2(1+c(F'))(1+ D\|\bar{q} - \bar{q}'\|MaxE)$$
The claim follows since by definition $c(F) \leq B-1$ for all
distributions $F$.  \proofmark

\medskip
Thus by the linearity of expectations we can conclude that for $1 \leq t \leq n$
\begin{equation}\label{EssPk}
E[ss(P_t)] \leq 2Bt\bigl(1 + D\|\bar{q} - \bar{q}'\| MaxE\bigr)
\end{equation}
and, by inequality (\ref{ebound}) in the proof of Lemma \ref{cauchylemm}, that
\begin{eqnarray*}
E[Max_t] \leq E\left[1+ \sum_{j=1}^{B-1}N_{P_t}(j)\right]
& \leq & 1+ \sqrt{B\cdot E[ss(P_t)]}\\
& \leq & 1+ \sqrt{2Bt\left(1+ D\|\bar{q} - \bar{q}'\| MaxE \right)}\\
& \leq & 2\sqrt{Bn\left(1+ D\|\bar{q} - \bar{q}'\| MaxE \right)}
\end{eqnarray*}
and hence
\begin{equation}\label{Emaxbound}
{\displaystyle MaxE \leq 2\sqrt{Bn\left(1+ D\|\bar{q} - \bar{q}'\| MaxE \right)}}.
\end{equation}

If $D\|\bar{q} - \bar{q}'\| MaxE \leq 1$,
we have $E[ss(P_n)] \leq 4Bn$ by (\ref{EssPk}).
So by Lemma \ref{cauchylemm} we have
$$E[W(P_n)] \leq \sqrt{B\cdot E[ss(P_n)]} \leq 2B\sqrt{n}$$
Otherwise we have by (\ref{Emaxbound}) that
$MaxE \leq 2\sqrt{2BDn\|\bar{q} - \bar{q}'\|}\sqrt{MaxE]}$.
But this implies $MaxE \leq 8BDn\|\bar{q} - \bar{q}'\|$, and consequently
by (\ref{EssPk})
$$
E[ss(P_n)] \leq 2Bn + 16(BD\|\bar{q} - \bar{q}'\| n)^2
$$
and hence by Lemma \ref{cauchylemm} that
$$
E[W(P_n)] \leq \sqrt{BE[ss(P_n)]} = O( n\|\bar{q} - \bar{q}'\|)
$$
for fixed $F$.
Thus $EW_n^{SS''}(F') = O(\max\{n\|\bar{q} - \bar{q}'\|,\sqrt{n}\})$
and Lemma \ref{wdiff} is proved.  \proofmark

\medskip
We can now address part (b) of Theorem \ref{learningtheo}.
Let us divide the waste created by $SS^*$ into three components.
Let $n_A$ denote the number of items seen before all sizes in $U_F$
have appeared.
\begin{itemize}
\item Waste in bins created during the packing of the first
$n_A$ items (during what we called {\em false} phases).
\item Waste in bins created after the first $n_A$ items have been
packed, either during the 0-phase or during an $i$-phase, $i>0$,
for which $\|\bar{q}-\bar{q}'\| > q_{min}$ in the terminology
of Lemma \ref{wdiff} ({\em Type 1} true phases).
\item Waste in the remaining bins ({\em Type 2} true phases).
\end{itemize}

For waste in bins created during false phases,
we first determine a bound on $E[n_A]$.
The analysis is similar to that used in the proof of Theorem \ref{sprimecor}.
The probability that we have not seen all item sizes after the
$h$th item arrives is $J \left( 1-p_{min} \right)^h$.
If we choose the smallest $t$ such that $J(1-p_{min})^t \leq 1/2$,
then for each integer $m > 0$, the probability that all the item sizes
have not been seen after $mt$ items have arrived is at most $1/2^m$.
Thus for each $i\geq 0$, the probability
that $n_A \in (mt,(m+1)t]$ is at most $1/2^m$.
Hence
$$E[n_A]
\leq \sum_{m=0}^\infty \Bigl( m+1)t \cdot p\bigl[n_A \in (mt,(m+1)t]\bigr]\Bigr)
\leq t \cdot\sum_{m=0}^\infty \frac{(m+1)}{2^m} = 4t.
$$
Thus the expected false phase waste resulting from bins that contain at least one
real item is bounded by $4t(B-1)/B$.

The only other possible waste during false phases
consists 1 unit of waste for each bin containing only imaginary items.
The expected number of imaginary items that arrive before all item
sizes have been seen is bounded by $(n_A+1)c(F_{max})$, where $F_{max}$
is the empirical distribution $F'$ that has the largest value of $c(F')$
among all those computed before all item sizes have been seen.
Since $c(F') \leq B-1$ for all distributions $F'$ this is at most $(4t+1)(B-1)$.
Moreover,
all but one of the bins containing only imaginary items that are
started during a given phase must
be completely full: as already remarked,
if there are any partially filled bins when
an imaginary item (of size 1) arrives, then placing it in a bin whose
level has the largest count (ties broken in favor of higher levels) will
cause a decrease in $ss(P)$ and hence is to be preferred to starting a new bin.
Thus the expected number of bins containing only imaginary items
is at most $(4t+1)(B-1)/B$ plus the expected number of false phases.
Since the number of false phases is clearly less than
$n_A/(10B)+J$, the total expected waste during false phases is at most
$8t + J+1 = O(1)$ for fixed $F$.

We now turn to the Type 1 true phases.
The first of these is the true 0-phase, which is Type 1 by definition.
In this phase the expected
number of real items packed is at most $10B$ and the expected waste
is at most $20B+2$ by an argument like that in the previous paragraph.

By a similar argument, if there is a true $i$-phase, $i>0$,
the number of real items packed in it is at most $30B\cdot 4^{i-1}$
and the expected waste
during the phase is at most $60B\cdot 4^{i-1} + 2 < 16B\cdot4^i$.
Whether this phase contributes to the Type 1 waste depends on the
empirical distribution $F'$ measured at the beginning of the phase.
In particular, we must have $\|\bar{q} - \bar{q}'\| > q_{min}$.

Now the distribution $F'$ is based on at least $10B \cdot 4^{i-1}$ samples
from $F$.
Thus by Lemma \ref{pdiff}(c), the probability that
$\|\bar{q} - \bar{q}'\| \geq \sqrt{2}JB\beta/\sqrt{2.5B4^i}$ is
bounded by $2Je^{-\beta^2}$.
Thus the probability that $\|\bar{q} - \bar{q}'\| \geq q_{min}$ is at most
$2Je^{-(1.25q_{min}^2/J^2B)4^i} = 2Jd^{-4^i}$ where
$d = e^{1.25q_{min}^2/J^2B} >1$ is a constant independent of $i$.
The expected waste that this phase can produce by being a Type 1
phase is thus at most
$(32BJ)(4^i/d^{4^i})$.
Summing over all true phases we conclude that the total expected
waste for Type 1 phases is at most
$$
20B+2 ~+~ 32BJ\sum_{i=1}^\infty\frac{4^i}{d^{4^i}} = O(1).
$$

Finally, let us turn to the waste during Type 2 true phases.
Suppose the true $i$-phase, $i>0$, is of Type 2,
and let $F'$ be the empirical distribution
at the beginning of the phase, with $\bar{p}'$ being its probability vector.
$F'$ must have been based on the observation of at least $10B4^{i-1}$
items generated according to $F$.
Thus by Lemma \ref{pdiff}(b) there are constants $\alpha$ and $\gamma$
depending on $F$ but independent of $i$ such that
$E\bigl[|c(F)-c(F')|\bigr] < \gamma/\sqrt{5B4^i} = \alpha 2^{-i}$.

Let $N_i$ be the number of real items packed during the true $i$-phase,
and recall that $N_i \leq 30B4^{i-1}$.
This means that the expected waste due to imaginary items
created during the phase is at most
$$
\frac{N_ic(F')}{B} \leq \frac{N_iB\bigl(c(F)+\alpha 2^{-i}\bigr)}{B}
\leq \frac{N_ic(F)}{B} + 7.5\alpha 2^i.
$$
Note that the total number of true phases
is at most $\lceil \log_4(n/10B) \rceil < \lfloor \log_4 n \rfloor
= \lfloor (1/2)\log_2 n \rfloor$.
Thus even if all such phases are of Type 2, we have that the expected total waste
during the Type 2 phases due to imaginary items is bounded by
$$
\frac{nc(F)}{B} + 7.5\alpha \sum_{i=1}^{\lfloor \log_4 n \rfloor}2^i
< \frac{nc(F)}{B} + 15\alpha \sqrt{n} = \frac{nc(F)}{B} + O(\sqrt{n})
$$

Now let us consider the waste caused by empty space in the bins
packed during true phases of Type 2.
First note that the set of items contained in open bins at the end
of the $i$-phase consists of all items packed during this phase plus possibly
items from immediately preceding true phases that operated with the
same value of $r$.
Even if all preceding true phases operated with the same value of $r$,
this could be no more than $10B4^i$ items.
Moreover, as argued above we know that the empirical distribution $F'$
computed at the beginning of the $i$-phase has
$E\bigl[|c(F)-c(F')|\bigr] < \alpha 2^{-i}$ for some fixed $\alpha$,
so that by Lemma \ref{qdiff},
$E\bigl[\|\bar{q}-\bar{q}'\|\bigr] < 2\alpha 2^{-i}$.
Since this is a Type 2 phase, we have be definition that
$\|\bar{q}-\bar{q}'\| \leq q_{min}$ and so Lemma \ref{wdiff} applies
and we can conclude that there is a constant $\gamma$ such that the
expected empty space in the packing is bounded by
$$
\gamma \max \left\{ (10B4^i)(2\alpha 2^{-i}),\sqrt{10B4^i}\right\}
= O(2^i).
$$

Thus the expected total empty space of this kind 
over all true phases of Type 2 is once again $O(\sqrt{n})$,
and so the expected total waste in bins started in Type 2 true
phases (empty space plus imaginary
items) is $nc(F)/B + O(\sqrt{n})$.
Given that the expected waste in false levels and in true levels of Type 1
was bounded, this means that
$$
EW_n^{SS^*}(F) = \frac{nc(F)}{B} + O(\sqrt{n})
$$
which by Theorem \ref{lptheo2} means that Claim (b) of Theorem \ref{learningtheo}
has been proved.

\medskip
It remains to prove Claim (e), that $EW_n^{SS^*}(F) = O(1)$ whenever
$EW_n^{OPT}(F) = O(1)$, i.e., whenever $F$ is a bounded waste distribution.
Suppose $F$ is a bounded waste distribution with size vector $\bar{s}$
and probability vector $\bar{p}$.
From the Courcoubetis-Weber Theorem, we know that
there is an $\epsilon > 0$ such that any distribution $F'$
over the same set of item sizes that has a probability vector $\bar{p}'$
satisfying $\|\bar{p}-\bar{p}'\| \leq \epsilon$ is a perfectly packable
distribution and hence has $c(F')=0$ by Theorem \ref{lptheo2}.

Once again, we can divide the waste produced in an $SS^*$ packing
of a list generated according to $F$ into three components, although
this division is somewhat different.
\begin{itemize}
\item Waste in bins created during false phases.
\item Waste in bins created in true phases through the last such phase
in which the starting empirical distribution $F'$ had $c(F')>0$.
\item Waste created in all subsequent phases.
\end{itemize}
As in the analysis of Claim (b), we can conclude that the total
expected waste for the false phases is bounded.

Consider now the waste created in true phases through the last
phase that started with $c(F') > 0$.
If this was the true 0-phase, the expected waste
is bounded by $20B+2$, again as argued in Claim (b).
If it was the true $i$-phase, $i>0$, then at most $10B4^i$ items
can have been packed in true phases through this point,
and so the expected waste would be at most $80B4^i+2<81B4^i$ by an
analogous argument.
Now the probability that the $i$-phase is the last phase with
$c(F')>0$ is clearly no more than the probability that it simply
had $c(F')>0$.
As remarked above, this can only have happened if
$\|\bar{p}-\bar{p}'\| > \epsilon$.
Since the empirical distribution at the start of the $i$-phase, $i>0$,
is based on at least $10B4^{i-1}$ samples from $F$,
by Lemma \ref{pdiff}(a), the probability that
$\|\bar{p}-\bar{p}'\| > \epsilon$ is at most
$2Je^{-(5B\epsilon^2/J^2)4^i} = 2Jd^{-4^i}$ for some $d>1$.
Thus the total expected waste through the last true phase with
$c(F')>0$ is at most
$$20B+2~+~ \frac{81B}{2J}\sum_{i=1}^\infty\frac{4^i}{d^{4^i}} = O(1).$$

Finally, if there are any phases after the last one
that had $c(F')>0$ and hence $r>0$, let the first such phase
be the $i_0$-phase.
This phase begins by closing all previously open bins because
$r$ has just changed from a positive value to 0.
From now on, however, no more bin closures will take place since
$r=0$ for all remaining phases and hence never changes.
Thus the packing beginning with the $i_0$-phase is simply an
$SS_{D(U_F)}^0 = SS_{D(U_F)}$ packing of items generated according
to $F$, and by Theorem \ref{sprimecor} has $O(1)$ expected waste.

Thus the total expected waste under $SS^*$ is $O(1)$, Claim (e) holds,
and Theorem \ref{learningtheo} is proved.  \proofmark

\section{$SS$ and Adversarial Item Generation}

The results for $SS^*$ in the previous section
are quite general with respect to the context
traditionally studied by papers on the average case analysis of 
bin packing algorithms: the standard situation
in which item sizes are chosen as independent samples from the same
fixed distribution $F$.
However, that context itself is somewhat limited, in that one can
conceive of applications in which some dependence exists between item sizes.
Perhaps surprisingly,
the arguments used to prove Theorems \ref{orlintheo} and \ref{boundedtheo}
imply that $SS$ itself can do quite well in some situations where there
is dependence and that dependence is controlled by an adversary.

Suppose that our item generation process works as follows:
Let $B$ be a fixed bin size.
For each item $x_i$, $i=1,2,\ldots$, the size of item $x_i$ is chosen
according to a discrete distribution $F_i$ with bin size $B$.
The choice of $F_i$, however, is allowed to be made by an adversary,
given full knowledge of all item sizes chosen so far, the current packing,
and the packing algorithm we are using.
It would be difficult to do well against such an adversary unless it were
somehow restricted, so to introduce a plausible restriction, let us
say that such an adversary is {\em restricted to $\cal F$},
where $\cal F$ is a set of discrete distributions,
if all the $F_i$ used must come from $\cal F$.
As a simple corollary of the proof of Theorem \ref{orlintheo}
we have the following.
\begin{theo}\label{Padversary}
Let $B$ be a given bin size and suppose items are generated by an adversary
restricted to the set of all perfectly packable distributions for bin size $B$.
Then the expected waste under $SS$ is $O(\sqrt{n})$.
\end{theo}

\noindent
{\bf Proof.}
By Lemma \ref{orlinlemm} we know that $E[ss(P)]$ increases
by less than 2 whenever $SS$ packs an item whose size is generated
by a perfectly packable distribution.
Thus we can conclude that if we pack $n$ items generated by our adversary,
we still must have $E[ss(P)] < 2n$.
The rest follows by Lemma \ref{cauchylemm}, as in the proof of
Theorem \ref{orlintheo}.  \proofmark

\medskip
Note that without the restriction to perfectly packable distributions,
the adversary could force the {\em optimal} expected waste to be linear,
so Theorem \ref{Padversary} is in a sense the strongest possible result
of this sort.
With even more severe restrictions on $\cal F$,
one can guarantee {\em bounded} expected
waste against an adversary.
\begin{theo} \label{Badversary}
Suppose $\cal F$ is a set of bounded waste distributions none
of which has nontrivial dead-end levels, and there is an $\epsilon>0$
such that every distribution that is within distance $\epsilon$ of
a member of $\cal F$ is perfectly packable distribution.
Then if items are generated by an adversary
restricted to $\cal F$, the expected waste under $SS$ is $O(1)$.
\end{theo} 

\noindent
{\bf Proof.}
This follows from the proof of Theorem \ref{boundedtheo}, since
the general hypothesis of Hajek's Lemma allows for adversarial item
generation.  Essentially the same proof as was used to show
Theorem \ref{boundedtheo} applies.  \proofmark

Theorem \ref{Badversary} seems very narrow, but it has an interesting corollary.

\begin{coro}
Suppose $\cal{F}$ $= \bigl\{ U\{j,k\}, 1 \leq j \leq k-2 \bigr\}$ for
some fixed $k>0$.
Then if items are generated by an adversary
restricted to $\cal F$, the expected waste under $SS$ is $O(1)$.
\end{coro}

\noindent
{\bf Proof.}
As shown in \cite{CCG98,CCG02}, $EW_n^{OPT}(F) = O(1)$ for all these distributions,
and so by the Courcoubetis-Weber theorem for each $j$, $1 \leq j \leq k-2$,
there is an $\epsilon_j > 0$ such that all distributions $F$
within distance $\epsilon_j$ of $J\{j,k\}$ are perfectly packable distributions.
We simply take $\epsilon = \min\{\epsilon_j: 1 \leq j \leq k-2\}$
and apply Theorem \ref{Badversary}.  \proofmark

\medskip
If we omit from Theorem \ref{Badversary} the requirement that
the distributions in $\cal{F}$ have no nontrivial dead-end levels,
then the best upper bound on the expected waste for $SS$
grows to $O(\log n)$, as follows from the proof of Theorem \ref{lognupperbound}.
Note that we cannot improve this to $O(1)$ by using $SS'$ instead of
$SS$ as we did in the non-adversarial case.
For example, the adversary could generate its first item using the distribution
that yields items of size 1 with probability 1, and then
switch to a bounded waste distribution with nontrivial dead-end levels.
$SS'$, having seen an item of size 1, would conclude that $1 \in U_F$
and hence that there are no dead-end levels.
So from then on it would pack exactly as $SS$ would and hence
would produce $\Omega(\log n)$ waste as implied by the lower bound
in Theorem \ref{lognupperbound}.

\section{The Effectiveness of Variants on $SS$}

In this section we return to the standard model for item generation,
and ask how much of the good behavior of $SS$
depends on the precise details of the algorithm.
It turns out that $SS$ is not unique in its effectiveness,
and we shall identify a variety of related algorithms $A$ that share one
or more of the following {\em sublinearity}
properties with $SS$ (where (a) is a weaker form of (b)):
\begin{enumerate}
\renewcommand{\theenumi}{\alph{enumi}}     
\renewcommand{\labelenumi}{(\alph{enumi})} 
\item {\rm [Sublinearity Property]}. If $EW_n^{OPT}(F) = O(\sqrt{n})$, then $EW_n^A(F) = o(n)$.
\item {\rm [Square Root Property]}. If $EW_n^{OPT}(F) = O(\sqrt{n})$, then $EW_n^A(F) = O(\sqrt{n})$.
\item {\rm [Bounded Waste Property]}. If $EW_n^{OPT}(F) = O(1)$ and $F$ has no
nontrivial dead-end levels, then $EW_n^A(F) = O(1)$.
\end{enumerate}

\subsection{Objective functions that take level into account}
One set of variants on $SS$ are those that replace the objective
function $ss(P)$ by a variant that multiplies the squared counts by
some function depending only on $B$ and the corresponding level,
and then packs items so as to minimize this new objective function.
Examples include
$$
\sum_{h=1}^{B-1}N_P(h)^2(B-h),~~\sum_{h=1}^{B-1}\left[N_P(h)(B-h)\right]^2,~
\mbox{\rm and}~\sum_{h=1}^{B-1}\frac{N_P(h)^2}{h}
$$

The first of the above three variants was proposed in 1996 by
David Wilson \cite{dbw96}, before we had invented the algorithm $SS$ itself.
Wilson's unpublished experiments with this algorithm
already suggested that it satisfied the Square Root and Bounded Waste Properties
for the $U\{j,k\}$ distributions, a claim we can now confirm as a consequence
of the following more general result.
\begin{theo}\label{wilsonalgs}
Suppose $f(h,B)$ is any function of the level and bin capacity,
and $A$ is the algorithm that packs items so as to minimize
$\sum_{h=1}^{B-1}N_P(h)^2f(h,B)$.
Then $A$ satisfies the Square Root and Bounded Waste Properties.
\end{theo}

\noindent
{\bf Proof.}
Such algorithms satisfy the Square Root Property,
since by Lemma \ref{orlinlemm}
the expected increase in the objective function at each step is still bounded
by a constant ($2\max\{f(h,B): 1 \leq h \leq B-1\}$).
They satisfy the Bounded Waste Property, since the proof of
Theorem \ref{boundedtheo} need only
be modified to change some of the constants used in the arguments.
Details are left to the reader.  \proofmark

\medskip
We conjecture that the
$EW_n^{SS}(F)=\Theta( \log n)$ result of Theorem \ref{lognupperbound}
for distributions $F$ with nontrivial
dead-end levels also carries over to these variants, but the length
and complexity of the proof of the original result makes verification
a much less straightforward task.

As to which of these variants performs best in practice, we performed
preliminary experimental studies
using the distributions studied in \cite{CJK99}, i.e.,
$U\{h,100\}$, $1 \leq h < 100$ (as defined in the Introduction),
and $U\{18,j,100\}$, $18 \leq j < 100$,
where $U\{h,j,k\}$ is the distribution in which the bin size is $k$,
the set of possible item sizes is $S = \{18,19,\ldots,h\}$,
and all sizes in $S$ are equally likely.
The distributions in the first class are all bounded waste distributions
except for $U\{99,100\}$, for which $EW_n^{OPT}(F) = \Theta(\sqrt{n})$.
The distributions in the second class include ones with all three possibilities
for $EW_n^{OPT}(F)$: $O(1)$, $\Theta(\sqrt{n})$, and $\Theta(n)$.
We also tested a few additional more idiosyncratic distributions.
The values of $n$ tested typically ranged from 100,000 to 100,000,000.
Our general conclusion was that
there is no clear winner among $SS$ and the variants describe above;
the best variant depends on the distribution $F$.

\subsection{Objective functions with different exponents}
A second class of variants that at least satisfy the
Sublinearity Property is obtained by
changing the exponent in the objective function.
\begin{theo}\label{powerroot}
Suppose $SrS$ denotes that algorithm that at each step attempts to minimize
the function $\sum_{h=1}^{B-1}(N_P(h))^r$.  Then for all perfectly packable distributions $F$,
$$
EW_n^{SrS}(F) ~~=~~ \left\{ \begin{array}{ll}
O \left( n^{\frac{1}{r}} \right ), & 1 < r \leq 2\\
\\
O \left( n^{\frac{r-1}{r}} \right ), & 2 \leq r < \infty
\end{array} \right.
$$
(Note that when $r=2$ both bounds equal $O(\sqrt{n})$, the known bound
for $SS = S2S$.)
\end{theo}

\noindent\textbf{Proof.}
Suppose $P$ is any packing and a random item $i$ is generated according
to $F$.
By the argument used in the proof of Lemma \ref{orlinlemm}, we know that
for there is an algorithm $A_F$ such that if $i$ is packed by $A_F$,
then for each $h$, $1 \leq h \leq B-1$,
the expected increase in $N_P(h)^r$
given that $N_P(h)$ changes
and that the current value $N_P(h) >0$, is bounded by
\begin{eqnarray*}
\frac{1}{2}\Bigl((N_P(h)+1)^r - N_P(h)^r \Bigr) & + & \frac{1}{2}\Bigl((N_P(h)-1)^r -N_P(h)^r\Bigr)\\[.2in]
& = & \frac{(N_P(h)+1)^r + (N_P(h)-1)^r}{2} -N_P(h)^r.
\end{eqnarray*}

Let $x = \max\{N_P(h): 1 \leq h \leq B-1\}$.
Given that at most two counts change when an item is packed
and that the expected increase for a zero-count is at most $1^r=1$,
the expected increase in $\sum_{h=1}^{B-1}(N_{P_n}(h))^r$ when $i$ is
packed is thus at most
\begin{equation}\label{ediff}
\max \Bigr\{2,~(x+1)^r + (x-1)^r -2x^r\Bigl\}.
\end{equation}
Since $SrS$ packs items so as to minimize $\sum_{h=1}^{B-1}(N_{P_n}(h))^r$,
the expected increase in this quantity when we pack $i$ using $SrS$
instead of $A_F$ can be no greater.

We thus need to bound (\ref{ediff}) when $r$ is fixed.
For $x \leq 2$, it is clearly bounded by a constant depending
only on $r$, so let us assume that $x > 2$. 
To bound (\ref{ediff}) in this case, we know by Taylor's Theorem that
there exist $\theta_1$ and $\theta_2$, $0 < \theta_1,\theta_2 < 1$, such that
\begin{eqnarray}
(x+1)^r &=& x^r + rx^{r-1} + \frac{r(r-1)}{2!}x^{r-2} + \frac{r(r-1)(r-2)}{3!}(x+\theta_1)^{r-3}\label{xplus1}\\
&&\nonumber \\
(x-1)^r &=& x^r - rx^{r-1} + \frac{r(r-1)}{2!}x^{r-2} - \frac{r(r-1)(r-2)}{3!}(x-\theta_2)^{r-3}\label{xminus1}
\end{eqnarray}
Substituting, we conclude that (\ref{ediff}) is bounded by the maximum of 2 and
\begin{equation}\label{ediffbound}
r(r-1)x^{r-2}
+ \frac{r(r-1)(r-2)}{6}\:\Bigl[(x+\theta_1)^{r-3}-(x-\theta_2)^{r-3}\Bigr]
\end{equation}

\noindent
If $1<r<2$, then (\ref{ediffbound}) has a fixed bound depending only on $r$
when $x>2$.
Thus if $P_n$ is the packing that exists after all $n$ items have been
packed by $SrS$, the expected value of $\sum_{h=1}^{B-1}(N_{P_n}(h))^r$ is $O(n)$.
If $r>2$, then (\ref{ediffbound}) grows as
$\Theta(x^{r-2}) = O(n^{r-2})$.
Thus in this case the expected value of
$\sum_{h=1}^{B-1}(N_{P_n}(h))^r$ is $O(n^{r-1})$.

\medskip
Let $C_i = \sum_{h=1}^{B-1} P[N_P(h)=i]$, $0 \leq i < n$.
Note that $\sum_{i=1}^nC_i = B$ and $\sum_{i=1}^niC_i$ is the expected number of
partially filled bins in the packing and hence an upper bound on the expected
waste.  We can bound this using Holder's Inequality:
\begin{equation} \label{holder}
\sum a_ib_i \leq \left( \sum a_i^p \right)^\frac{1}{p} \left( \sum b_i^q \right)^\frac{1}{q} \mbox{ when } \frac{1}{p} + \frac{1}{q} = 1
\end{equation}

\noindent
Set $\displaystyle{
a_i = i(C_i)^\frac{1}{r},~~b_i = (C_i)^\frac{r-1}{r},~~
p = r,~~\mbox{and}~~q = \frac{r}{r-1}
}$.
In the case where $1 < r < 2$, we have concluded that there is a $d$ such that
$\sum_{i=1}^nC_ii^r \leq dn$.
Thus Holder's Inequality yields
$$
E[W(P_n)] ~<~
\sum iC_i \leq \left(\sum C_ii^r \right)^\frac{1}{r}
\left( \sum C_i \right)^\frac{r-1}{r}
\leq (dn)^\frac{1}{r}B^\frac{r-1}{r} = O(n^\frac{1}{r})
$$
as claimed.
On the other hand, if $r>2$ we have $\sum_{i=1}^nC_ii^r \leq dn^{r-1}$ for some
constant $d$ and so Holder's Inequality yields
$$
E[W(P_n)] ~<~
\sum iC_i \leq \left(\sum C_ii^r \right)^\frac{1}{r}
\left( \sum C_i \right)^\frac{r-1}{r}
\leq d^\frac{1}{r}n^\frac{r-1}{r}B^\frac{r-1}{r} = O(n^\frac{r-1}{r})
$$
as claimed.  \proofmark

\smallskip
Despite the differing qualities of the bounds in Theorem \ref{powerroot},
limited experiments with the $SrS$ for $r =$ 1.5, 3, and 4 revealed no
consistent winner among these variants and $SS$.
Indeed, they suggest that these algorithms, and perhaps all the algorithms
$SrS$ with $r>1$, might satisfy the
Square Root and Bounded Waste Properties as well as
the Sublinearity Property.
Although we currently do not see how to prove these conjectures in general,
we can show that the algorithms $SrS$ satisfy the Bounded Waste
Property when $r \geq 2$.

\begin{theo}\label{boundedSrS}
If $r \geq 2$ and $F$ is a bounded waste distribution with no nontrivial
dead-end levels, then $EW_n^{SrS}(F) = O(1)$.
\end{theo}

\noindent
{\bf Proof.}
As in the proof of Theorem \ref{boundedtheo}, we apply Hajek's Lemma.
By an argument analogous to the one used in that proof,
it is straightforward to show that the desired conclusion will follow
if Hajek's Lemma can be shown to apply to the potential function
$$
\phi(\bar{x}) = \left(\sum_{h=1}^{B-1}x_i^r\right)^{1/r}
$$

For this potential function, the Initial Bound Hypothesis applies
since we begin with the empty packing.
The Bounded Variation Hypothesis applies since for a given value
$y$ of $\phi(\bar{x})$, the maximum possible change in $\phi$ occurs when
a single entry in $\bar{x}$ equals $y$ and all the rest are 0,
in which case $\phi$ can increase to at most $y+1$ and decrease
to no less than $y-1$.

The main challenge in the proof is proving that the Expected Decrease
Hypothesis applies.
For this we need the following results, analogues of Lemmas \ref{orlinlemm},
\ref{squarelemm}, and \ref{goodmovelemm}, used in the proof of
Theorem \ref{boundedtheo}.

\begin{lemm}\label{r-orlinlemm}
Let $F$ be a perfectly packable distribution and $r \geq 2$.
Then there is a constant $d$, depending only on $r$, such that if
$P$ is an arbitrary packing into bins of size $B$ whose
profile is given by the vector $\bar{x}$ with $\phi(\bar{x}) > 0$,
$i$ is an item randomly generated according to $F$,
and $\bar{x}'$ is the profile of the packing
resulting if $i$ is packed into $P$ according to $SrS$,
$$
E\left[\phi(\bar{x}')^r:x\right] < \phi(\bar{x})^r +d\phi(\bar{x})^{r-2}.
$$
\end{lemm}

\noindent
{\bf Proof.}
Note that for all $x_h \leq \phi(\bar{x})$, $1 \leq h \leq B-1$ by definition.
The result thus follows by (\ref{ediffbound}) in the proof
of Theorem \ref{powerroot}.  \proofmark

\begin{lemm}\label{powerlemm}
Let $y$ and $a$ be positive and $r \geq 2$.  Then
\begin{equation}\label{rpowerbound}
y-a \leq \frac{y^r-a^r}{ra^{r-1}}.
\end{equation}
\end{lemm}

\noindent
{\bf Proof.}
Consider the functions
$f_a(y) = (y-a) - (y^r-a^r)/(ra^{r-1})$, $a>0$.  We need to show that
for all $a>0$, $f_a(y) \leq 0$ whenever $y > 0$.
But observe that the derivative
$$f_a'(y) = 1 - \frac{ry^{r-1}}{ra^{r-1}}$$
is greater than 0 if $y<a$, equals 0 if $y=a$, and is less than 0
if $y>a$.
Thus $f_a(y)$ takes on its maximum value when $y=a$, in which case
it is 0, as desired.  \proofmark

\begin{lemm} \label{r-goodmovelemm}
Suppose $F$ is a distribution with no nontrivial dead-end levels
and $r \geq 2$.
Let $P$ be any packing that can be created by applying $SrS$
to a list of items all of whose sizes are in $U_F$.
If $\bar{x}$ is the profile of $P$ and $\phi(\bar{x}) > r^2B^{r+1/r}$
where $B$ is the bin size,
then there is a size $s \in U_F$ such that if an item of size $s$ is packed
by $SrS$ into $P$, the resulting profile
$\bar{x}'$ satisfies
$$\phi(\bar{x}')^r \leq \phi(\bar{x})^r
- \frac{\phi(\bar{x})^{r-1}}{B^{(r^2-1)/r}}$$
\end{lemm}

\noindent
{\bf Proof.}
Let $x_h$ be the largest level count.
By the definition of $\phi$ we have $\phi(\bar{x})^r \leq Bx_h^r$
and hence $x_h \geq \phi(\bar{x})/B^{1/r} \geq r^2B^r$.
Thus $h$ cannot be a nontrivial dead-end level and as in the proof
of Lemma \ref{goodmovelemm}, there must be some $h'\geq h$ and size
$s \in U_F$ such that $h'+s \leq B$ and
$$
\Delta \equiv x_{h'} - x_{h'+s} \geq x_h/B \geq
\frac{\phi(\bar{x})}{B^{1+1/r}} \geq r^2B^{r-1}.
$$
Let $y$ denote $x_{h'+s}$.
Then if an item of size $s$ were to be packed, we could reduce
$\sum_{h=1}^{B-1}x_h^r$ by at least
$$
(y+\Delta)^r - (y+\Delta-1)^r + y^r - (y+1)^r.
$$
Using Taylor's Theorem as in the proof
of Theorem \ref{powerroot} but with one fewer term in the
expansions than in (\ref{xplus1}) and (\ref{xminus1}),
we conclude the reduction is at least
$$
\left[r(y+\Delta)^{r-1} - \frac{r(r-1)(y+\Delta-\theta_1)^{r-2}}{2}\right]
- \left[ry^{r-1} +  \frac{r(r-1)(y+\theta_2)^{r-2}}{2}\right]
$$
where $0 < \theta_1,\theta_2 < 1$.
But note that the amount we must subtract due to
the two lower order terms is less than
$$
r(r-1)(y+\Delta)^{r-2} \leq \frac{r(y+\Delta)^{r-1}}{(y+\Delta)/(r-1)}
\leq \frac{r(y+\Delta)^{r-1}}{\Delta/r} \leq
\frac{r(x_h)^{r-1}}{rB^{r-1}} = \left(\frac{x_h}{B}\right)^{r-1}
\leq \Delta^{r-1}
$$
Since the higher order terms are
$r(y+\Delta)^{r-1} - ry^{r-1} \geq r\Delta^{r-1}$,
we can conclude that $\phi$ must decrease by at least
$$
(r-1)\Delta^{r-1} \geq (r-1)\left(\frac{\phi(\bar{x})}{B^{1+1/r}}\right)^{r-1}
\geq \frac{\phi(\bar{x})^{r-1}}{B^{(r^2-1)/r}}
$$
as claimed.  \proofmark

To prove that $\phi$ satisfies the Expected Decrease Hypothesis of Hayek's
Lemma, we argue much as in the proof of Theorem \ref{boundedtheo}.
Since $F$ is a bounded waste distribution, there is an $\epsilon > 0$
such that the process of generating items according to $F$ is equivalent
to generating items of the size $s$ specified in Lemma \ref{r-goodmovelemm}
with probability $\epsilon$ and otherwise generating items according to
a slightly modified perfectly packable distribution $F'$.
By Lemmas \ref{r-orlinlemm} and \ref{r-goodmovelemm},
the expected increase in $\phi(\bar{x})^r$ is then at most
$$(1-\epsilon)d\phi(\bar{x})^{r-2}
- \frac{\epsilon\phi(\bar{x})^{r-1}}{B^{(r^2-1)/r}}
$$
which, assuming $\phi(\bar{x})$ is sufficiently large,
is less than $-b\phi(\bar{x})^{r-1}$ for some constant $b>0$
depending only on $F$ and $r$.
By Lemma \ref{powerlemm} we thus have
$$
E\left[\phi(\bar{x}') - \phi(\bar{x})\right] \leq
- \frac{b\phi(\bar{x})^{r-1}}{r\phi(\bar{x})^{r-1}} = -\frac{b}{r}
$$
and so the Bounded Decrease Hypothesis holds for $\phi$, Hajek's Lemma
applies, and we can conclude as in Theorem \ref{boundedtheo} that
$EW_n^{SrS}(F) = O(1)$.  \proofmark

\subsection{Combinatorial variants}
In this section we consider satisfying the Sublinearity Property
with algorithms that don't depend on powers of counts.
As our first two candidates, consider the algorithms that are in a sense
the limits of the $SrS$ algorithms as $r \rightarrow 1$ and
$r \rightarrow \infty$, a promising approach since the $SrS$ algorithms
all satisfy the Sublinearity Property and may even satisfy the
Square Root Property.

An obvious candidate for a limiting algorithm when $r \rightarrow 1$
is $S1S$, the algorithm that always
tries to minimize $\sum_{h=1}^{B-1}N_P(h)$, i.e., the number of
partially filled bins.
To do this, we simply must never start a new bin if that can
be avoided and must always perfectly pack a bin when possible (i.e., if the size
of the item to be packed is $s$ and there is a partially full
bin with level $B-s$, we must place the item in such a bin).
By itself this is not a completely defined algorithm, since
one needs to provide a tie-breaking rule.
If we use our standard tie-breaking rule (always
chooses a bin with the highest acceptable level),
note that $S1S$ reduces to the classic Best Fit algorithm.
As already observed in the Introduction, Best Fit
provably has linear expected waste for the bounded waste
distributions $U\{8,11\}$ and $U\{9,12\}$, and empirically seems to
behave just as poorly for many other such distributions \cite{CJS93}.
We doubt that any other tie-breaking rule will do better.
For instance, if we always choose the lowest available level when the item won't
pack perfectly, we typically do much worse than Best Fit.
Thus no $S1S$ algorithm is likely to satisfy the Sublinearity Property.

Taking the limit of $SrS$ as $r \rightarrow \infty$ seems more
promising.
Assume by convention that $N_P(B)$ is always 0.
Then $S\infty S$ is the algorithm that places an item of size $s$
into a bin of level $h$ for that $h$ with the maximum value of $N_P(h)$ in
$\{h: 1\leq h \leq B-s, \mbox{ and } N_P(h)>N_P(h+s)\}$, should that
set be non-empty, and otherwise places the item in a bin with
level $h \geq 0$ for that $h$ with the minimum value of $N_P(h+s)$,
ties always broken in favor of the higher level.
It is easy to see that for any fixed packing these are the choices
that will be made by $SrS$ for all sufficiently large values of $r$.

Experiments suggest that $S\infty S$ has bounded expected waste
for $U\{8,11\}$ and $U\{9,12\}$ as well as all the bounded waste
distributions $U\{h,100\}$, $1\leq h \leq 98$.
It still violates the Sublinearity Property, however.
For example, $EW_n^{OPT}(U\{18:27,100\}) = \Theta(\sqrt{n})$ but
experiments clearly indicate that
$S\infty S$ has linear waste for this distribution.
A simpler distribution exhibiting the behavior is $F$ with $B=51$,
$U_F = \{11,12,13,15,16,17,18\}$, and all sizes equally likely.
Experiments convincingly suggest that $EW_n^{S\infty S}(F) = \Theta(n)$,
but it is easy to see that this is a perfectly packable distribution,
since both the first four and the last three item sizes sum to $B=51$.
Moreover, if one modifies $F$ to obtain a distribution $F'$ in which
items of size 1 are added, but with only $1/10$ the probability of the
other items, one obtains a bounded waste distribution for which
$S\infty S$ continues to have linear waste.
Using other tie-breaking rules, such as preferring the lower level bin,
appears only to make things worse.
So no $S\infty S$ algorithm is likely to satisfy the Sublinearity
Property.

Not surprisingly, the simpler combinatorial variants obtained by using 
just one of the two rules from the definition of $S\infty S$ also fail.
In the first of these, $Smaxh$, we always place an item $x$ in 
a bin whose level has maximum count among all levels no greater than $B-s(x)$,
assuming that the count for empty bins is by definition 0.
In the second, $Sminh$, we place the item so as to
minimize the count of the resulting level, assuming that the count
for full bins is by definition 0.
$Smaxh$ has linear waste for $U\{8,11\}$ and $U\{9,12\}$, perhaps
not surprising since even if the item to be packed would perfectly
fill a bin, $Smaxh$ may well choose not to do this.
$Sminh$ is better, seeming to handle the $U\{j,k\}$ appropriately.
However, it has linear waste on the same three perfectly packable/bounded
waste distributions mentioned above on which $S\infty S$ also failed.
Perhaps surprisingly, its constants of proportionality appear to be
better than those for $S\infty S$ on these distributions.
This may be because, unlike the latter algorithm, it will choose
a placement that perfectly packs a bin when this is possible.

Indeed, perfectly packing a bin when that is possible would seem
like an inherently good idea.
We know that it is not {\em necessary} to do this, since $SS$ 
doesn't always do it, but how could it hurt?
Let {\em perfectSS} be the algorithm that places the current item so
as to perfectly pack a bin if this is possible, but otherwise places
it so as to minimize $ss(P)$.
Surely this algorithm should do just as well as $SS$.
Surprisingly, there are cases where
this variant too violates the Sublinearity Property.

Consider the distribution $F$ with bin size $B=10$,
$U_F=\{1,3,4,5,8\}$,
$p(1)=p(3)=p(5)= 1/4$, and $p(4)=p(8)=1/8$.
This is a perfectly packable distribution, as the probability vector
can be viewed as a convex combination of the perfect packing configurations
$(8,1,1)$, $(4,3,3)$, and $(5,5)$.
However, experiments show that {\em perfectSS} has linear waste for
this distribution (as does $Sminh$ but not $S\infty S$).
Why does this happen?
Note that essentially all the items of size 1 must be used to fill the
bins that contain items of size 8.
Thus whenever a 1 arrives and there is a bin of level 8, we need to place
the 1 in such a bin.
Unfortunately, {\em perfectSS} will prefer to put that 1 in a bin
with level 9 if such a bin exists, and
bins with level 9 can be created in other ways than simply
with an 8 and a 1.
Three 3's or a 5 and a 4 will do.
On average this happens enough times to ruin the packing.
(The count for level 9 never builds up to inhibit the
nonstandard creation of such bins because
level 9 bins keep getting filled by 1's.)
Standard $SS$ avoids this problem and has $\Theta(\sqrt{n})$
expected waste because it allows the counts for
levels 8 and 9 to grow roughly as $\sqrt{n}$, with the latter being
roughly half the former.
This means that placing a 1 in a bin with level 8 is a downhill
move, but creating a level 9 bin by any other means is an uphill move.

\subsection{Variants designed for speed}
Our final class of alternatives to $SS$ are designed to improve the running
time, possibly at the cost of packing quality.
Recall that $J$ denotes the number of item sizes under $F$.
The $\Theta(nB)$ running time for the naive implementation of $SS$
can be improved to $\Theta(nJ)$
by maintaining for each item size $s \in U_F$ the list-of-lists data
structure we introduced to handle items of size 1
in the implementation of algorithm
$SS^*$ described in Section \ref{tuned}.
This approach unfortunately will not be much of an improvement over
the naive algorithm for distributions $F$ with large numbers of item sizes,
and it remains an open problem
as to whether $SS$ (or any of the variants described above that
satisfy the Sublinearity Property)
can be implemented to run in $o(nB)$ time in general.
However, if one is willing to alter the algorithm itself, rather than
just its implementation, one can obtain more significant speedups.
Indeed, we can devise algorithms that satisfy both the Square Root
and Bounded Waste Properties and yet run in time $O(n\log B)$ or
even $O(n)$ (although there will of course be a tradeoff between
running time and the constants of proportionality on the expected waste).

We shall first describe the general algorithmic approach and
prove that algorithms that follow it will satisfy the two properties.
We will then show how algorithms of this type can be implemented in
the claimed running times.
The key idea is to use data structures for each item size, as in
the $O(nJ)$ implementation mentioned above, but only require that
they be approximately correct (so that we need not spend so much time
updating them).
In particular, we maintain for each item size $s$ a set of local values
$N_{P,s}(h)$ for the counts $N_P(h)$, and only require these local
counts satisfy
\begin{equation}\label{approxbound}
\bigl|N_P(h)-N_{P,s}(h)\bigr| \leq \delta
\end{equation}
for some constant $\delta$.
When an item of size $s$ arrives, we place it so as to minimize
$ss_s(P) = \sum_{i=1}^{B-1}N_{P,s}(h)^2$, subject only to the additional
constraint that we cannot place the item in a bin with local count $\delta$ or less,
since there is no guarantee that such bins exist.
Let {\em Approx}$SS_\delta$ be an algorithm that operates in this way.

\begin{lemm}\label{approxSSlemm}
Suppose $F$ is a perfectly packable distribution with bin size $B$,
$P$ is a packing into bins of size $B$, $\delta \geq 0$, and $x$ is
an item randomly generated according to $F$.
Then if $x$ is packed according to {\it Approx}$SS_\delta$, the
expected increase in $ss(P)$ is at most $10\delta+3$.
\end{lemm}

\noindent
{\bf Proof}.
We first need a generalization of Claim \ref{orlinclaim} from
the proof of Lemma \ref{orlinlemm}:

\begin{claim}\label{approxorlinclaim}
Suppose $F$ is a perfectly packable distribution with bin size $B$
and $\delta \geq 0$.
Then there is an algorithm $A_F$ such for any packing $P$ into bins
of size $B$, if an item $x$ is randomly generated according to $F$,
$A_F$ will pack $x$ in such a way that $x$ does not go in a bin with a level $h$
for which $N_P(h) \leq \delta$ and yet for each level $h$ with
$N_P(h) > \delta$, $1 \leq h \leq B-1$, the probability that $N_P(h)$
increases is no more than the probability that it decreases.
\end{claim}

\noindent
This is proved by a simple modification of the proof of 
Claim \ref{orlinclaim} to require that for each optimal bin the
items are ordered so that all the levels $S_1$ through $S_{last(Y)}$ have
counts greater than $\delta$ and none of the levels
$S_{last(y)}+s(y_i)$ do for $i>last(y)$.

\medskip
Claim \ref{approxorlinclaim} implies that the expected increase
in $ss(P)$ under $A_F$ is at most $2\delta +2$:
If a count greater than $\delta$ changes, the proof of Lemma \ref{orlinlemm}
implies that the expected increase in $ss(P)$ is at most 1.
Counts of $\delta$ or less can only increase, but in
this case $ss(P)$ can increase by no more than $2\delta+1$.
At most two counts can change during any item placement,
and at most one of them can be a count of $\delta$ or less.
Thus the expected change in $ss(P)$ obeys the claimed bound,
and if $SS_\delta$ is the algorithm that places items so as to
minimize $ss(P)$ subject to the constraint that no item can
be placed in a partially filled bin whose level's count is $\delta$
or less, we can conclude that the expected increase in $ss(P)$ when $SS_\delta$
places an item generated according to $F$ is also at most $2\delta+2$.

So consider what happens when $SS_\delta$ packs an item with size $s \in U_F$.
Suppose that placement is into a bin of level $h$, and
that $N_P(h+s)-N_P(h) = d$.
Note that by Lemma \ref{deltalemm} the smallest increase in $ss(P)$ this can
represent is $2d+1$.
Now by (\ref{approxbound}) we must
have $N_{P,s}(h+s)-N_{P,s}(h) \leq d+2\delta$
and so the move chosen by {\em Approx}$SS_\delta$ must
place the item in a bin of level $h'$ satisfying
$N_{P,s}(h'+s)-N_{P,s}(h') \leq d + 2\delta$.
But then, again by (\ref{approxbound}), we must have
$N_P(h'+s)-N_P(h') \leq d + 4\delta$ and hence, again by
Lemma \ref{deltalemm}, $ss(P)$ can increase by at most
$2d+8\delta+2$, or at most $8\delta + 1$ more than the increase under
$SS_\delta$.
Since the expected value for the latter was at most $2\delta +2$, the
Lemma follows.  \proofmark

\begin{theo}\label{approxSStheo}
For any $\delta \geq 0$,
\begin{itemize}
\item [{\rm (a)}]  If $F$ is a perfectly packable distribution, then
$EW_n^{ApproxSS_\delta}(F) = O(\sqrt{n})$.
\item [{\rm (b)}]  If $F$ is a bounded waste distribution with no
nontrivial dead-end levels, then
$EW_n^{ApproxSS_\delta}(F) = O(1)$.
\item [{\rm (c)}]  Suppose {\it Approx}$SS_\delta'$ is the
algorithm that mimics {\it Approx}$SS_\delta$ except that it
never creates a bin that, based on the item sizes seen so far,
has a dead-end level, unless this is unavoidable, in which case it
starts a new bin.  Then this algorithm has 
$EW_n^{ApproxSS_\delta'}(F) = O(1)$ for {\em all} bounded waste distributions,
as well as $EW_n^{ApproxSS_\delta'}(F) = O(\sqrt{n})$ for all perfectly
packable distributions.
\end{itemize}
\end{theo}

\noindent
{\bf Proof}.
Note that for any fixed $\delta$, $10\delta + 6$ is a constant,
and having a constant bound on the expected increase in $ss(P)$ was
really all we needed to prove the above results for $SS$ and $SS'$.
Thus the above three claims all follow by essentially the same arguments
we used for $SS$ and $SS'$, with constants increased appropriately
to compensate for property (\ref{approxbound}).  \proofmark

\medskip
Let us now turn to questions of running time.

\begin{lemm}\label{approxSStimelemm}
Suppose $t \geq 1$ and $J \geq 1$ are integers.
Then there are implementations of {\it Approx}$SS_{tJ}$
and {\it Approx}$SS_{tJ}'$ that work for all instances with
$J$ or fewer item sizes and run in time
$O(n(1+(\log B)/t))$.
\end{lemm}

\noindent
{\bf Proof}.
We shall describe an implementation for {\em Approx}$SS_{tJ}$.
The implementation for {\em Approx}$SS_{tJ}'$ is almost identical
except for the requirement that we keep track of the dead-end levels and
avoid creating bins with those levels when possible, which we already
discussed in Section \ref{boundedSS'}.

Our implementations maintain a data structure for each item size $s$
encountered, the data structure being initialized when the
size is first encountered.
We are unfortunately unable to use the list-of-list data structure involved in
the implementation of $SS^*$, since the efficiency of that data structure
relied on the fact that counts could only change by 1 when they were updated.
Now they may change by as much as $tJ$.
Therefore we use a standard priority queue for the up to $B$
possible levels $h$ of bins into which an item of size $s$ might be placed.
Here the ``possible levels'' for $s$ are 0 together with all
those $h$ such that $h+s\leq B$ and $N_{P,s}(h) > tJ$.
The levels are ranked by the increase
in $ss_s(P)$ that would result if an item of size $s$ were packed in
a bin of level $h$.
We can use any standard priority queue implementation that takes $O(1)$ time
to identify an element with minimum rank and $O(\log B)$ to delete
or insert an element.
Initially, the only element in each priority queue is the one for
level 0, i.e., the representative for starting a new bin.

When we pack an item of size $s$, we first
identify the ``best'' level $h$ for it as specified by the priority queue
for $s$.  We then place $x$ in a bin of level $h$ and update
the global counts $N_P(h)$ and $N_P(h+s)$.
This all takes $O(1)$ time.
Local counts are not immediately changed when an item is packed.
Local count updates are performed more sporadically,
and initiated as follows.
We maintain a counter $c(h)$ for each level $h$.
This counter is incremented by 1 every time $N_P(h)$ changes
and reset to 1 whenever it reaches the value $tJ+1$.
Suppose the item sizes seen so far are $s_1,s_2,\ldots,s_j$, $j \leq J$.
The local count $N_{P,s_i}(h)$ is updated only when
the new value of $c(h)$ satisfies $c(h) \equiv O(\mbox{mod } t)$
and $i = c(h)/t$.
Note that this means that $N_P(h)$ changes only $tJ$ times
between any two updatings of $N_{P,s_i}(h)$ and so
(\ref{approxbound}) is satisfied for $\delta = tJ$.

Whenever $N_{P,s}(h)$ is updated, we make up to two changes in the
priority queue for $s$, each of which involves one or two
insertions/deletions and hence takes $O(\log B)$ time:
First, if $h+s\leq B$ we may need to update the priority queue
entry for $h$.
If $h$ is in the queue but now $N_{P,s}(h) \leq tJ$, then we must
delete it from the queue.
If it is not in the queue but now $N_{P,s}(h) > tJ$ we must insert it.
Finally, if it is in the queue and $N_{P,s}(h) > tJ$, but its rank
is not the correct value (with respect to $N_{P,s}(h)$ and $N_{P,s}(h+s)$),
then it must be deleted and reinserted with the correct value.
Similarly, if $h-s \geq 0$, then we may have to update the entry
for $h-s$.

It is easy to verify that the above correctly implements {\em Approx}$SS_{tJ}$.
The overall running time is $O(n)$ for packing and
updating the true counts $N_P(h)$ and $O((n/t)\log B)$ for
updating local counts and priority queues, as required.  \proofmark

\begin{theo}\label{approxSStimetheo}
There exist algorithms $A1SS$, $A2SS$, $A1SS'$ and $A2SS'$
such that
\begin{itemize}
\item [{\rm (a)}]  All four satisfy the Square Root and Bounded Waste Properties.
\item [{\rm (b)}]  $A1SS'$ and $A2SS'$ have bounded expected waste for
{\em all} bounded waste distributions.
\item [{\rm (c)}]  $A1SS$ and $A1SS'$ run in time $O(n \log B)$.
\item [{\rm (c)}]  $A2SS$ and $A2SS'$ run in time $O(n)$.
\end{itemize}
\end{theo}

\noindent
{\bf Proof}.
Given Theorem \ref{approxSStheo},
it is easy to get algorithms with the above
properties from Lemma \ref{approxSStimelemm} assuming we know $J$ in advance:
If we take $t=1$ we get running time $O(n\log B)$ and
if we take $t = \log B$ we get running time $O(n)$.
(The tradeoffs only involve the constants of proportionality
on the expected waste.)
Moreover, it is really not necessary to know $J$ in advance,
as there are adaptive algorithms that learn $J$ in the
process of constructing their packings, still run in time
$O(n\log B)$ or $O(n)$, and have the desired average case performance.
For instance, we can start by running
{\em Approx}$SS_5$ ({\em Approx}$SS_{5\log B}$) as long as the number $J$
of item sizes seen so far is no more than 5.
Thereafter, whenever we see a new item size, we close all partially filled
bins, start running {\em Approx}$SS_{J+1}$ ({\em Approx}$SS_{(J+1)\log B}$),
and then set $J = J+1$.
Since by the analysis used in Section \ref{boundedSS'} the expected
number of items packed before all item sizes have been seen must be
bounded by a constant for any $F$, the bins constructed before
we start running the correct algorithm contain only bounded expected
waste and so cannot endanger our conclusions about asymptotic
expected waste rates.  \proofmark

\medskip
We can also devise fast analogues of Section \ref{tuned}'s distribution-specific
algorithms $SS^F$ that always have $ER_\infty^A(F) = 1$,
even for distributions whose optimal expected waste is linear.
This however involves more than just applying the
approximate data structures described above.
The $O(nB)$ running times for the $SS^F$ algorithms derive from two sources,
only one of which (the need for $\Theta(B)$ time to pack an item)
is eliminated by using the approximate data structures.
The second source of $\Theta(nB)$ time is the need to possibly
pack $\Theta(nB)$ imaginary items of size 1.

To avoid this obstacle, we need an additional idea.
Recall that $SS^F$ attains $ER_\infty^{SS^F}(F) = 1$ by simulating the
application of $SS$ to a perfectly packable distribution $F'$ derived from $F$.
The modified distribution $F'$ was constructed using the optimal value $c(F)$
for the linear program of Section \ref{lpsect}.
Distribution $F'$ was equivalent
to generating items according to $F$ with probability $1/(1+c(F))$ and otherwise
generating an (imaginary) item of size 1.

Our new approach uses more information from the solution to the LP.
Let $v(j,h)$, $1 \leq j \leq J$ and
$0 \leq h \leq B-1$, be the variable values in an optimal solution
for the LP for $F$.
For $1 \leq h \leq B-1$ define
$$\Delta_h \equiv \sum_{j=1}^J v(j,h-s_j) -  \sum_{j=1}^J v(j,h).$$
Note that $\Delta_h$ is essentially the percentage of partially filled
bins in an optimal packing whose gap is of size $B-h$.
Let $T = \sum_{h=1}^{B-1}\Delta_h$ and note that we must have $T \leq 1$.
Our new algorithm uses $SS$ to pack the modified distribution $F''$
obtained as follows.
With probability $1/(1+T)$ we generate items according to the original
distribution $F$.
Otherwise (with probability $T/(1+T)$) we generate
``imaginary'' items according to the distribution
in which items of size $s$ have probability $\Delta_{B-s}$.
It is not difficult to show that this is a perfectly packable distribution
and that the expected total size of the imaginary items is $c(F)$,
as in $SS^F$.
Now, however, the number of imaginary items
is bounded by $n$, so the time for packing them is no more than
that for packing the real items, and hence can be $O(n\log B)$
or $O(n)$ as needed.

One can construct a learning algorithm $SS^{**}$ based on these variants just
as we constructed the learning algorithm $SS^*$ based on the original
$SS^F$ algorithms.  We conjecture that $SS^{**}$ will satisfy the
same general conclusions as listed for $SS^*$ in Theorem \ref{learningtheo}.
The proof will be somewhat more complicated, however, and so we leave
the details to interested readers.

\medskip
We should note before concluding the discussion of fast variants of $SS$ that
our results on this topic are probably of theoretical interest only.
A complicated $O(n\log B)$ algorithm like {\em Approx}$SS_J$ would be preferable
to an $O(nB)$ or $O(nJ)$ implementation of $SS$ only when $J$ is fairly
large, presumably well over 100.
However, the constants involved in the expected waste produced by
{\em Approx}$SS_J$ are substantial in this case.

For instance, consider the bounded waste distribution $U\{400,1000\}$.
For $n = 100,000$, {\em Approx}$SS_{400}$ typically uses 100,000 bins,
i.e., one per item and
roughly 5 times the optimal number, even though Theorem \ref{approxSStimetheo}
says that the expected waste is asymptotically $O(1)$.
On the other hand, Best Fit, which also runs in time $O(n\log B)$ but
is conjectured to have linear expected waste for this distribution,
uses roughly 0.3\% more bins than necessary.
($SS$ uses roughly 0.25\%.)
Things have improved by the time $n=10,000,000$, but not enough
to change the ordering of algorithms.
Now {\em Approx}$SS_J$ uses only roughly 9.8\% more bins than necessary,
while Best Fit uses roughly 0.28\%.
$SS$ is down to an average excess of 0.0025\%.
This consists of roughly 50 excess bins (as compared to 45
for $n=100,000$) and should be compared to the roughly 200,000 excess bins
for {\em Approx}$SS_J$.
Admittedly the latter algorithm could be modified to significantly
lower its expected waste, but it is unlikely that it could be made
competitive with Best Fit except for much larger values of $n$.
 
\section{Conclusions and Open Questions}\label{conclusion}

In this paper we have discussed a collection of
new, nonstandard, and surprisingly effective algorithms for the
classical one-dimensional bin packing problem.
We have done our best to leave as few major open problems as
possible, but several interesting ones do remain:

\begin{itemize}
\item Can $SS$ itself be implemented to run in time $o(nB)$, so
that we aren't forced to use the approximate versions described in
the previous section?
\item What is $\max \{ER_\infty^{SS}(F):\:F \mbox{ is a discrete distribution}\}$?
The results of Section 4 only show that this maximum is at least 1.5 and
no more than 3.0.
A related question is what is the asymptotic worst-case performance ratio
for $SS$.
Here the results of \cite{Vli92a} for arbitrary on-line algorithms
imply a lower bound of 1.54, but the best upper bound is still the
abovementioned 3.0.
\item Is our conjecture correct that $SrS$ satisfies both the Square
Root and Bounded Waste Properties for all $r > 1$?
Is there any polynomial-time
algorithm that satisfies the Sublinearity Property
and does not involve at least implicitly computing
the powers of counts?
\item Can one obtain a meaningful theoretical
analysis of the constants of proportionality involved in the
expected waste rates for particular distributions
and the various bin packing algorithms we have discussed?
Empirically we have observed wide differences in these constants
for algorithms that, for example, both have bounded expected waste for a
given distribution $F$, so theoretical insights here may well be
of practical value.
\item Is there an effective way
to extend the Sum-of-Squares approach to continuous distributions while
preserving its ability to get sublinear waste when the optimal
waste is sublinear?
\end{itemize}

Finally, there is the question of the extent to which approaches like that
embodied in the Sum-of-Squares algorithm can be applied to other problems.
A first step in this direction is the adaptation of $SS$ to the
bin covering problem in \cite{CJK01}.
In bin covering we are given a set of items
and a bin capacity $B$, and must assign the items to bins so that each
bin receives items whose total size is {\em at least} $B$
and the number of bins packed is maximized.
Here ``waste'' is the total {\em excess} over $B$ in the bins
and the class of ``perfectly packable distributions'' is the same as
for ordinary bin packing.
The interesting challenge here becomes to construct algorithms
that have good worst- and average-case behavior for distributions
that aren't perfectly packable, while still having $O(\sqrt{n})$
expected waste for perfectly packable distributions.  For details,
see \cite{CJK01}.

The results for bin covering suggest that the Sum-of-Squares approach
may be more widely applicable, but bin covering is still quite
close to the original bin packing problem.
Can the Sum-of-Squares approach (or something like it) be extended to
problems a bit further away?

{\small
\bibliographystyle{alpha}
\bibliography{bp}

\newcommand{\etalchar}[1]{$^{#1}$}
\begin{thebibliography}{CCG{\etalchar{+}}02}

\bibitem[ABD{\etalchar{+}}]{ABDJS02}
D.~L. Applegate, L.~Buriol, B.~Dillard, D.~S. Johnson, and P.~W. Shor.
\newblock The cutting-stock approach to bin packing: Theory and experiments.
\newblock Draft, 2002.

\bibitem[AJKL84]{AJK84}
S.~B. Assman, D.~S. Johnson, D.~J. Kleitman, and J.~Y-T. Leung.
\newblock On a dual version of the one-dimensional bin packing problem.
\newblock {\em J. Algorithms}, 5:502--525, 1984.

\bibitem[AM98]{AM98}
S.~Albers and M.~Mitzenmacher.
\newblock {Average-case analyses of first fit and random fit bin packing}.
\newblock In {\em Proc. Ninth Annual ACM-SIAM Symposium on Discrete
  Algorithms}, pages 290--299, Philadelphia, 1998. Society for Industrial and
  Applied Mathematics.

\bibitem[AS92]{AS92}
N.~Alon and J.~H. Spencer.
\newblock {\em The Probabilistic Method}.
\newblock Wiley-Interscience, New York, 1992.

\bibitem[CCG{\etalchar{+}}91]{CCG91}
E.~G. {Coffman,~Jr.}, C.~Courcoubetis, M.~R. Garey, D.~S. Johnson, L.~A.
  McGeoch, P.~W. Shor, R.~R. Weber, and M.~Yannakakis.
\newblock Fundamental discrepancies between average-case analyses under
  discrete and continuous distributions.
\newblock In {\em Proceedings 23rd Annual ACM Symposium on Theory of
  Computing}, pages 230--240, New York, 1991. ACM Press.

\bibitem[CCG{\etalchar{+}}00]{CCG98}
E.~G. {Coffman, Jr.}, C.~Courcoubetis, M.~R. Garey, D.~S. Johnson, P.~W. Shor,
  R.~R. Weber, and M.~Yannakakis.
\newblock Bin packing with discrete item sizes, {P}art {I}: Perfect packing
  theorems and the average case behavior of optimal packings.
\newblock {\em SIAM J. Disc. Math.}, 13:384--402, 2000.

\bibitem[CCG{\etalchar{+}}02]{CCG02}
E.~G. {Coffman, Jr.}, C.~Courcoubetis, M.~R. Garey, D.~S. Johnson, P.~W. Shor,
  R.~R. Weber, and M.~Yannakakis.
\newblock Perfect packing theorems and the average-case behavior of optimal and
  online bin packing.
\newblock {\em SIAM Review}, 44:95--108, 2002.
\newblock Updated version of \cite{CCG98}.

\bibitem[CJK{\etalchar{+}}99]{CJK99}
J.~Csirik, D.~S. Johnson, C.~Kenyon, P.~W. Shor, and R.~R. Weber.
\newblock A self organizing bin packing heuristic.
\newblock In M.~Goodrich and C.~C. McGeoch, editors, {\em Proceedings 1999
  Workshop on Algorithm Engineering and Experimentation}, pages 246--265,
  Berlin, 1999. Lecture Notes in Computer Science 1619, Springer-Verlag.

\bibitem[CJK{\etalchar{+}}00]{sumsq2000}
J.~Csirik, D.~S. Johnson, C.~Kenyon, J.~B. Orlin, P.~W. Shor, and R.~R. Weber.
\newblock On the sum-of-squares algorithm for bin packing.
\newblock In {\em Proceedings of the 32nd Annual ACM Symposium on the Theory of
  Computing}, pages 208--217, New York, 2000. ACM.

\bibitem[CJK01]{CJK01}
J.~Csirik, D.~S. Johnson, and C.~Kenyon:.
\newblock Better approximation algorithms for bin covering.
\newblock {\em SODA 2001:}, pages 557--566, 2001.

\bibitem[CJM{\etalchar{+}}]{CJS3}
E.~G. {Coffman,~Jr.}, D.~S. Johnson, L.~A. McGeoch, P.~W. Shor, and R.~R.
  Weber.
\newblock Bin packing with discrete item sizes, {P}art~{III}: Average case
  behavior of {FFD} and {BFD}.
\newblock (In preparation).

\bibitem[CJSW93]{CJS93}
E.~G. {Coffman,~Jr.}, D.~S. Johnson, P.~W. Shor, and R.~R. Weber.
\newblock Markov chains, computer proofs, and average-case analysis of {B}est
  {F}it bin packing.
\newblock In {\em Proceedings 25th Annual ACM Symposium on Theory of
  Computing}, pages 412--421, New York, 1993. ACM Press.

\bibitem[CJSW97]{CJS97}
E.~G. {Coffman, Jr.}, D.~S. Johnson, P.~W. Shor, and R.~R. Weber.
\newblock Bin packing with discrete item sizes, part {II}: Tight bounds on
  first fit.
\newblock {\em Random Structures and Algorithms}, 10:69--101, 1997.

\bibitem[CW90]{CW90}
C.~Courcoubetis and R.~R. Weber.
\newblock Stability of on-line bin packing with random arrivals and long-run
  average constraints.
\newblock {\em Probability in the Engineering and Informational Sciences},
  4:447--460, 1990.

\bibitem[GG61]{GG61}
P.~C. Gilmore and R.~E. Gomory.
\newblock {A linear programming approach to the cutting stock problem}.
\newblock {\em Oper. Res.}, 9:849--859, 1961.

\bibitem[GG63]{GG63}
P.~C. Gilmore and R.~E. Gomory.
\newblock {A linear programming approach to the cutting stock program ---
  Part~II}.
\newblock {\em Oper. Res.}, 11:863--888, 1963.

\bibitem[GJ79]{GJ79}
M.~R. Garey and D.~S. Johnson.
\newblock {\em Computers and Intractability: A Guide to the Theory of
  NP-completeness}.
\newblock W. H. Freeman, New York, New York, 1979.

\bibitem[Haj82]{Haj82}
B.~Hajek.
\newblock Hitting-time and occupation-time bounds implied by drift analysis
  with applications.
\newblock {\em Adv. Appl. Prob.}, 14:502--525, 1982.

\bibitem[KM00]{KM00}
C.~Kenyon and M.~Mitzenmacher.
\newblock Linear waste of best fit bin packing on skewed distributions.
\newblock In {\em 41st Annual Symposium on Foundations of Computer Science},
  pages 582--589. IEEE Computer Society Press, 2000.
\newblock To appear in {\em Random Structures and Algorithms}.

\bibitem[KRS98]{KRS98}
C.~Kenyon, Y.~Rabani, and A.~Sinclair.
\newblock Biased random walks, {L}yapunov functions, and stochastic analysis of
  best fit bin packing.
\newblock {\em J. Algorithms}, 27:218--235, 1998.
\newblock Preliminary version under the same title appeared in {\em Proc.
  Seventh Annual ACM-SIAM Symposium on Discrete Algorithms}, pages 351--358,
  1996.

\bibitem[Rhe88]{Rhe88}
W.~T. Rhee.
\newblock Optimal bin packing with items of random sizes.
\newblock {\em Math. Oper. Res.}, 13:140--151, 1988.

\bibitem[RT93a]{RT93b}
W.~T. Rhee and M.~Talagrand.
\newblock On line bin packing with items of random size.
\newblock {\em Math. Oper. Res.}, 18:438--445, 1993.

\bibitem[RT93b]{RT93c}
W.~T. Rhee and M.~Talagrand.
\newblock {On line bin packing with items of random sizes -- II}.
\newblock {\em SIAM J. Comput.}, 22:1251--1256, 1993.

\bibitem[Vai89]{V89}
P.~M. Vaidya.
\newblock Speeding-up linear programming using fast matrix multiplication.
\newblock In {\em Proceedings, The 30th Annual Symposium on Foundations of
  Computer Science}, pages 332--337, Los Alamitos, CA, 1989. IEEE Computer
  Society Press.

\bibitem[{Val}99]{Carvalho99}
J.~M. {Val\'{e}rio de Carvalho}.
\newblock Exact solutions of bin-packing problems using column generation and
  branch and bound.
\newblock {\em Annals of Operations Research}, 86:629--659, 1999.

\bibitem[vV92]{Vli92a}
A.~van Vliet.
\newblock An improved lower bound for on-line bin packing algorithms.
\newblock {\em Inf. Proc. Lett.}, 43:277--284, 1992.

\bibitem[Wil]{dbw96}
D.~B. Wilson.
\newblock Personal communication, 2000.

\end{thebibliography}
}

\end{document}